\documentclass[a4paper,11pt]{article}
\pdfoutput=1
\usepackage[T1]{fontenc}
\usepackage{xcolor}
\usepackage{jheppub}

\usepackage{graphicx}
\usepackage[utf8]{inputenc}
\usepackage{epsfig}
\usepackage[inline]{enumitem}
\usepackage{mathrsfs}
\usepackage{hyperref}
\usepackage{breakurl}
\usepackage{tensor}
\usepackage{bm}
\usepackage{xspace}
\usepackage{color}
\usepackage{tikz}
\usepackage{subfig}
\usepackage{cleveref}

\usepackage{orcidlink}

\usetikzlibrary{decorations}
\usetikzlibrary{trees}
\usetikzlibrary{decorations.pathmorphing}
\usetikzlibrary{decorations.markings}
\usetikzlibrary{external}
\usetikzlibrary{intersections}
\usetikzlibrary{shapes,arrows}
\usetikzlibrary{arrows.meta}
\usetikzlibrary{calc}
\usetikzlibrary{shapes.misc}
\usetikzlibrary{decorations.text}
\usetikzlibrary{backgrounds}
\usetikzlibrary{tikzmark,calc,arrows,shapes,decorations.pathreplacing}

\tikzset{
	graviton/.style={decorate,line width=0.25mm, decoration={snake,amplitude=.5mm, segment length=2mm}},
	massive/.style={postaction={decorate},
		line width=0.4mm,
	},
}

\newcommand{\ThreePointAmp}{
\begin{tikzpicture}
    \pgfmathsetmacro{\l}{2.5};
    \draw [massive] (0,0) node[below=0pt]{$p_1,S,K$} -- (\l,0);
    \draw [graviton] (\l/2,0) -- (\l/2,\l/2) node[above=0pt]{$\kk_3,\pol_3$};
    \filldraw [fill=gray!50!white] (\l/2,0) circle (0.25cm);
\end{tikzpicture}}
\newcommand{\FourPointAmp}{
\begin{tikzpicture}
    \pgfmathsetmacro{\l}{2.5};
    \draw [massive] (0,0) node[below=0pt]{$p_1,S,K$} -- (\l,0);
    \draw [graviton] (\l/2,0) -- ++(135:\l/2) node[above=0pt]{$\kk_3,\pol_3$};
    \draw [graviton] (\l/2,0) -- ++(45:\l/2) node[above=0pt]{$\kk_4,\pol_4$};
    \filldraw [fill=gray!50!white] (\l/2,0) circle (0.25cm);
\end{tikzpicture}}
 
\makeatletter
\def\simgt{\mathrel{\lower2.5pt\vbox{\lineskip=0pt\baselineskip=0pt
           \hbox{$>$}\hbox{$\sim$}}}}
\def\simlt{\mathrel{\lower2.5pt\vbox{\lineskip=0pt\baselineskip=0pt
           \hbox{$<$}\hbox{$\sim$}}}}
\makeatother

\crefname{equation}{Eq.}{Eqs.}
\crefname{figure}{Fig.}{Figs.}

\crefname{section}{Sec.}{Secs.}

\def\sect#1{Sec.~\ref{#1}}
\def\instate{\mathrm{i}}
\def\outstate{\mathrm{f}}

\def\com{COM\xspace}
\def\cco{\hyperlink{targetCCO}{CCO}\xspace}
\def\gco{\hyperlink{targetGCO}{GCO}\xspace}

\def\ccos{\hyperlink{targetCCO}{CCOs}\xspace}
\def\gcos{\hyperlink{targetGCO}{GCOs}\xspace}

\def\covM{\mathcal{M}}
\def\canM{\mathscr{M}}
\def\wpp{w_{++}}
\def\wpm{w_{-+}}
\def\ypp{y_{++}}
\def\ypm{y_{-+}}

\def\yy{\sigma}

\def\gS{\mathsf{S}}

\def\qftK{K}
\def\kK{\mathsf{k}}

\def\qftKrf{\bm{K}}

\def\qftS{S}
\def\aa{\mathsf{a}}

\def\qftSrf{\bm{S}}

\def\clS{\mathbf{S}}

\def\clK{\mathbf{K}}

\def\covclK{\mathrm{K}}
\def\covclk{\mathrm{k}}

\def\rl{\mu}
\def\rls{\tilde\alpha}

\def\eqns#1.#2{Eqs.~\eqref{#1} and~\eqref{#2}}
\def\spa#1.#2{\left\langle#1\,#2\right\rangle}
\def\spb#1.#2{\left[#1\,#2\right]}
\def\sand#1.#2.#3{%
\left\langle#1{\vphantom1}\right|{#2}\left|#3\right]}%
\def\sandmp#1.#2.#3{%
\left\langle#1{\vphantom1}\right|{#2}\left|#3\right]}%
\def\sandpm#1.#2.#3{%
\left[#1{\vphantom1}\right|{#2}\left|#3\right\rangle}%
\def\sandmm#1.#2.#3{%
\left\langle#1{\vphantom1}\right|{#2}\left|#3\right\rangle}%
\def\sandpp#1.#2.#3{%
\left[#1{\vphantom1}\right|{#2}\left|#3\right]}%

\def\LL{\mathcal{L}}
\def\phis{\phi_{s}}
\def\CC{C_{2}}
\def\df{\tilde{f}}
\def\PLS{\mathbb{S}}

\def\order{{\cal O}}

\def\hdelta{{\hat\delta}}

\def\nn{\nonumber}

\def\pol{\epsilon}
\def\polM{\mathcal{E}}

\def\kk{k}

\newcommand{\al}[3]{\alpha_{{#1}}^{\{{#2},{#3}\}}}
\newcommand{\aC}[3]{a_{{#1},\textrm{can}}^{\{{#2},{#3}\}}}

\def\opH{\mathcal{H}}
\def\opp{\bm p}
\def\opr{\bm r}
\def\opL{\left(\opr \times \opp\right)}
\def\Lq{i \left(\opp \times \clq\right)}
\newcommand{\spinStr}[1]{\Sigma_{#1}}

\def\ampEFT{\mathbb{M}}
\newcommand{\en}[1]{\mathsf{E}_{#1}}

\def\doe{\partial}
\def\bs{\boldsymbol}
\newcommand{\bsh}[1]{\hat{\bs #1}}

\def\clp{\bm p}

\def\clq{\bm q}
\def\pder{\mathcal{D}}

\newcommand{\sym}[1]{\{#1\}}

\def\nn{\nonumber}

\newcommand{\be}{\begin{equation}}
\newcommand{\ee}{\end{equation}}

\renewcommand{\imath}{\mathrm{i}}

\newcommand{\PM}[1]{{G^{#1}}}

\def\topbotatom#1{\hbox{\hbox to 0pt{$#1\bot$\hss}$#1\top$}}

\def\thirdMass{m_2}
\def\boundMassTotal{m_1}
\def\boundMassReduced{\mu_B}
\def\boundMassOne{m_{B, 1}}
\def\boundMassTwo{m_{B, 2}}
\def\thirdMomentumVector{{\bm p}_2}

\begin{document}

\title{
    Observables
    and Unconstrained Spin Tensor Dynamics in General Relativity from Scattering Amplitudes
}

\author[a]{Mark Alaverdian,}
\emailAdd{mxa895@psu.edu}
\affiliation[a]{Institute for Gravitation and the Cosmos,
	Pennsylvania State University,
	University Park, PA 16802, USA}
\author[b]{Zvi~Bern,}
\emailAdd{bern@physics.ucla.edu}
\affiliation[b]{
	Mani L. Bhaumik Institute for Theoretical Physics,
	University of California at Los Angeles,
	Los Angeles, CA 90095, USA}
\author[c]{Dimitrios~Kosmopoulos\orcidlink{0000-0001-9976-3435},}
\emailAdd{dimitrios.kosmopoulos@unige.ch}
\affiliation[c]{
         D\'epartement de Physique Th\'eorique, Universit\'e de Gen\`eve, 24 quai E. Ansermet, CH-1211 Geneva, Switzerland
}		
\author[d]{Andres~Luna,}
\emailAdd{andres.luna@nbi.ku.dk}
\affiliation[d]{
	Niels Bohr International Academy,
	Niels Bohr Institute, University of Copenhagen,
	Blegdamsvej 17, DK-2100, Copenhagen \O , Denmark}
\author[a,e]{Radu~Roiban,}
\emailAdd{radu@phys.psu.edu}
\affiliation[e]{Institute for Computational and Data Sciences,
Pennsylvania State University,
University Park, PA 16802, USA}
\author[f]{Trevor~Scheopner,}	
\emailAdd{trevor.scheopner@aei.mpg.de}
\affiliation[f]{
	Max Planck Institute for Gravitational Physics (Albert Einstein Institute), Am M\"uhlenberg 1, Potsdam D-14476, Germany}
\author[g]{Fei~Teng}
\emailAdd{f\_teng@fudan.edu.cn}
\affiliation[g]{Department of Physics and Center for Field Theory and Particle Physics, Fudan University, Shanghai 200438, China}

\abstract{

In a previous Letter, we showed that physical scattering observables for compact spinning objects in general relativity can depend on additional degrees of freedom beyond those described by the spin vector, which can be packaged into an unconstrained spin tensor. 
In this paper, we provide further details on the physics of these additional degrees of freedom, whose commutation relations and Poisson brackets are inherited from the underlying Lorentz symmetry, and on their consequence on observables. 
In particular, we give the waveform at leading order in Newton's constant and up to second order in the components of the spin tensor, and the conservative impulse, boost and spin kick, exhibiting spin magnitude change, through next-to-leading-order in Newton's constant and third order in the components of the spin tensor.
We provide explicit examples---a Newtonian two-particle bound system and a certain black-hole solution in an exotic matter-coupled gravitational theory---that exhibit these degrees of freedom and are described by our four-dimensional and worldline field theories. We also discuss connections between these degrees of freedom and dynamical worldline multipole moments.
We construct effective two-body Hamiltonians, we demonstrate explicitly that the extra degrees of freedom beyond the spin vector are necessary to describe the complete dynamics, and we explicitly remove certain unphysical singularities. 
Moreover, we show that the previously proposed eikonal (or radial action) formula correctly captures observables derived from the classical Hamiltonian.
Finally, we comment on possible descriptions of the additional degrees of freedom from the perspective of Goldstone's theorem. 
}

\maketitle

\section{Introduction}

The detection of gravitational waves by the LIGO/Virgo collaboration~\cite{LIGOScientific:2016aoc, LIGOScientific:2017vwq} promises major new discoveries in astronomy, cosmology, and potentially even particle physics. As gravitational-wave detectors continue to enhance their sensitivity~\cite{Punturo:2010zz, LISA:2017pwj, Reitze:2019iox}, the role of the spin of compact astrophysical objects, along with its complex three-dimensional dynamics, will become increasingly critical for accurate signal identification and interpretation. For further details, see Refs.~\cite{Blanchet:2006zz, Porto:2016pyg, Levi:2018nxp, Buonanno:2022pgc} and references therein.

The study of spinning systems in general relativity has a long history (see, e.g., Refs.~\cite{Lense:1918zz, Papapetrou:1951pa, Mathisson:1937zz, Dixon:1970zza}). In traditional worldline approaches~(see, e.g.,~Refs.~\cite{Porto:2005ac,  Steinhoff:2015ksa, Porto:2006bt, Faye:2006gx, Damour:2007nc, Levi:2015msa, Liu:2021zxr}) as well as in worldline quantum field theory~\cite{Jakobsen:2021zvh}, a spin supplementary condition (SSC)~\cite{Fleming} identifies the spin vector as the only physical degrees of freedom in the spin tensor and restricts its interactions. The SSC has been interpreted~\cite{Steinhoff:2015ksa, Levi:2015msa, Vines:2016unv} in terms of a spin-gauge symmetry encoding the freedom to locally shift the worldline in the ambient space. Imposing an SSC while assuming a variational principle in terms of only the worldline trajectory, linear momentum, and spin vector leads to a Hamiltonian that depends solely on those minimal degrees of freedom, all of which commute with the magnitude of the spin and therefore keep it fixed under conservative evolution.

In a previous Letter~\cite{Alaverdian:2024spu}, we showed that physical scattering observables for compact spinning objects in general relativity can depend on the additional degrees of freedom in the spin tensor beyond those described by the spin vector alone.  This conclusion emerges from resolving a puzzle in the scattering-amplitude-based quantum-field-theory (QFT) approach to spin as formulated in Ref.~\cite{Bern:2020buy}.  Calculations within this framework, beginning at the third order in spin, exhibited the unexpected feature that physical observables, such as the scattering angle, nontrivially depend on a larger number of independent Wilson coefficients and associated operators than those present in conventional worldline descriptions where a spin supplementary condition is imposed~\cite{Bern:2022kto}. The description with the extra Wilson coefficients is built according to the basic principles of effective field theory (EFT) to be the most general one that captures any system subject to general covariance, the desired separation of scales, and the assumption of conservative dynamics. A key observation in  
Ref.~\cite{Bern:2022kto} is that one can match the standard worldline description with an appropriate choice of Wilson coefficients, i.e., that the QFT construction contained the standard worldline description as a subset.   However, this left unresolved the question regarding the origin of the additional Wilson coefficients and whether meaningful physical systems could actually be described within this extended framework,  with choices of Wilson coefficients leading to dynamics beyond the conventional worldline system where an SSC is imposed.   
In both worldline and QFT formalisms, the Wilson coefficients characterize the interactions of gravity with compact objects, with different physical compact objects corresponding to distinct sets of Wilson coefficients. Consequently, the emergence of these extra Wilson coefficients can be viewed within the broader question of identifying the interactions that are consistent with Lorentz and general coordinate invariance, given the relevant separation of scales and where absorptive effects are not included.

To investigate the origin of the additional independent Wilson coefficients, Ref.~\cite{Bern:2023ity} examined electrodynamics, which is a simpler theory to analyze than general relativity because the Wilson coefficients appear at lower orders in spin. The study employed various variants of three distinct formalisms: field theory, worldline methods, and an effective two-body Hamiltonian approach. Several insights emerged from this investigation: 
\begin{enumerate*}[label=\textbf{(\arabic*)}]
\item \hypertarget{targetCCO}{Imposing the spin supplementary condition on the worldline can be 
realized
in field theory by enforcing that only a single quantum spin $s$ propagates. We refer to such a system as a \emph{conventional compact object} (CCO), since it is standard practice to impose an SSC.}
\item \hypertarget{targetGCO}{Allowing multiple quantum spins to propagate in field theory corresponds to a worldline theory with an extra vector degree of freedom and thereby leads to the additional independent operators and associated Wilson coefficients. We refer to such a system as a \emph{generic compact object} (GCO).} It is possible to describe this object in terms of a conventional SSC satisfying worldline that carries nonmininal degrees of freedom~\cite{Dixon:1970zza}. Here, instead, we combine an extra vector degree of freedom with the spin vector into an unconstrained spin tensor that no longer satisfies an SSC.
\item Results for a \gco, obtained from a worldline theory lacking an SSC or a field theory with multiple propagating spins, can be reduced to corresponding ones for a \cco by selecting specific values for the Wilson coefficients. As a particular example, this was demonstrated for the case of the root-Kerr charge distribution in electrodynamics.
\item For more general choices of Wilson coefficients, the resulting dynamics reveal additional intrinsic ``boost'' degrees of freedom. The corresponding two-body Hamiltonian that aligns with this extended theory explicitly incorporates these extra degrees of freedom beyond those described by the spin vector in a way consistent with the underlying Lorentz algebra. 
\item A key characteristic of the conventional worldline theory with the SSC imposed, of the field theory with a single propagating spin state, and of the Hamiltonian that depends on the spin vector alone is that the magnitude of the spin vector is conserved. For a \gco, where these restrictions are relaxed, only the magnitude of the spin tensor is conserved.
\end{enumerate*}
Ref.~\cite{Alaverdian:2024spu} reached similar conclusions 
for gravitationally-interacting spinning particles through $\order(G^2)$ and second order in the spin tensor. That study also demonstrated that the waveform generated by the system exhibits a nontrivial dependence on the additional degrees of freedom when the Wilson coefficients are generic.  

If we decompose the spin tensor, $\gS^{\mu\nu}$, into a spin vector, $\qftS^\mu$, and an intrinsic boost vector, $\qftK^\mu$, the usual covariant SSC sets this boost vector to zero.  A system without an SSC imposed has additional degrees of freedom captured by $\qftK^\mu$.\footnote{As we discussed in Ref.~\cite{Alaverdian:2024spu} and we review in \cref{sec:basics}, the three-dimensional vectors $\qftSrf$ and $\qftKrf$ are related to $\qftS^\mu$ and $\qftK^\mu$ as described below \cref{eq:gStensor}, while $\clK$ is given in terms of $\qftKrf$ via the 
identification explained below \cref{eq:SpinTensorConservation}. In the effective Hamiltonian, we use $\clS = \qftSrf$ for notational consistency.}
We investigate what these additional degrees of freedom correspond to in a physical system in two explicit examples: 
\begin{itemize}
    \item \textbf{A spinless particle scattering off a bound state.} Taking the inner bound system as Newtonian and the effect of the fly-by of the third body on the bound system as small and treatable in perturbation theory, we can relate $\bm K$ with the Laplace-Runge-Lenz vector, which points along the semi-major axis of the ellipse and is proportional to the eccentricity of the ellipse.  The fly-by of the third object can change both $\bm S$ and $\bm K$. 
    \item \textbf{A spinless particle scattering off the Rasheed-Larsen black hole~\cite{Rasheed:1995zv,Larsen:1999pp}.} This black hole has a qualitatively different multipolar structure than that of the Kerr black hole~\cite{Bena:2020uup}. We find that we can capture this structure in terms of combinations of $\clS$ and $\clK$.
\end{itemize}
These examples reveal two aspects of the interpretation of $K$: 
Firstly, $K$ describes light degrees of freedom in the system, which can evolve without changing the energy.
Secondly, $K$ is related to certain multipole moments. In particular, the $K$ vector induces additional quadrupolar (and also higher multipolar) couplings that are linear in the Riemann tensor, including also parity-odd contributions. Such interactions could be added by hand~\cite{Porto:2005ac, Steinhoff:2021dsn}, but a key difference here is that the underlying Lorentz algebra constrains them.

We extend the results of Ref.~\cite{Alaverdian:2024spu}: 
We present the field-theory Lagrangian that describes the most general gravitational interactions of a \gco through 4\textsuperscript{th} power in the spin tensor. From this field theory, we obtain gravitational Compton amplitudes for a \gco through $\order(G \, \gS^4)$; we verify that they reduce to corresponding expressions for a compact object with an SSC imposed~\cite{Haddad:2023ylx} (see also Refs.~\cite{Ben-Shahar:2023djm, Scheopner:2023rzp}) and for a Kerr black hole~\cite{Levi:2015msa, Vines:2017hyw, Chen:2021kxt, Bern:2022kto, Bautista:2022wjf, Cangemi:2023bpe, Bohnenblust:2024hkw}, under appropriate restrictions on the Wilson coefficients. We provide these Compton amplitudes in the ancillary files~\cite{anc}. We similarly expand on the extended gravitational worldline with no imposed SSC. This worldline formulation allows for the computation of a gravitational Compton amplitude that matches the one coming from field theory.
We extend our Hamiltonian construction through the third power in $\clS$ and $\clK$ and connect it to our field theory by matching the corresponding two-body scattering amplitudes. We solve Hamilton's equations to obtain the momentum impulse and the variations of $\clS$ and $\clK$ during a scattering event.
An alternative way for doing so at $\order(G^2)$ is in terms of the eikonal phase~\cite{Bern:2020buy, Gatica:2023iws, Bern:2023ity, Luna:2023uwd, Alaverdian:2024spu, Gatica:2024mur}. We verify that the impulse and change in the spin tensor derived from the eikonal formulas from both Refs.~\cite{Bern:2023ity, Alaverdian:2024spu} and Ref.~\cite{Luna:2023uwd}, match the observables obtained from the two-body Hamiltonian through $\order(G^2 \clS^a \clK^b)$ with $a+b=3$.

To further display the physical effect of the $K$ vector, we analytically compute the  $\order(G^2\, \gS^2)$ 
gravitational waveform emitted in a scattering event between a \gco and a Schwarzschild black hole. This waveform is directly related to the Fourier transform of the corresponding five-point tree-level scattering amplitude with a single outgoing graviton~\cite{Cristofoli:2021vyo}. To recognize the inequivalence between the waveforms sourced by \ccos and \gcos, we find it convenient to compare the so-called gravitational memory, which follows from the soft limit of the five-point amplitude~\cite{Saha:2019tub, Sahoo:2021ctw}.
As demonstrated by the numerical analysis of Ref.~\cite{Alaverdian:2024spu}, if we do not decouple the dynamics of $K$ by choosing special values for the Wilson coefficients, the effect of $K$ is encoded in the gravitational waveform.
Ref.~\cite{Alaverdian:2024spu} also showed that, as long as each order in the spin tensor can be treated separately, the above conclusion holds true even if redefinitions between $S$ and $K$ are allowed.
Here, we relax this assumption and also allow for redefinitions mixing different orders in the spin tensor; we again confirm that the extra degrees of freedom lead to measurable effects.
We conclude that only for special values of Wilson coefficients do the extra degrees of freedom decouple, recovering the standard results with the SSC imposed~\cite{Steinhoff:2015ksa, Levi:2015msa, Vines:2016unv}.

Finally, we comment on the appearance of zero-velocity singularities in the two-body Hamiltonian. We show that selecting a minimal set of terms for the Hamiltonian can result in coefficients that exhibit spurious zero-velocity singularities. These singularities can be interpreted as gauge artifacts, as they can be easily eliminated by adding redundant terms to the Hamiltonian. In contrast, when attempting to match the dynamics of a \gco to a Hamiltonian without the dynamical $\clK$ degrees of freedom, we find that the Hamiltonian coefficients exhibit physical zero-velocity singularities. This provides further evidence that the dynamics of $\clK$ must be considered for the accurate modeling of such systems.

This paper is organized as follows:
In \sect{sec:basics}, we give the basic QFT setup, explaining how the classical spin tensor emerges in a form that makes it unnatural to impose an SSC. This section also comments on the connection of an SSC in the classical limit and the restriction that only quantum states of a single spin magnitude exist in the theory, or that transitions between states of different spin magnitude are not allowed. 
In \sect{sec:toymodel}, we analyze two physical systems that exhibit both a $K$ and an $S$ vector:
A bound Newtonian two-body system and the Rasheed-Larsen black hole.
In \sect{sec:HigherOrderFT}, we expand on the gravitational interactions of a \gco by providing the effective Lagrangian through the 4\textsuperscript{th} power in the spin tensor and generating Compton amplitudes through $\order(G \,\gS^4)$. 
In \sect{sec:wl}, we describe the worldline theory without an SSC imposed, corresponding to the above field theory. 
In \sect{sec:waveform}, we derive and study gravitational waveforms, showing that the contribution from the dynamics of $K$ cannot be reabsorbed into redefinitions of the Wilson coefficients of systems where $K$ decouples.
In \sect{sec:TwoBodyAmplitude} we provide two-body amplitudes and explicitly confirm formulas for extracting observables from the eikonal phase through $\order(G^2 \clS^a \clK^b)$ with $a+b=3$.
In \sect{sec:Hamiltonian} we derive a two-body Hamiltonian that matches to both the extended field theory and the worldline theory with the extra $K$ degrees of freedom. We obtain the Hamiltonian in a gauge with a minimal number of terms in momentum space and in a second gauge where the coefficients are free of spurious zero-velocity singularities. 
Finally, in \sect{sec:conclusion}, we present our conclusions and outlook.  
We include two appendices, one containing our conventions and the other the four-point $R^2$ contact interactions involving the $K$ degrees of freedom.


\section{Basic field theory setup and summary of previous results}
\label{sec:basics}

Our basic field-theory setup follows the approach of Refs.~\cite{Bern:2020buy,Kosmopoulos:2021zoq, Bern:2022kto, Bern:2023ity, Alaverdian:2024spu}, which we summarize below.  
This construction includes only conservative dynamics, and therefore, the spin-vector magnitude change is due to conservative effects.  For further details on the formalism, the reader may consult  Refs.~\cite{Bern:2020buy, Bern:2023ity}. 
Radiative and absorptive effects can be included along the lines of Refs.~\cite{Jones:2023ugm, Chen:2023qzo, Aoude:2024jxd, Bautista:2024emt}.

In this formalism, the interactions are categorized by local operators containing a symmetric and traceless tensor field $\phis$ of arbitrary rank, used to describe the spinning objects, providing a natural way to organize the interactions. The minimal interaction of a spinning field with Einstein's gravity is\footnote{We use the mostly-minus $(+\!-\!--)$ signature throughout. We suppress an overall $(-1)^s$ factor in the Lagrangian that makes the norm of the spin-$s$ component positive. Namely, throughout the paper our Lagrangian is in fact $(-1)^s\mathcal{L}$.} 
\begin{align}\label{eq:Lmin}
    \LL_{\rm min} &= - \frac{1}{2}\phis(\nabla^2+m^2)\phis + \frac{H}{8}R_{abcd}\phis M^{ab}M^{cd}\phis \,,
\end{align}
where we are using a local frame defined by the vierbein $e^{a}_{\mu}\,$, $M^{ab}$ are the Lorentz generators acting on the $\phis$, 
and $\nabla_{c}\phi_s\equiv e_c^\mu(\partial_{\mu}\phi_s+\frac{i}{2}\omega_{\mu ab}M^{ab}\phi_s)$
with the spin connection $\omega_{\mu ab}$. The exact definitions for these quantities are given in \cref{sec:conventions_appendix}. 
The simple form of the kinematic term in this Lagrangian allows multiple spin states to propagate, some of which are ghost states with negative norms. While such ghost states would be problematic for defining a complete quantum theory of arbitrary spin, this Lagrangian is meant to be used in the classical limit only.  
Ref.~\cite{Alaverdian:2024spu} compared results in the classical limit and concluded that this system is equivalent to a technically more complicated setup where only positive-norm states of a given spin magnitude propagate and transitions between states of different spin magnitude are allowed.
Other amplitudes-based approaches to spin may be found in, for example, Refs.~\cite{Holstein:2008sx, Vaidya:2014kza, Arkani-Hamed:2017jhn, Guevara:2018wpp, Chung:2018kqs, Guevara:2019fsj, Maybee:2019jus, Chung:2019duq,  Damgaard:2019lfh, Kosmopoulos:2021zoq,
Aoude:2022thd, Cangemi:2022bew, FebresCordero:2022jts, Aoude:2023vdk, Cangemi:2023ysz, Cangemi:2023bpe, Akpinar:2024meg, Akpinar:2025bkt}. 

In addition to the minimal interactions, we need nonminimal interactions to describe compact spinning objects, including Kerr black holes.  These additional contributions are organized by increasing powers of Lorentz generators.  The lowest of these start at $\order(M_{ab}^2)$, and are given by
\begin{align}
\label{eq:ExtraInteractions}
    \LL_{\text{non-min}} & = - \frac{\CC}{2m^2} R_{a f_1 b f_2}\nabla^a\phis \PLS^{(f_1}\PLS^{f_2)}\nabla^b\phis  + \frac{D_2}{2m^2}R_{abcd} \nabla_{i}\phis\sym{M^{ai}M^{cd}}\nabla^b\phis\\[3pt]
    & \quad + \frac{E_2-2D_2}{2m^4}R_{abcd}\nabla^{(a}\nabla^{i)}\phis\sym{M\indices{^b_i}M\indices{^d_j}}\nabla^{(c}\nabla^{j)}\phis\,,\nonumber
\end{align}
where $\PLS^{a} \equiv {-i}\varepsilon^{abcd}M_{bc}\nabla_{d}/(2m)$.\footnote{The Levi-Civita tensor is defined with $\varepsilon^{0123}=1$.}
In the field-theory formalism, it is natural to include all interactions pertinent to 
classical physics. The four Wilson coefficients $(H, C_2, D_2, E_2)$ appear in only three distinct combinations in physical observables, which allows us to fix one of them to an arbitrary value. We choose to set $H=1$, leaving the other three coefficients as the independent ones.
The specific choice of $E_2$ and $D_2$ follows from Ref.~\cite{Alaverdian:2024spu}, aiming to streamline the amplitudes and observables.   It is also possible to introduce higher-dimension operators that involve more Lorentz generators. In \cref{sec:HigherOrderFT},  we discuss interactions up to the 4\textsuperscript{th} order in the Lorentz generators.  

To apply quantum-field-theory methods in the classical limit, we need a dictionary to interpret the expectation values of Lorentz generators in terms of the classical spin of particles. Here, we follow Refs.~\cite{Bern:2023ity, Alaverdian:2024spu}, which distinguishes between the spin tensor and the spin vector.   In usual formulations, these are taken to be equivalent, and indeed, all evidence points to this being true for interacting Kerr black holes. On the other hand, for generic compact objects with no special constraints on Wilson coefficients, Ref.~\cite{Alaverdian:2024spu} demonstrated that observables, including waveforms, depend on the details of the treatment of the degrees of freedom in the spin tensor.   In field theory, this difference is captured by whether a single quantum spin is used to model the classical object or a combination of quantum spins is employed.  In the worldline language, this difference corresponds to whether or not the SSC is imposed. 

In the field-theory formulation, the spin tensor emerges as the expectation value of the Lorentz generator in the spin coherent states~\cite{Bern:2023ity} describing the asymptotic states. We may represent a spin coherent state with momentum $p$ by the normalized polarization tensor $\polM(p)$, such that, 
\begin{align}
\label{eq:generalSa}
\polM(p) \cdot \{M^{\mu_1\nu_1 } , \dots , M^{\mu_n\nu_n } \} \cdot \polM(p)
=   \gS(p)^{\mu_1\nu_1}\dots \gS(p)^{\mu_n\nu_n} \,.
\end{align}
The notation $\{ \ldots \}$ signifies the symmetric product of the Lorentz generators weighed by the number of terms.
Antisymmetric combinations of the spin tensor are reduced to terms with fewer Lorentz generators using the Lorentz algebra. 
Furthermore, the factorization of the expectation value of the product of Lorentz generators into the product of individual expectation values in \cref{eq:generalSa} reflects the classical nature of the asymptotic states $\polM$.
Without loss of generality, the spin tensor $\gS^{\mu\nu}$ in four dimensions may be decomposed into a spin vector and a boost vector,\footnote{
The relation between spin tensor and spin vector used here differ by a sign from the one used in Ref.~\cite{Bern:2020buy}, which is accounted for by the different convention of the Levi-Civita tensor.} 
\begin{align}
\gS(p)^{\mu\nu} &= - \frac{1}{m}\varepsilon^{\mu\nu\rho\sigma}p_\rho \qftS_\sigma + \frac{i}{m}(p^{\mu} \qftK^{\nu}-p^{\nu} \qftK^{\mu}) \nonumber\\
&= - \frac{1}{m}\varepsilon^{\mu\nu\rho\sigma}p_\rho \qftS_\sigma + \frac{1}{m}(p^{\mu} \covclK^{\nu}-p^{\nu} \covclK^{\mu}) \,,
\label{eq:gStensor}
\end{align}
where $\covclK^{\mu}=i\qftK^{\mu}$. The spin tensor in a generic frame can be obtained by boosting the rest-frame spin tensor,
\begin{align}
\gS(p)^{\mu\nu} = L(p_0,p)^{\mu}{}_{\rho}L(p_0,p)^{\nu}{}_{\lambda}\gS(p_0)^{\rho\lambda}\,, \qquad \gS(p_0)^{\mu\nu} = \polM(p_0) \cdot M^{\mu\nu} \cdot \polM(p_0) \,,
\end{align}
where $p_0=(m,0,0,0)$ and $L(p_0,p)$ is the Lorentz boost from $p_0$ to a generic momentum $p$.
The four-vectors $\qftS$, $\qftK$ and $\covclK$ in \cref{eq:gStensor} are related to their rest frame counterpart $\qftS_0=(0,\qftSrf)$, $\qftK_0=(0,\qftKrf)$ and $\covclK_0=(0,\clK)$ by the same Lorentz boost.
Therefore, these two vectors automatically satisfy the transversality condition $p^{\mu}\qftS_{\mu}=p^{\mu}\qftK_{\mu}=0$, giving a total of six degrees of freedom, corresponding to the number of degrees of freedom in the spin tensor.  
We note that $\qftS$ and $\covclK$ are the expectation values of the Lorentz generators $M^{ij}$ and $M^{0i}$ in the spin coherent state.
In the spirit of the relation between quantum commutators and classical Poisson brackets, we endow $\qftS$ and $\covclK$ in \cref{eq:gStensor}  with Poisson brackets so that they also generate a Lorentz algebra.
It is, however, more natural to express results from field-theory calculations in terms of $\qftS$ and $\qftK$. When we construct the effective Hamiltonian, we 
identify $i \qftKrf \equiv \clK$, i.e., we make this replacement without changing the commutation relations of the algebra. We return to this point later in this section and also in subsequent sections.

We now consider a scattering process in which a spinning particle has an incoming particle with momentum $p_1$ and an outgoing particle with momentum $p_4$, and $q=-p_1-p_4$ as the momentum transfer. Note that we use the all-outgoing convention here. \Cref{eq:generalSa} generalizes to~\cite{Bern:2023ity}
\begin{align}
\label{eq:generalS}
& \polM_1 \cdot \{M^{\mu_1\nu_1 } , \dots , M^{\mu_n\nu_n } \} \cdot \polM_4
=   \gS_1^{\mu_1\nu_1}\dots \gS_1^{\mu_n\nu_n} 
\polM_1 \cdot \polM_4 + \order(q^{1-n}) \, ,
\end{align} 
where $\polM_i\equiv\polM(p_i)$ and $\gS_1\equiv\gS(p_1)$.\footnote{At one loop $\gS((p_1+p_4)/2)$ is a more convenient choice when matching the field-theory amplitudes with those obtained from the Hamiltonian. 
The difference between $\gS(p_1)$ and $\gS((p_1+p_4)/2)$ is not relevant at tree level.}
The terms of a higher order in $q$ are suppressed in the classical limit.
In \cref{eq:generalS}, the product of polarization tensors is given by 
\begin{align}
    \polM_1 \cdot \polM_4
    = \exp\left[\frac{1}{m_1} \bs{q}\cdot \qftKrf_1   \right]
    \exp\left[ -   \frac{ i (\bs{p}_1\times\bs{q})\cdot\qftSrf_1}{m_1(m_1 + \en{1})}+\order(q^2) \right] + \order(q) \, ,
    \label{epdotep}
\end{align}
where we choose $p_1^{\mu}=-(\en{1},\bs{p}_1)$ with $\en{1}=\sqrt{m_1^2+\bs{p}_1^2}$. Note that the $K$-dependent factor can be rewritten in the covariant form $\exp\left[-\frac{q\cdot K_1}{m_1}\right]$. 
The second exponential factor in \cref{epdotep} can be identified as the contribution from the fixed-spin polarization tensors~\cite{Chung:2018kqs},
\begin{align}
    \polM_1^{(s)} \cdot \polM_4^{(s)} = \exp\left[ -   \frac{ i (\bs{p}_1\times\bs{q})\cdot\qftSrf_1}{m_1(m_1 + E_1)} \right]\,.
\end{align}
\Cref{epdotep} captures the leading terms in the classical limit where we have dropped a possible overall sign that depends on the representation chosen for the fields (see Ref.~\cite{Bern:2023ity} for further details). The polarization product~\eqref{epdotep} can be factored out in the classical amplitudes,
\begin{align}\label{eq:ampSK}
\canM(\qftS_1,\qftK_1) = \polM_1 \cdot \polM_4\, \widetilde{\mathcal{M}}(\qftS_1,\qftK_1) = \exp\left[-\frac{q\cdot \qftK_1}{m_1}\right]
\exp\left[ - \frac{ i (\bs{p}_1\times\bs{q})\cdot\qftSrf_1}{m_1(m_1 + E_1)} \right]
\widetilde{\mathcal{M}}(\qftS_1,\qftK_1) \,, 
\end{align}
where $\widetilde{\mathcal{M}}$ is determined by the Lagrangian.
Under the Fourier transform to position space, these exponential factors introduce a shift to the particle's worldline that depends on the $S$ and $K$ vectors.
Throughout this paper, we expose the complete $K$ dependence from the polarization products. On the other hand, the shift introduced by the spin-dependent factor effectively switches between covariant and canonical spin.
The amplitude with covariant spin is thus given by 
\begin{align}\label{eq:M_def}
    \covM(\qftS_1,\qftK_1) \equiv \exp\left[-\frac{q\cdot\qftK_1}{m_1}\right] \widetilde{\mathcal{M}}(\qftS_1,\qftK_1) \,.
\end{align}
The spin vector in $\canM$ is then identified as the canonical spin. The amplitudes with canonical and covariant spin are related through
\begin{align}\label{eq:Mcan_def}
    \canM(\qftS_1,\qftK_1) = \polM_1^{(s)} \cdot \polM_4^{(s)} \covM(\qftS_1,\qftK_1) = 
    \exp\left[ - \frac{ i (\bs{p}_1\times\bs{q})\cdot\qftSrf_1}{m_1(m_1 + E_1)} \right]\covM(\qftS_1,\qftK_1) \,.
\end{align}
These relations apply to the scattering processes involving one spinning particle in the incoming and outgoing states, which is the case for the Compton amplitudes to be studied in \cref{sec:HigherOrderFT}.
We call $\covM$ the \emph{covariant-spin amplitude} and $\canM$ the \emph{canonical-spin amplitude}. 
The generalization of \cref{eq:M_def,eq:Mcan_def} to more spinning particles is straightforward. We study the two-body case in \cref{sec:TwoBodyAmplitude}.
The amplitudes computed from field theory are more naturally given as $\covM$, while $\canM$ is used to derive the canonical effective Hamiltonian through matching.

By imposing the covariant SSC, $p_\mu \gS^{\mu\nu} = 0$, the $K$ dependence  vanishes~\cite{Porto:2005ac,  Steinhoff:2015ksa, Porto:2006bt, Faye:2006gx, Damour:2007nc, Levi:2015msa, Liu:2021zxr, Jakobsen:2021zvh}, reducing the degrees of freedom to those of the spin vector. Here, this follows from \cref{eq:generalS} and the transversality condition $p^{\mu}\qftK_{\mu}=0$. The ability to impose an SSC has been interpreted as a gauge freedom that amounts to a shift of the worldline~\cite{Steinhoff:2015ksa, Levi:2015msa, Vines:2016unv}.

On the other hand, the field-theory formulation does not refer to worldlines, nor is there a natural interpretation of an SSC acting directly on a Lorentz generator.   Indeed, a puzzle arose in Ref.~\cite{Bern:2022kto} by starting from the minimal Lagrangian \eqref{eq:Lmin} and following the usual procedure of adding higher-dimension operators consistent with Lorentz symmetry. Imposing the covariant SSC in the final results, attempting to match the worldline results, starting at $\mathcal{O}(S^3)$, one finds that the results deviate fundamentally so that even the number of independent Wilson coefficients is larger in field theory than in the worldline theory. 
Both the field-theory and the worldline formalisms reproduce the observables for Kerr black holes, at least through the explicitly checked orders, $\order(G^2\gS^3)$, indicating their consistency.
Indeed, more generally, in all cases where a corresponding worldline result is available, one could find a match to the field theory by choosing the additional Wilson coefficients appropriately.  This suggests that the basic field theory setup is self-consistent.

A helpful clue towards understanding the origin of a larger space of field-theory solutions is that if one uses projectors to force the propagating field to be of a single quantum spin, then the number of independent Wilson coefficients of the worldline theory agrees with that of the field theory~\cite{Kim:2023drc, Haddad:2023ylx}.  With a single quantum spin, the spin-vector length cannot change under conservative evolution, which agrees with the worldline situation where the SSC forces the spin-vector length to be constant under the dynamics.  However, this immediately leads to another puzzle: Why would one be forced to interpret a classical spin as a single quantum spin, rather than a superposition of quantum spins with transitions allowed between them?

A simpler setting to analyze the origin of this puzzle is electrodynamics. The basic puzzle is identical:  Even after imposing the SSC on the final results, the number of independent Wilson coefficients is larger in field theory than in the worldline theory.   In this simpler theory, Ref.~\cite{Bern:2023ity} tracked the origin of the additional Wilson coefficients using variants of three distinct formalisms: field theory, worldline, and an effective two-body Hamiltonian.   In Ref.~\cite{Alaverdian:2024spu}, the same study was carried out for gravity through $\order(G^2 \gS^2)$, with the addition of studying the consequences for waveforms. The conclusions from these papers through the checked orders are:
\begin{enumerate}
\item A quantum field theory with a single quantum spin-$s$ field propagating gives results matching conventional worldline constructions where the SSC is imposed. This case describes a \cco. 

\item A quantum field theory with multiple propagating quantum spins is equivalent to a \gco worldline theory with no imposed SSC.   These extended systems contain additional Wilson coefficients and associated operators. 

\item Without the SSC, the dynamics conserve the spin-tensor magnitude $\gS_{\mu\nu}\gS^{\mu\nu}$ so that, 
\begin{align}
\frac{d}{dt}\gS^{\mu\nu}\gS_{\mu\nu} = \frac{d}{dt}(-\qftS^2 - \qftK^2) = \frac{d}{dt}(\qftSrf^2 + \qftKrf^2)  = \frac{d}{dt}(\qftSrf^2 - \clK^2) = 0  \,,
\label{eq:SpinTensorConservation}
\end{align}
implying that the spin-vector magnitude is not conserved generically. The spin tensor $\gS$ is defined in \cref{eq:gStensor}, and $i \qftKrf \equiv \clK$ is the identification of $\qftKrf$ and $\clK$ that absorbs the $i$ in front of $\qftK$ in \cref{eq:gStensor} (as used in subsequent sections). 
We emphasize that we treat both $\clK$ and $\qftKrf$ as real. We further comment on this point in \cref{sec:conclusion}.
By imposing the covariant SSC on the worldline, $K$ vanishes, and the spin-vector magnitude conservation holds as usual. See Ref.~\cite{Bern:2023ity} for a discussion.  

\item For the scattering of a \gco and a scalar particle, two-body scattering amplitudes do not exhibit $\qftKrf$-dependent dipole terms. In impact-parameter space and through second order in $\gS$, their structure is
\begin{align}\label{eq:general_2body}
\covM^{\text{2 body}}  &= A_0 + A_1 \bm{L}\cdot\qftSrf + A_{2,1}\qftSrf^2 + A_{2,2}\qftKrf^2 + A_{2,3}\qftSrf\cdot\qftKrf + A_{2,4}(\bm{b}\cdot\qftSrf)^2 \nonumber\\
&\quad + A_{2,5}(\bm{p}\cdot\qftSrf)^2 + A_{2,6}(\bm{b}\cdot\qftKrf)^2 + A_{2,7}(\bm{p}\cdot\qftKrf)^2 + A_{2,8}(\bm{b}\cdot\qftSrf)(\bm{p}\cdot\qftSrf)\nonumber\\
&\quad + A_{2,9}(\bm{L}\cdot\qftSrf)(\bm{b}\cdot\qftKrf) + A_{2,10}(\bm{L}\cdot\qftSrf)(\bm{p}\cdot\qftKrf) + A_{2,11}(\bm{b}\cdot\qftSrf)(\bm{L}\cdot\qftKrf) \nonumber\\
&\quad + A_{2,12} (\bm{p}\cdot\qftSrf)(\bm{L}\cdot\qftKrf)  + A_{2,13}(\bm{b}\cdot\qftKrf)(\bm{p}\cdot\qftKrf) + \order(\gS^3)\,,
\end{align}
where the $A$'s are theory-dependent coefficients. 
We omit the subscripts in $\qftSrf$ and $\qftKrf$, since only particle 1 is a spinning object in this case.
Starting from $\order(\gS^2)$, the $A_{2,i}$ coefficients also depend on all the Wilson coefficients in the Lagrangian. See \cref{sec:TwoBodyAmplitude} for further details on the independent operators.
In both \cref{eq:general_2body} and \cref{sec:TwoBodyAmplitude}, we express the amplitudes in the center-of-mass frame, where $\bm{b}$ is the impact parameter, $\pm\bm{p}$ are the three-momenta of the scattering particles, and $\bm{L}=\bm{b}\times\bm{p}$ is the orbital angular momentum. 

\item For \gcos, physical observables, including waveforms, depend nontrivially on the additional Wilson coefficients of the extended worldline or field-theory systems. This implies that \ccos and \gcos are physically distinct.

\item  The effective two-body Hamiltonian that aligns with the \gco field theory or worldline systems includes the additional boost degrees of freedom, $\clK$, alongside the standard spin degrees of freedom comprised by $\clS$. The associated Poisson brackets are derived from the Lorentz algebra, as described in \cref{sec:Hamiltonian}.

\item By selecting specific values for the Wilson coefficients, the \gco worldline or field theories reduce to their \cco counterparts. This equivalence arises because the additional degrees of freedom nontrivially decouple under these special choices. Notable examples include Kerr black holes in general relativity~\cite{Alaverdian:2024spu} and the root-Kerr solution in electrodynamics~\cite{Bern:2023ity}.

\end{enumerate}

In summary, for generic values of the Wilson coefficients, the conservative system without enforcing the SSC exhibits physically distinct observables compared to the system where the SSC is imposed. The key difference lies in the fact that, in the former case, the magnitude of the spin vector is not conserved. We elaborate further on these points here, providing additional details complementary to the Letter~\cite{Alaverdian:2024spu} and explicitly demonstrate that the above conclusions hold up to $\order(G^2 \gS^3)$. As part of our analysis, we compute the gravitational Compton amplitude through $\order(G^2 \gS^4)$.


\section{Examples of physical systems exhibiting a \texorpdfstring{$K$}{K} vector}
\label{sec:toymodel}

To help explain the physical meaning of $K$, in this section, we describe 
examples of physical systems exhibiting such degrees of freedom. 
Our first example is that of a spinless particle scattering off a bound state. 
Three-body systems were discussed in the literature with effective field theory methods in e.g.~\cite{Jones:2022aji, Kuntz:2022onu, Loebbert:2020aos, Solon:2024zhr}. 
We demonstrate that it exhibits a variable analogous to the $K$ vector introduced in Ref.~\cite{Bern:2023ity, Alaverdian:2024spu} and reviewed above.
We demonstrate this by comparing a scattering-amplitude-like quantity in this model to the generic structure of amplitudes involving $S$ and $K$ vectors, as inferred from general principles and outlined in Eq.~\eqref{eq:general_2body} and further elaborated on in Sec.~\ref{sec:TwoBodyAmplitude}.
While this analysis focuses on the existence of the $K$ vector as a means for conservative changes in the length of spin, it also exposes more general requirements when this can occur.

The second example is that of a spinning black hole that also carries dilaton, electric, and magnetic charges, usually referred to as the Rasheed-Larsen black hole~\cite{Rasheed:1995zv, Larsen:1999pp}. By scattering an uncharged scalar probe particle off this metric and comparing with the general structure of the two-body amplitude~\eqref{eq:general_2body}, we demonstrate that this metric is indeed sourced by a stress tensor with multipoles depending on a vector with the same properties as our $K$ vector. 

\begin{figure}
    \centering
    \includegraphics[width=4in]{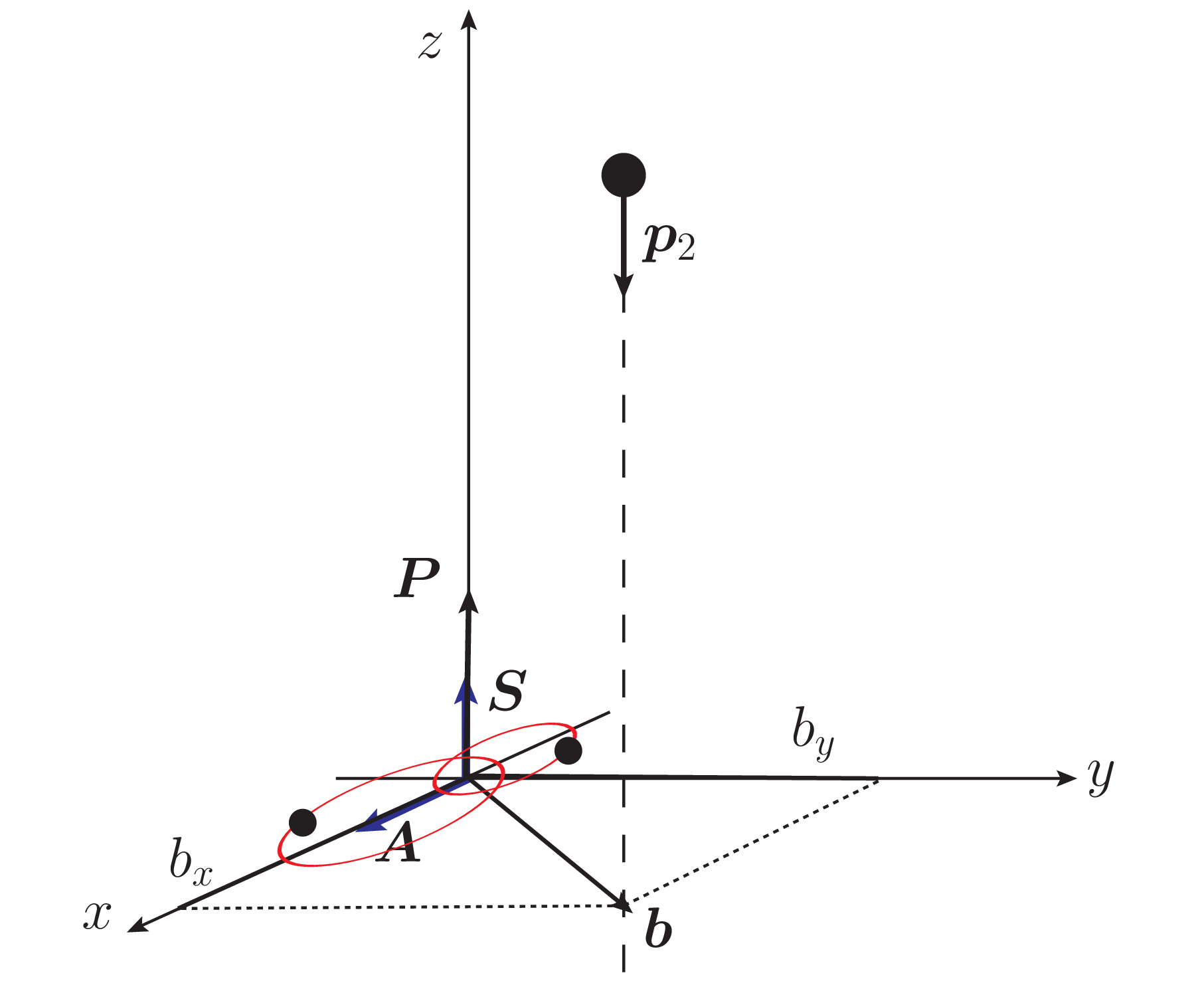}
    \caption{Scattering of a particle of momentum $\thirdMomentumVector$ off a two-body Newtonian bound state of internal angular momentum $\bm S$ and Laplace-Runge-Lenz vector $\bm A$ pointing along the $\hat {\bm x}$ axis. Both the particle and the bound state are moving along the $\hat {\bm z}$ direction.}
    \label{toymodelfigure}
\end{figure}

\subsection{Scattering off a Newtonian bound state}

To interpret $\bm K$, we consider an example of a three-body system as shown in \cref{toymodelfigure}, in which a body of mass $\thirdMass$ scatters off a bound two-body system with constituents of masses $\boundMassOne$ and $\boundMassTwo$ and total mass $\boundMassTotal=\boundMassOne+\boundMassTwo$.
We first study this system as a quantum-mechanical model and then analyze the classical limit of this quantum-mechanical system, assuming that we are in the regime where Newtonian gravity governs both the internal dynamics of the bound state and its interaction with the scattering (third) body (i.e., in the regime of large separation and small velocity); relativistic corrections for the latter can also be straightforwardly included, though we do not do so here.

\subsubsection{Quantum mechanical scattering of elliptic states \label{sec:elliptic}}

Assuming the size of the bound state is much smaller than its minimal separation from the scattering particle, the Hamiltonian can be naturally expressed as:
\begin{align}\label{eq:H_3b}
H &= \frac{\thirdMomentumVector^2}{2 \thirdMass} 
+ \frac{\bm P^2}{2 \boundMassTotal}
+H_0({\bm p}, {\bm r}) 
- \frac{ G \boundMassOne {\thirdMass}}{|\bm R+\frac{\boundMassTwo}{\boundMassTotal}\bm r|} 
- \frac{ G \boundMassTwo {\thirdMass}}{|\bm R-\frac{\boundMassOne}{\boundMassTotal}\bm r|} \,,
\end{align}
where $\bm P$ is the momentum of its center-of-mass,
$H_0$ is the Hamiltonian of the bound system in its center-of-mass frame, 
\begin{equation}\label{eq:H0}
H_0 = \frac{{\bm p}^2}{2\boundMassReduced} - \frac{G \boundMassTotal \boundMassReduced}{|\bm r|} \, ,
\end{equation}
and 
$\boundMassReduced=\boundMassOne\boundMassTwo/\boundMassTotal$ is the reduced mass of the bound system. The distances between the scattering particle with mass $\thirdMass$ and the two constituent particles in the bound system are $|\bm R+(\boundMassTwo/\boundMassTotal) \bm r|$ and $|\bm R-(\boundMassOne/\boundMassTotal) \bm r|$, where $\bm R$ is the distance between the scattering particle and the center-of-mass of the bound system. 
We identify the internal orbital angular momentum with the spin of the bound system. 

Because of the rotational symmetry of $H_0$, the energy levels of the bound system are degenerate, being independent of the projection of its spin on the spin-quantization axis, which we choose to be the $\hat {\bm z}$ axis. 
In our field-theory setup, an essential feature is that we are allowed to change the spin magnitude without changing the energy of the system. Its counterpart here is a system for which the energy levels are degenerate with respect to the orbital quantum number $l$. Thus, we choose the binding potential to be $V\propto 1/|\bm r|$ that exhibits such a degeneracy.

In the limit $|\bm R|\gg |\bm r|$ we may expand \cref{eq:H_3b} as
\begin{align}
H &= \frac{\thirdMomentumVector^2}{2 \thirdMass} + \frac{\bm P^2}{2\boundMassTotal} 
- \frac{ G \boundMassTotal {\thirdMass}}{|\bm R|} +  H_0({\bm p}, {\bm r}) 
- \frac{3 G \boundMassReduced {\thirdMass}}{2|\bm R|^5} 
\left( (\bm r\cdot\bm R)^2 - \frac{1}{3}|\bm r|^2 |\bm R|^2\right) \nonumber\\
&\quad -\frac{1}{2}\sqrt{1-4\frac{\boundMassReduced}{\boundMassTotal}} \frac{G\boundMassReduced \thirdMass}{|{\bm R}|^7}  {\bm R}\cdot{\bm r} \left(5 \left(\bm R\cdot \bm r \right)^2 - 3 \bm r^2 \bm R^2\right)+\dots \nonumber\\
&= \frac{\thirdMomentumVector^2}{2 \thirdMass} + \frac{\bm P^2}{2\boundMassTotal} 
- \frac{ G\boundMassTotal {\thirdMass}}{|\bm R|} +  H_0({\bm p}, {\bm r}) 
- \frac{3 G \boundMassReduced {\thirdMass}}{2|\bm R|^5} Q_{jk}(\bm r)Q^{jk}(\bm R) 
\nonumber\\
&\quad -\frac{5}{2}\sqrt{1-4\frac{\boundMassReduced}{\boundMassTotal}} \frac{G \boundMassReduced \thirdMass}{|{\bm R}|^7}  Q^{jkl}(\bm r)Q_{jkl}(\bm R)+ \dots \,,
\label{eq:Hq}
\end{align}
%
where $Q_{jk}(\bm r)= r_j r_k - ({\bm r}^2/3)\delta_{jk}$ and $Q_{jkl}(\bm r) =  r_j r_k r_l - (\bm r^2/5) (r_j\delta_{kl} + r_k \delta_{lj} + r_l \delta_{kj})$ are the quadrupole and octupole operators, respectively and the ellipsis stands for terms that drop off faster that $|\bm r|/|\bm R|^3$.

The first three terms of \cref{eq:Hq} describe the scattering of a particle with mass $\thirdMass$ and a particle with mass $\boundMassTotal$ interacting via Newton's potential $-\frac{G \boundMassTotal \thirdMass}{|\bm R|}$.
The fourth term, $H_0$, assigns some internal dynamics to the composite particle of mass $\boundMassTotal$, while keeping this internal structure decoupled from the scattering process.
For the choice~\eqref{eq:H0}, the internal structure is a bound state with the $1/|\bm r|$ type potential.

The last two terms displayed in the Hamiltonian are crucial and describe the interaction between the quadrupole and octupole moments of the bound system, $Q_{jk}(\bm r)$ and  $Q_{jkl}(\bm r)$, and those of the scattering system, $Q^{jk}(\bm R)$ and $Q^{jkl}(\bm R)$. In this interaction, the bound state is treated as structureless, aligning with the philosophy of the effective field theory framework, which separates scales by focusing on the relevant degrees of freedom at a given scale while neglecting finer details of internal structure at smaller scales.
The bilinear-in-spin Hamiltonian presented in Ref.~\cite{Alaverdian:2024spu} has a similar interpretation, with the bound-state quadrupole being replaced with the $S$- and $K$-induced quadrupoles.

We treat these last two terms as a perturbation that couples the internal structure of the bound system to the scattering process. They introduce two types of contributions: on the one hand, they describe how the internal dynamics of the bound state affect the final-state asymptotic scattering observables, and on the other hand, they describe how the presence of the scattering particle affects the internal state of the bound system.
At leading order, which is our focus, these effects are additive, so we can treat them separately. Moreover, since we want to explore the circumstances leading to a change in the spin of the bound state, we focus on the latter.

In the classical limit of quantum field theory, the typical separation of the particles is large. It is, therefore, consistent to treat the last two displayed terms of \cref{eq:Hq}, which are suppressed by the separation between the bound system ($\boundMassTotal$) and the third body ($\thirdMass$), as perturbations to the dynamics of two otherwise free particles.
Thus, at leading order, we take $\thirdMass$ and $\boundMassTotal$ to be free, with $\boundMassTotal$ being at rest and $\thirdMass$ moving along the straight line separated by the impact parameter $\bm{b}$,
\be
{\bm R} = (b_x,b_y,-v_0 t)
~~
\Longrightarrow
~~
|\bm R| = \sqrt{|\bm{b}|^2+v_0^2 t^2}\,, \hskip 1 cm   |\bm{b}|^2 = b_x^2+b_y^2\, .
\label{INI}
\ee
For the bound system described by $H_0$, the particle $\thirdMass$ can cause transitions between different bound states. 

We discuss in detail the analysis of the quadrupole interaction and simply quote the result for the octupole interaction. 
The transition amplitude between initial and final states $|\,\instate\,\rangle$ and $|\,\outstate\,\rangle$ of the bound system, with energies $\en{\instate}^B$ and $\en{\outstate}^B$, is given by 
first-order time-dependent perturbation theory in quantum mechanics, 
\begin{align}\label{eq:A3bLO}
{\cal A}_{\instate\rightarrow\outstate}^\text{LO}\Big|_{Q_{jk}} & =  \int_{-\infty}^{+\infty} dt \, e^{i(\en{\outstate}^B-\en{\instate}^B)t}   \left\langle \outstate \left| \frac{3 G \boundMassReduced {\thirdMass}}{2|\bm R|^5} \left( (\bm{r}\cdot \bm{R})^2 - \frac{1}{3} |\bm{r}|^2 |\bm{R}|^2\right) \right|\instate \right\rangle  \nonumber\\
& =  3 G \boundMassReduced {\thirdMass}
\int_{-\infty}^{+\infty} dt \, e^{i(\en{\outstate}^B-\en{\instate}^B)t}  
\frac{ Q^{jk}(\bm{R}(t)) }{2|\bm R(t)|^5} \langle \,\outstate\, | Q_{jk}(\bm{r}) | \,\instate\, \rangle \,.
\end{align}
The transition probabilities are exponentially suppressed if the initial and final states have different energies, 
\be
{\cal P}_{\instate\rightarrow\outstate}^\text{LO} = | \mathcal{A}_{\instate\rightarrow\outstate}^{\text{LO}} |^2 \propto e^{-b|\en{\outstate}^B-\en{\instate}^B|/v_0} \quad \text{if} \quad \frac{b|\en{\outstate}^B-\en{\instate}^B|}{v_0}\gg 1 \, ;
\ee
such transitions describe absorption or (stimulated) emission. Thus, in agreement with physical intuition, the higher the relative velocity of the bound state and the scattering particle (i.e., the higher the energy in the scattering process), the higher the absorption probability. 
Our goal is to describe conservative processes, so we assume that the initial and final internal energies of the bound state are exactly equal, $\en{\outstate}^B= \en{\instate}^B$. We further assume that we are in the regime in which post-Minkowskian perturbation theory is applicable, i.e., that the minimal approach distance is much larger than the typical size of the bound state $r_{\text{cl}, n}$ introduced below, $b v_0^2 \gg r_{\text{cl}, n}$.

With these assumptions and in the Cartesian coordinates, the integration over time in \cref{eq:A3bLO} gives
\begin{align}
\label{eq:scattering_quadrupole}
\int_{-\infty}^{+\infty} dt  \frac{ Q^{jk}(\bm{R}(t)) }{2|\bm R(t)|^5} = 
\frac{1}{3 |\bm b|^4 v_0} 
\left(\begin{array}{ccc}
b_x^2-b_y^2 & 2 b_x b_y&0\cr
2b_x b_y&b_y^2-b_x^2 &0\cr
0&0&0
\end{array}
\right) \,,
\end{align}
and the matrix element of the quadrupole operator becomes a linear combination of 
\begin{align}\label{eq:x2y2}
    \langle \,\outstate\, | (x^2 - y^2) | \,\instate\, \rangle
    ~~\text{and}~~
    \langle \,\outstate\, | x y | \,\instate\, \rangle \, .
\end{align}
To evaluate these matrix elements, we need to specify the states $|\,\instate\,\rangle$ and $|\,\outstate\,\rangle$.

While the standard energy and angular-momentum eigenstates in the $V\sim 1/{|{\bm r}|}$ potential reproduce the correct energy and angular momentum in the limit of large quantum numbers, they are not immediately suitable for our purpose because they do not reproduce, e.g., the classical motion of a bound system.\footnote{\label{FT3footnote} They may, however, be used in the sense of Field Theory~3 of Ref.~\cite{Bern:2023ity}, which results in a change in the classical spin through transitions between quantum states with different spins.}
They also do not minimize the dispersion, so the expectation values of the products of operators do not exhibit the expected classical factorization. For that to be incorporated, we need to consider a coherent state.

We use the elliptic orbit coherent state $|\alpha\rangle$  from Ref.~\cite{DBhaumik_1986}  to describe the state of the bound system in the classical limit. This coherent state has the probability distribution sharply peaked at 
\begin{align}
\en{\instate}^B = - \frac{G \boundMassReduced \boundMassTotal}{2 a_0 n^2} = - \frac{G \boundMassReduced \boundMassTotal}{2 r_{\text{cl},n}} \, .
\end{align}
The characteristic radius of this state is $r_{\text{cl},n} = a_0 n^2$, where $a_0=\hbar^2/(G\boundMassReduced \boundMassTotal)$ is the Bohr radius of the bound system and the principal quantum number $n$ is taken to be large\footnote{In the clasical limit $n\rightarrow\infty, \hbar\rightarrow 0$ with fixed $n \hbar \sim S$, both $r_{\text{cl}, n}$ and $\en{\instate}^B$ are effectively independent of~$\hbar$.}. The expectation value of the position operator follows an elliptic trajectory, 
\begin{align}
    \langle \alpha | x | \alpha \rangle &= r_{\text{cl},n} \Big[\cos(2\omega_{\text{cl}}t) + \sin(2\chi)\Big]\,, \nonumber\\
    \langle \alpha | y | \alpha \rangle &= r_{\text{cl},n} \sin(2\omega_{\text{cl}}t)\cos(2\chi)\,, 
    \label{eq:trajectory}\\
    \langle \alpha | z | \alpha \rangle &= 0 \, , \nonumber
\end{align}
where for simplicity we put the motion into the $x$-$y$ plane with the semi-major axis along the $\hat {\bm x}$ direction. Thus, the eccentricity vector $\bm{e}$, which can be identified as the suitably-normalized Laplace-Runge-Lenz vector $\bm{A}$, is 
\begin{align}
\bm{A} \equiv \bm e=\sin(2\chi) {\hat {\bm x}}\,.
\end{align}
Since the motion is in the $x$-$y$ plane, the orbital angular momentum, which we identify as the spin of the bound state, is in the $\hat{\bm z}$ direction. 
The orbital frequency, $\omega_{\text{cl}} = 1/ (2\boundMassReduced a_0^2 n^3)$, is much higher than the inverse time scale of the scattering process. As a result, one can perform a time average over the orbital motion, simplifying the analysis by averaging out the rapid oscillations of the bound system's internal dynamics,
\begin{align}
    \langle \,\outstate\, | (x^2 - y^2) | \,\instate\, \rangle &= \frac{\omega_{\text{cl}}}{\pi} \int_{0}^{\pi/\omega_{\text{cl}}} dt \left(\langle \alpha | x | \alpha \rangle^2 - \langle \alpha | y | \alpha \rangle^2\right) = \frac{3 r_{\text{cl},n}^2}{2} \sin^2(2\chi)\, , 
    \cr 
    \langle \,\outstate\, | x y  | \,\instate\, \rangle &= \frac{\omega_{\text{cl}}}{\pi} \int_{0}^{\pi/\omega_{\text{cl}}} dt \, \langle\alpha|x|\alpha\rangle\langle\alpha|y|\alpha\rangle = 0 \,,
\end{align}
where we used the dispersion-minimizing property of the elliptic states to infer that the difference between $\langle \alpha | x | \alpha \rangle^2$ and $\langle \alpha | x^2 | \alpha \rangle$ is quantum-suppressed in the classical limit. This leads to the final result for the transition amplitude,
\begin{align}
    \mathcal{A}^{\text{LO}}_{\instate\rightarrow\outstate}(\bm{b})\Big|_{Q_{jk}} &= \frac{3G\boundMassReduced \thirdMass r_{\text{cl},n}^2}{2|\bm b|^2 v_0} \left(\frac{2 b_x^2}{|\bm b|^2}-1\right) \sin^2(2\chi)= \frac{3G\boundMassReduced \thirdMass r_{\text{cl},n}^2}{2|\bm b|^2 v_0} 
 \left[ \frac{2(\bm{b}\cdot\bm{A})^2}{|\bm b|^2} - |\bm A|^2 \right] \,,
\label{eq:transition_toy_model}
\end{align}
where we used the fact that the Laplace-Runge-Lenz vector $\bm A$ equals the eccentricity vector, which, for our choice of kinematics, points along the $\hat {\bm x}$ axis, as indicated in \cref{toymodelfigure}.
Importantly, this combination has a rational Fourier transform to momentum space,  
\begin{align}
\label{eq:transition_toy_model_q_space}
    \mathcal{A}^{\text{LO}}_{\instate\rightarrow\outstate}(\bm q)\Big|_{Q_{jk}} &= \int d^2\bm b \, \mathcal{A}^{\text{LO}}_{\instate\rightarrow\outstate}(\bm b) \, e^{i \bm q\cdot\bm b} = - \frac{3 \pi G \boundMassReduced  \thirdMass r_{\text{cl},n}^2}{v_0} \frac{(\bm q\cdot\bm A)^2}{|\bm q|^2} \,.
\end{align}
Note that we enforce the transversality of the impact parameter $\bm b$ on the momenta of both scatterers, so the Fourier transform is two-dimensional.   We encounter analogous features in the off-shell relativistic context in the construction of the Hamiltonian in Sec.~\ref{sec:Hamiltonian}, where the Fourier transform is three-dimensional.

Repeating the analysis for the octupole interaction term in the Hamiltonian \cref{eq:Hq}, we find that the ${\cal O}(\bm A^3)$ contributions to the impact parameter space and momentum space amplitudes are
\begin{align}
\label{eq:octupoleB}
{\cal A}_{\instate\rightarrow\outstate}^\text{LO}(\bm{b})  \Big|_{Q_{jkl}} &= \frac{5 G \boundMassReduced \thirdMass \, r_{\text{cl},n}^3}{3|\bm b|^3 v_0} 
\sqrt{1-4\frac{\boundMassReduced}{\boundMassTotal}} \frac{\bm b\cdot \bm A}{|\bm b|} \left(4 \frac{(\bm b\cdot \bm A)^2}{|\bm b|^2} - 3|\bm A|^2\right)\,,
\\
{\cal A}_{\instate\rightarrow\outstate}^\text{LO}(\bm{q})  \Big|_{Q_{jkl}}  &
= -i \frac{5 \pi G \boundMassReduced \thirdMass \, r_{\text{cl},n}^3}{ 3v_0} 
\sqrt{1-4\frac{\boundMassReduced}{\boundMassTotal}} \,
\frac{(\bm q\cdot \bm A)^3}{|\bm q|^2} \ .
\end{align}
As for the quadrupole interaction, the particular impact parameter-dependent combination in \cref{eq:octupoleB} is required for the Fourier transform to be rational.

\subsubsection{Comparison with the general \texorpdfstring{$\bm{K}$}{K}-dependent two-body amplitude}

To compare \cref{eq:transition_toy_model} with the expected general structure of two-body amplitude to second order in $\bm S$ and $\bm K$ outlined in \cref{eq:general_2body}, we must first restrict the latter to the special kinematic configuration chosen in \cref{sec:elliptic}, but in the center-of-mass frame of $m_1$ and $m_2$. The amplitude $\mathcal{A}^{\text{LO}}_{\instate\rightarrow\outstate}$ is invariant under spatial rotations and translations, so we can directly use \cref{eq:transition_toy_model} in the comparison with the field-theory amplitudes in the center-of-mass frame, with $v_0$ identified as the relative velocity.

We work in the low-velocity limit, which does not include any of the terms containing the momentum $\thirdMomentumVector \sim \bm p$ of the particle passing by the bound system.  
To match the kinematic configuration considered in \cref{sec:elliptic} 
we assume that the internal angular momentum $\bm S$ of the bound state points in the $\hat {\bm z}$ direction and that it is parallel with the momentum of the center of mass of the bound state, see \cref{toymodelfigure}.
Thus, in the kinematic configuration we are focusing on, the general form \eqref{eq:general_2body} of the two-body amplitude simplifies to 
\begin{align}
\label{eq:general}
{\cal M}^{\text{2 body}} 
&= A_0 
+A_{2,1} \bm S^2
+A_{2,2} \bm K^2
+A_{2,3} \bm S\cdot \bm K
+A_{2,6} (\bm b\cdot \bm K)^2 + \order(\gS^3) \, .
\end{align}
At $\order(\gS^3)$, we have $(\bm b\cdot \bm K)^3$ and $(\bm b\cdot \bm K)\bm K^2$ and further terms proportional to $\bm S\cdot \bm K$ which, as we demonstrate, turn out to be unimportant.
The center-of-mass momentum $\bm p$ is given in terms of the relative velocity $v_0$ by $|\bm{p}|=\frac{m_1m_2}{m_1+m_2}v_0$. We note that, since this is an impact-parameter space amplitude, all coefficients in \cref{eq:general}
include a $|\bm p|^{-1}\propto v_0^{-1}$ factor, arising from the transversality constraints of the momentum transfer, see e.g.~\cref{eq:eikonal_Fourier}. 

Comparing Eq.~\eqref{eq:transition_toy_model} with the expected general structure \eqref{eq:general} allowed by our kinematic configuration and making use of the fact that the spin (identified as the orbital angular momentum of the bound state) is in the $\hat{\bm z}$ direction so that $\bm S\cdot \bm A = 0$, it is easy to see that ${\cal A}_{\instate\rightarrow\outstate} \big|_{Q_{jk}}+{\cal A}_{\instate\rightarrow\outstate} \big|_{Q_{jkl}}={\cal M}^{\text{2 body}} $ through third order in $\bm K$ provided that we map the $\bm K$ vector into a multiple of the Laplace-Runge-Lenz vector, 
\begin{align}
\bm K = i\,G \boundMassTotal^2 \,\frac{\boundMassReduced}{\boundMassTotal} 
\sqrt{\frac{\boundMassReduced}{2 |\en{\instate}^B|}}\,\bm A \, .
\label{eq:AvsK}
\end{align}
The imaginary factor relates the $\text{SO}(4)$ algebra of the Newtonian bound state with the $\text{SO}(1,3)$ 
algebra of the relativistic field theory, effectively removing the imaginary factor on the first line of \cref{eq:gStensor}. 
We chose the prefactor on the right-hand side so that the Newtonian relation between the eccentricity, internal angular momentum and the energy, 
\begin{align}\label{eq:NewtonianE}
    e = \sqrt{ 1 
    - \frac{2 |\en{\instate}^B| }{G^2 \boundMassReduced^3 \boundMassTotal^2} |\bm S|^2 } \,,
\end{align}
becomes $\bm S^2-\bm K^2 = \text{const}$, which is the analog of~\cref{eq:SpinTensorConservation} appropriate for an $SO(4)$ symmetry and follows from the first line of \cref{eq:gStensor} without the imaginary factor. We also enforced that the internal energy of the bound state $\en{\instate}^B$ is negative.
When carrying out a comparison with the field-theory calculations in, e.g., Sec.~\ref{sec:TwoBodyAmplitude} we need to recall that the amplitudes listed there are given in terms of rescaled $K$ (and $S$),
\begin{align}
\kK = 
\frac{1}{\boundMassTotal} {\bm K} \, ,
\label{eq:rescalingK}
\end{align}
see Eq.~\eqref{eq:rescaledKandS}. We also note that the factor of $G\boundMassTotal^2$ in \cref{eq:AvsK} is reminiscent of the analogous factors extracted from the spin vector in the physical counting of the PM order, see e.g.~Ref.~\cite{Buonanno:2024vkx} as well as comments in \cref{sec:TwoBodyAmplitude,sec:Hamiltonian}. In comparing amplitude expressions in \cref{sec:TwoBodyAmplitude}, this factor must be dropped because it is not manifested there.

The relation between $\bm S^2$ for $\bm K^2$ stemming from the conservation of the spin tensor, i.e., \cref{eq:NewtonianE} in this model, allows us to, in principle, trade $\bm S^2$ for $\bm K^2$. The ensuing constant is a combination of other variables and conserved quantities. 
The geometry of our scattering process has an average quadrupole moment only in the $x$-$y$ plane, cf. \cref{eq:scattering_quadrupole}, so $Q^{jk}S_{k} = 0$. Thus, the absence of the spin vector in the amplitude $\mathcal{A}^{\text{LO}}_{\instate\rightarrow\outstate}$ is expected.

Being already written in impact parameter space, the amplitude in Eq.~\eqref{eq:transition_toy_model} is the analog in our example of the tree-level two-body field theory leading-order eikonal phase $\chi_1$ in Eq.~\eqref{eq:LOeikonal}, including the factor of $(2 \en{} |\bm p|)^{-1}\propto v_0^{-1}$.\footnote{We note that the basis of $\bm S$- and $\bm K$-dependent operators that is convenient for the toy model discussed here is slightly different from the basis of operators used in \cref{sec:TwoBodyAmplitude}. } 
\footnote{Alternatively, we may also compare the elastic cross section or the scattering probability per unit time that follows from \cref{eq:QFTamplitude1PM} on the one hand and that given by the nonrelativistic quantum mechanics amplitude \cref{eq:transition_toy_model} or \cref{eq:transition_toy_model_q_space} on the other. In the former, the normalization of the initial states provides a factor of 
$1/(4\en{1}\en{2})$ and the state normalization and the phase-space integration of the final state yields a factor of $1/(4 m_1 m_2 \sqrt{(u_1\cdot u_2)^2-1})\sim (4 m_1 m_2 v_0)^{-1}$. Further including the factor of the velocity that is present in the nonrelativistic quantum-mechanical probability per unit time and taking the square root leads to the same relation as when equating the
probability amplitude \eqref{eq:transition_toy_model} and the tree-level two-body field theory leading-order eikonal phase $\chi_1$ in Eq.~\eqref{eq:LOeikonal}. The relevant terms are given by the coefficient $\alpha_1^{\{2,5\}}$ and $\alpha_1^{\{3,11\}}$ in the latter equation. 
} 
Since the Wilson coefficient $E_2$ in the Lagrangian~\eqref{eq:ExtraInteractions} governs the bilinear-in-$\bm K$ interactions, we should be able to read off its value for our bound two-body system by matching Eq.~\eqref{eq:transition_toy_model} against the nonrelativistic expansion of the two-body scattering amplitude following from this Lagrangian.
With foresight on this amplitude, which is discussed in Sec.~\ref{sec:TwoBodyAmplitude}, this comparison yields 
%
%
\begin{align}
E_2^{\text{bound 2-body}} &= \frac{3 |\en{\instate}^B| \boundMassTotal}{\boundMassReduced^2 } \, 
\left(\frac{2 r_{\text{cl}, n}}{r_s}\right)^2  = 
\frac{3 \boundMassTotal}{4 |\en{\instate}^B|}
 \,,
\\
F_3^{\text{bound 2-body}} &= 
10\sqrt{\frac{2 |\en{\instate}^B|}{\boundMassReduced}}\,
\sqrt{1-\frac{4\boundMassReduced}{\boundMassTotal}}\, 
\left(\frac{2 r_{\text{cl}, n}}{r_s}\right)^3 =10 \left(\frac{\boundMassReduced}{2 |\en{\instate}^B|}\right)^{5/2}
\sqrt{1-\frac{4\boundMassReduced}{\boundMassTotal}}\,,
\end{align}
%
%
where $r_s = 2 G \boundMassTotal$ is the Schwarzschild radius of the bound state, $\en{\instate}^B <0$ for the bound state, and $E_2^{\text{bound 2-body}}$ and $F_3^{\text{bound 2-body}}$ are respectively the $E_2$ and $F_3$ Wilson coefficients for the bound two-body system.
As discussed below \cref{eq:rescalingK}, to read off this value, we included the rescaling in \cref{eq:rescalingK}.

Classical observables following from the amplitude~\eqref{eq:transition_toy_model} can be evaluated via the same methodology as for the relativistic amplitudes, i.e., either by constructing a Hamiltonian and then solving its equations of motion or by constructing an eikonal phase and using its relation to observables.
We postpone this analysis to \cref{sec:TwoBodyAmplitude}.


The identification in \cref{eq:AvsK} respects all the properties of the quantum-field-theory $K$ vector specialized to our example. From the field theory standpoint, the magnitude of $\bm K$ is fixed in terms of other quantities present in the theory (the spin and the magnitude of the spin tensor) while the magnitude of the Laplace-Runge-Lenz vector (i.e., the eccentricity) is determined in terms of the internal orbital angular momentum $\bm S$ of the bound system and its internal energy. 
The Laplace-Runge-Lenz vector obeys one additional constraint, $\bm A \cdot \bm S = 0$, compared to a general $\bm K$ vector. All in all, $\bm A$ has one degree of freedom, which in our initial kinematic configuration was fixed by demanding that the semi-major axis of the initial bound trajectory points in the positive $\hat{\bm x}$ direction. 
In the presence of the scattering body, the internal angular momentum of the bound state is no longer conserved. However, the relation~\eqref{eq:NewtonianE} implies that it is possible to trade internal angular momentum for eccentricity without changing the energy of the bound state.
Moreover, the orientation of the semi-major axis of the bound orbit can also be changed without altering its energy. Thus, the Laplace-Runge-Lenz vector effectively encodes gapless degrees of freedom, in line with the general expectation that $\bm K$ has this property.

In this example, the vector $\bm K$ characterizes the ``shape'' and ``orientation'' of a composite particle realized as a two-body bound state in a $1/|\bm r|$ potential and is a manifestation of the independence of the energy and the spin (i.e., internal angular momentum) of the bound state. More complicated multi-particle bound states in the same potential exhibit similar properties.
We note that the $\bm K$ vector has nontrivial Poisson brackets with the spin of the bound state, which are inherited from the  
commutation relations of the Laplace-Runge-Lenz vector and the internal orbital angular momentum of the bound state.\footnote{These commutation relations generate an $\text{SO}(4)$ algebra.} 
In \cref{sec:eikonal} we connect two-body amplitudes of the type shown in Eq.~\eqref{eq:general}, so in particular the scattering amplitude~\eqref{eq:transition_toy_model}, to observables and we  see that the presence of $\bm K$ implies that time evolution leads to a 
conservative change in the magnitude of its spin and in \cref{sec:Hamiltonian} we reach the same conclusion from a Hamiltonian perspective.

The analysis above can be enhanced by considering relativistic corrections to the interaction between the third (scattering) body and the constituents of the bound state. 
A starting point is the three-body post-Minkowskian Hamiltonians discussed in Ref.~\cite{Loebbert:2020aos, Jones:2022aji}, which are subsequently expanded to 0PN order in the relative velocity and relative separation of two of the three particles so as to enforce the Newtonian nature of the bound state. 
An alternative starting point, which is perhaps closer to the QFT treatment of the particle exhibiting the $K$ vector as pointlike, is to simply interpret the potential as the integral over the bound-state particle distribution of the 1PM two-body potential while promoting the kinetic energy part of the center of mass to its relativistic form, $\bm P^2/2\boundMassTotal \rightarrow \sqrt{\bm P^2+\boundMassTotal^2}$. 
In both cases, it is necessary to judiciously choose the center of mass---and consequently the canonical coordinates---so that the resulting Hamiltonian for $|\bm r|\ll |\bm R|$ does not lead to a linear mass-dipole interaction.
We, however, do not pursue this here.

While this particular realization of light degrees of freedom can be applied immediately to certain classes of astrophysical systems, such as planets and their satellites, it would be important to understand whether they can occur in compact systems relevant to gravitational waves. We anticipate that this is possible and we comment along these lines in \cref{sec:conclusion} from the perspective of symmetry breaking.

We close this section by mentioning that the example discussed here exhibits certain similarities with a bound analog of a three-body system that exhibits the so-called Kozai-Lidov effect~\cite{Kozai:1962zz, Lidov:1962wjn} (see, e.g., Ref.~\cite{Naoz_2016} for a review). This system consists of a bound or unbound two-body system under the influence of a bound, distant, and heavy third body, which is treated as a perturbation. 
In such a three-particle configuration, the inner (bound) system exhibits an accidental conservation of the component of the angular momentum of the inner bound system along the orbital angular momentum of the outer motion. 
Moreover, this component of the angular momentum depends both on the eccentricity and orbital inclination of the orbit of the inner bound system, which can, therefore, be traded dynamically for each other. 
This leads to oscillations of eccentricity and inclination over time scales that are much larger than the periods of inner and outer motions. 
The Kozai-Lidov effect was proposed as a mechanism for producing sufficiently many observationally-relevant eccentric binary systems in our Universe, see e.g.~\cite{Samsing:2017xmd, Randall:2018qna} which also discuss the role of post-Newtonian corrections, even though such systems radiate away their eccentricity on relatively short time scales.\footnote{We thank Zihan Zhou for emphasizing the similarity of features of our toy model with the Kozai-Lidov effect and for pointing out  references~\cite{Naoz_2016, Samsing:2017xmd, Randall:2018qna}.}

\subsection{A metric sourced by a \texorpdfstring{$\bm{K}$}{K}-dependent stress tensor}
\label{sec: metricWK}

As a second example of a system that may be characterized as a \gco,
we consider the Rasheed-Larsen black hole~\cite{Rasheed:1995zv, Larsen:1999pp}, which is a spinning black hole that also carries dilaton, electric and magnetic charges and solves the equations of motion of a certain supergravity theory.  This provides an explicit example of an exotic astrophysical object that naturally involves $K$. The multipole moments of this system were analyzed in Ref.~\cite{Bena:2020uup}, where it was shown that they are qualitatively different from those of the Kerr black hole: In the terminology of Ref.~\cite{Bena:2020uup}, for the Kerr black hole, only the even mass multipoles and the odd current multipoles are non-zero. In contrast, for the Rasheed-Larsen black hole all multipoles are non-zero. Hence, it is natural to expect that the Rasheed-Larsen black hole cannot be described by the spin vector alone. Indeed, we find that the leading-order amplitude for an uncharged scalar probe particle scattering off a Rasheed-Larsen black hole can be described by the amplitude obtained by our QFT construction of \cref{sec:HigherOrderFT}, which involves $K$.   

The metric is given by
\begin{align}
    ds^2 = \frac{H_3}{\sqrt{H_1 H_2}} (dt+B d\phi)^2 &- \sqrt{H_1 H_2} \left( \frac{dr^2}{\Delta} + d\theta^2 + \frac{\Delta}{H_3} \sin^2 \theta d\phi^2 \right) ,
\end{align}
with
\begin{align}
    B &= \frac{2 G \rls \sqrt{\rl_1 \rl_2} (\rl_0^2 + \rl_1 \rl_2)}{\rl_0 (\rl_1 + \rl_2) \left(\rls^2 \cos^2 \theta + r^2 \right)} r \sin^2 \theta  \,,\qquad \Delta = \rls^2 + r(r-2G\rl_0) \,, \nonumber\\
    H_1 &= -2G \rls\, \rl_1 \cos\theta \frac{\sqrt{(\rl_1^2-\rl_0^2)(\rl_2^2-\rl_0^2)}}{\rl_0(\rl_1+\rl_2)} + \rls^2 \cos^2 \theta + r (r+2G(\rl_1-\rl_0)) \,,  \nn \\
    H_2 &= 2G \rls\, \rl_2 \cos\theta \frac{\sqrt{(\rl_1^2-\rl_0^2)(\rl_2^2-\rl_0^2)}}{\rl_0(\rl_1+\rl_2)} + \rls^2 \cos^2 \theta + r (r+2G(\rl_2-\rl_0)) \,,  \nn \\
    H_3 &= \rls^2 \cos^2 \theta + r(r-2G\rl_0) \,.
\end{align}
The complete solution to the supergravity equations of motion also contains non-zero dilaton and gauge-field configurations; these are not important for our purposes, as we take the scalar probe particle to not be charged under these fields. This matter-coupled four-dimensional gravitational solution was obtained by dimensional reduction from five-dimensional Einstein gravity~\cite{Rasheed:1995zv, Larsen:1999pp}, with the dilaton and gauge fields originating in this reduction.
The metric is characterized by 4 parameters, one length scale $\rls$ and 3 mass scales $\rl_n$, $n=0,1,2$. It reduces to the Kerr black hole for the following choice of the parameters,
\begin{align}
    \hspace{-.5cm} \text{Kerr limit:} \hspace{1.3cm} \rl_0 = \rl_1 = \rl_2 = M\,, \hspace{1.3cm} \rls = a\,, 
    \label{eq:rlKerrLimit}
\end{align}
where $M$ and $a$ are the mass and ring radius of the Kerr black hole, respectively. Away from this limit, we have $\rl_1 \geq \rl_0$ and $\rl_2 \geq \rl_0$, while the black hole has a magnetic and electric charge, respectively,
\begin{align}
    Q_m^2 = \frac{\rl_1\left(\rl_1^2-\rl_0^2\right)}{\rl_1+\rl_2}\,, \hspace{1.5cm}
    Q_e^2 = \frac{\rl_2\left(\rl_2^2-\rl_0^2\right)}{\rl_1+\rl_2}\,.
\end{align}

We employ the formalism of Ref.~\cite{Kosmopoulos:2023bwc} to compute the amplitude for the scattering of an uncharged probe particle off the Rasheed-Larsen black hole to leading order in $G$. This formalism allows for the systematic calculation of amplitudes in curved backgrounds and was developed to tackle the problem of gravitational self-force~\cite{Mino:1996nk, Quinn:1996am, Detweiler:2000gt, Detweiler:2002mi, Gralla:2008fg, Pound:2009sm, Rosenthal:2006iy, Detweiler:2011tt, Pound:2012nt, Gralla:2012db, Barack:2018yvs, Pound:2021qin, Galley:2006gs, Galley:2008ih} within quantum field theory; for similar approaches from the worldline perspective see Refs.~\cite{Galley:2008ih, Cheung:2023lnj, Wilson-Gerow:2023syq, Jakobsen:2023tvm, Cheung:2024byb}. The connection between black-hole metrics and Feynman diagrams goes back to the work of Duff~\cite{Duff:1973zz} and was explored further with modern methods in Refs.~\cite{Neill:2013wsa, Mougiakakos:2020laz, Jakobsen:2020ksu, DOnofrio:2022cvn, Mougiakakos:2024lif, Mougiakakos:2024nku}, while calculations of amplitudes in curved backgrounds have also been considered in Refs.~\cite{Adamo:2020qru, Cristofoli:2020hnk, Adamo:2021rfq, Adamo:2022mev, Adamo:2022rmp, Adamo:2022qci, Adamo:2023cfp, Adamo:2023fbj, Aoki:2024bpj}. We take the probe particle to not be charged under the dilaton and electromagnetic charges of the black hole, such that only the gravitational interactions are relevant.

Following Ref.~\cite{Kosmopoulos:2023bwc}, the amplitude in question is given by 
\begin{align}
    \mathcal{M}_\text{RL} = 2GM \, \delta \tilde{g}^{\mu\nu}(\bm q) p_{2\mu} p_{2\nu}  + \mathcal{O}\left(G^2\right),
    \label{eq:curvedAmpDef}
\end{align}
where $M$ is the physical mass of the black hole, $p_2$ is the initial momentum of the probe and $\delta \tilde{g}^{\mu\nu}(\bm q)$ is the Fourier transform of $\delta g^{\mu\nu}(\bm r)$ (see \cref{eq:FTMetric}), which is defined in terms of the background metric $g_{\mu\nu}(\bm r)$ as 
\begin{align}  
    g_{\mu\nu}(\bm r) = \eta_{\mu\nu} + G \delta g_{\mu\nu}(\bm r) + \mathcal{O}\left(G^2\right),
\end{align}
and $\eta_{\mu\nu}$ is the flat metric. We note that the factor $2M$ in \cref{eq:curvedAmpDef} accounts for the non-relativistic normalization inherent in the construction of Ref.~\cite{Kosmopoulos:2023bwc}.
In order to perform the Fourier transform, we switch to Cartesian coordinates using
\begin{align}
    x = \sqrt{r^2+\rls^2} \sin \theta \cos \phi \,, \qquad 
    y = \sqrt{r^2+\rls^2} \sin \theta \sin \phi \,, \qquad 
    z = r \cos \theta \,.
\end{align}
Furthermore, in this coordinate system, the black hole has an apparent mass dipole~\cite{Bena:2020uup}, which we remove by performing the coordinate transformation
\begin{align}
    z \rightarrow z - \rls\, \frac{(\rl_2-\rl_1)\sqrt{(\rl_1^2-\rl_0^2)(\rl_2^2-\rl_0^2)}}{\rl_0(\rl_1+\rl_2)^2} \,.
\end{align}
Were the apparent mass dipole not removed, it would manifest in the amplitude as a linear-in-$K$ term. Such terms, however, are expected to vanish in general relativity due to the freedom of choice of the origin of the coordinate system, as we can explicitly see in this example.

To demonstrate that this amplitude is indeed described by the QFT amplitudes of \cref{sec:TwoBodyAmplitude}, we find it sufficient to compare at linear order in $G$ and up to cubic order in $S$ and $K$. The order of the multipoles of the Rasheed-Larsen black hole is controlled by $\rls$~\cite{Bena:2020uup}. Hence, it suffices to expand the above amplitude to cubic order in $\rls$. For the QFT amplitudes, we use the kinematics
\begin{align}
    \bar{v}_1 = (1,0) \,, \qquad
    p_2 = (\en{2}, \bm p) \,, \qquad
    q = (0,\bm q) \,,
\end{align}
where $p_1 = M \bar{v}_1 + \mathcal{O}\left(\bm q\right)$.

We obtain the mass $M$ and spin vector $\clS$ of the black hole by matching at zeroth and first order in $\rls$. We find
\begin{align}
    M = \frac{\rl_1+\rl_2}{2} \,, \hspace{1.5cm}
    \clS = (0,0,\rls)\, \frac{\sqrt{\rl_1 \rl_2}(\rl_0^2+\rl_1 \rl_2)}{\rl_0 (\rl_1+\rl_2)} \,.
\end{align}
These agree with the mass monopole and current dipole obtained in Ref.~\cite{Bena:2020uup}. Given the universal coupling of gravity at these orders, the above matching equations do not contain free Wilson coefficients. To extract the $\clK$ vector that describes the black hole, we need to match at quadratic order in $\rls$. We have
\begin{align}
    D_2 \clK = (0,0,\rls)\, \frac{(\rl_2-\rl_1)\sqrt{(\rl_1^2-\rl_0^2)(\rl_2^2-\rl_0^2)}}{2\rl_0(\rl_1+\rl_2)} \,.
    \label{eq:matchToRLD2}
\end{align}
We observe that $\clK$ is parallel to $\clS$ for the Rasheed-Larsen black hole, as we can guess from the axial symmetry of the metric. The complete matching at quadratic and cubic orders in $\rls$ yields
\begin{align}
    C_2 \clS^2 - E_2 \clK^2 = \rls^2 \frac{f_1}{4\rl_0^2(\rl_1+\rl_2)^2} \,, \quad & \quad
    C_3 \clS^2 - 3 E_3 \clK^2 = \rls^2\,\frac{f_2}{4\rl_0^2(\rl_1+\rl_2)^2} \,, \nn \\
    \Big( 3 (D_3-C_2) \clS^2 - F_3 \clK^2 \Big) \clK &= \rls^2 (0,0,\rls)\,\frac{(\rl_2-\rl_1)f_3}{4\rl_0^3(\rl_1+\rl_2)^3} \,,
    \label{eq:rlD2K}
\end{align}
with
\begin{align}
    \label{eq:rlCEF}
    f_1 &= (\rl_0^4+\rl_1^2\rl_2^2)(\rl_1^2-6\rl_1\rl_2+\rl_2^2) + 2\rl_0^2 \rl_1 \rl_2 (3\rl_1^2-2\rl_1\rl_2+3\rl_2^2) \,, \\
    f_2 &= (\rl_0^4+\rl_1^2\rl_2^2)(-3\rl_1^2+2\rl_1\rl_2-3\rl_2^2) + 2\rl_0^2 (2 \rl_1^4 - \rl_1^3 \rl_2 + 2\rl_1^2\rl_2^2 - \rl_1\rl_2^3 + 2\rl_2^4) \,, \nn \\
    f_3 &= \sqrt{(\rl_1^2-\rl_0^2)(\rl_2^2-\rl_0^2)} \left( (\rl_0^4+\rl_1^2\rl_2^2)(\rl_1-\rl_2)^2 + 2\rl_0^2 \rl_1 \rl_2 (3 \rl_1^2 + 2\rl_1\rl_2 + 3\rl_2^2) \right) \nn \,.
\end{align}

Importantly, in order to write the above expressions, we identified $i \qftKrf \equiv \clK$ with $\qftSrf = \clS$, as described below \cref{eq:SpinTensorConservation}. Here, we see that this identification is necessary for matching to the amplitude obtained from the Rasheed-Larsen black hole, whose parameters are real. For more details on this identification, see Ref.~\cite{Bern:2023ity}.

We observe that since $\clK$ is parallel to $\clS$, we cannot resolve the individual Wilson coefficients. Instead, we are only able to obtain constraints among them. This is to be expected since all the mass and current multipoles of the Rasheed-Larsen black hole are aligned~\cite{Bena:2020uup}, and hence their contributions are degenerate. 
In principle, one should be able to obtain $\clS$ and $\clK$ by computing the spin tensor corresponding to the Rasheed-Larsen black hole. The latter may be derived from the stress-energy tensor sourcing the black hole, following the analysis of Ref.~\cite{Vines:2017hyw}. Together with \cref{eq:matchToRLD2}, this would allow for the complete specification of $D_2$. 
Fully determining the remaining Wilson coefficients would require a generalization of the Rasheed-Larsen black hole where $\clK$ and $\clS$ are not parallel. 

In the Kerr limit of \cref{eq:rlKerrLimit}, all Wilson coefficients appearing in \cref{eq:rlD2K,eq:rlCEF} are zero. While we cannot establish that the individual Wilson coefficients are zero due to the aforementioned degeneracy, we see that the combinations appearing in these equations vanish. In this limit, we also recover the known amplitude for a scalar probe particle scattering off a Kerr black hole.

As expected, the dynamics of $\clK$ decouple in the Kerr limit. We observe in \cref{eq:rlD2K,eq:rlCEF} that this does not mean that $\clK$ itself must be zero; rather, its dynamics can decouple due to the vanishing of the corresponding Wilson coefficients. This is in line with the analysis of Refs.~\cite{Bern:2022kto, Bern:2023ity,Alaverdian:2024spu} and the present work: Results for a \gco may be reduced to those for a \cco by appropriate choice of the Wilson coefficients.

Matching our QFT amplitude to that coming from the Rasheed-Larsen black hole at order $G$ establishes that $\clK$ is a viable description of a \gco. 
Here, we analyzed the static limit of the black hole. It would be interesting to explore this relation at higher orders in $G$, where the backreaction of the black hole becomes important. To take backreaction into account, we would need to include the dynamics of the Goldstone bosons associated with the spacetime spontaneous symmetry breaking pattern induced by the Rasheed-Larsen black hole~\cite{Kosmopoulos:2023bwc}. 
Matching the dynamical evolution of the Rasheed-Larsen black hole to our QFT would be a rather non-trivial test of our construction. 


\section{Field Theory: Higher-order interactions}
\label{sec:HigherOrderFT}

In this section, we extend the Lagrangian~\eqref{eq:Lmin} and~\eqref{eq:ExtraInteractions}, which describes a \gco with both $S$ and $K$ degrees of freedom, to the fourth order in the Lorentz generator. We also present the classical Compton amplitudes to the fourth order in the spin tensor $\gS$.
Before the detailed discussion, we first summarize the result. The Lagrangian up to $\order(M_{ab}^4)$ is given by
\begin{align}\label{eq:full_L}
    \mathcal{L}_{\text{\gco}} = - \frac{1}{2} \phis (\nabla^2+m^2) \phis & + \underset{\eqref{eq:L3BH}}{\mathcal{L}^{(3)}_{\text{BH}}}  + \underset{\eqref{eq:L4BH}}{\mathcal{L}^{(4)}_{\text{BH}}} \nonumber\\
    & + \underset{\eqref{eq:L3gen}}{\mathcal{L}^{(3)}_{\text{gen}}} + \underset{\eqref{eq:LS4_cont}}{\mathcal{L}^{(4)}_{S^4}} + \underset{\eqref{eq:LS3K_cont}}{\mathcal{L}^{(4)}_{S^3K}} + \underset{\eqref{eq:LS2K2_cont}}{\mathcal{L}^{(4)}_{S^2K^2}} + \underset{\eqref{eq:LSK3_cont}}{\mathcal{L}^{(4)}_{SK^3}} + \underset{\eqref{eq:LK4_cont}}{\mathcal{L}^{(4)}_{K^4}} \,,
\end{align}
with the exact definition of each term given in the equation listed. To describe a \cco, we further impose the conditions~\eqref{eq:NS_cond} and~\eqref{eq:NS_cont_cond} on the Wilson coefficients such that the remaining Wilson coefficients agree with Ref.~\cite{Levi:2015msa},
\begin{align}\label{eq:L_NS}
    \mathcal{L}_{\text{\cco}} = \mathcal{L}_{\text{full}}\Big|_{\text{impose \cref{eq:NS_cond,eq:NS_cont_cond}}}\,.
\end{align}
These constraints effectively impose the covariant SSC such that the spin vector is the only internal degree of freedom of a \cco.
The Wilson coefficients are chosen such that we specialize to the case of Kerr black holes by setting all of them to zero,
\begin{align}
    \mathcal{L}_{\text{BH}} = - \frac{1}{2} \phis (\nabla^2+m^2) \phis + \underset{\eqref{eq:L3BH}}{\mathcal{L}^{(3)}_{\text{BH}}}  + \underset{\eqref{eq:L4BH}}{\mathcal{L}^{(4)}_{\text{BH}}} \,.
\end{align}
The above Lagrangians are derived by matching to on-shell classical Compton amplitudes at three and four points, as discussed in this section.
We take the large-spin limit and factor out $\polM_1\cdot\polM_2$ using \cref{eq:generalS}. We expose the full $K$ dependence and express the result in terms of the covariant spins. Therefore, the Compton amplitudes given in this section are the covariant spin amplitudes $\covM$ defined in \cref{eq:M_def}.
We also omit the subscripts on $\qftS$, $\qftK$, and mass in this section because there is only one massive spinning particle in the Compton amplitudes.
The spin-dependent factor in \cref{epdotep} is then used for the conversion to canonical spin variables when constructing the eikonal in \cref{sec:eikonal} and the Hamiltonian in \cref{sec:Hamiltonian}. 
%

\begin{figure}
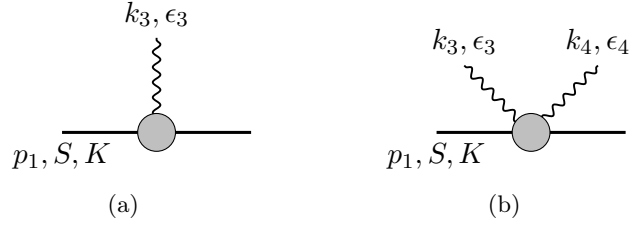

    \centering
    \subfloat[]{\label{fig:3pComp}\ThreePointAmp}
    \qquad\qquad
    \subfloat[]{\label{fig:4pComp}\FourPointAmp}
    \caption{The three- and four-point Compton amplitudes. All momenta are outgoing. }
\end{figure}

\subsection{Three-point interactions}\label{sec:3pCompAmp}

At three points, there is a unique gauge-invariant structure at each order in $S$ and $K$. The most generic amplitude is given by 
\begin{align}\label{eq:M3_full}
\covM^{\text{3pt}}_{\text{\gco}} &= -(\pol_3\cdot p_1)^2\sum_{n=0}^{\infty}\sum_{l=0}^{\infty}\frac{\mathcal{C}_{2n,l}}{(2n)! \, l!}\bigg[\frac{\kk_3\cdot \qftS}{m}\bigg]^{2n}\bigg[\frac{\kk_3\cdot \qftK}{m}\bigg]^{l} \nonumber\\
& \quad + \frac{(\pol_3\cdot p_1)\df_3(p_1,S)}{m}\sum_{n=0}^{\infty}\sum_{l=0}^{\infty}\frac{\mathcal{C}_{2n+1,l}}{(2n+1)! \, l!}\bigg[\frac{\kk_3\cdot \qftS}{m}\bigg]^{2n}\bigg[\frac{\kk_3\cdot \qftK}{m}\bigg]^{l}\,,
\end{align}
where $\df_i(p,S) = i\,\varepsilon_{\mu\nu\rho\sigma}p^{\mu}S^{\nu}k_i^{\rho}\pol_i^{\sigma}$, and $\mathcal{C}_{n,l}$ is the Wilson coefficient at $\order(S^n K^l)$. We do not allow contractions such as $S^2$, $K^2$, or $S\cdot K$ because, from dimensional grounds, these terms are accompanied by vanishing Mandelstam variables at three points. In addition, terms such as $|S| = \sqrt{-S^2}$ are not allowed at three points because there are no kinematics invariants that scale as a single power of the momentum transfer.

By choosing $\mathcal{C}_{n,l=0}=1$ and $\mathcal{C}_{n,l\neq 0}=0$ for the Wilson coefficients in Eq.~\eqref{eq:M3_full}, all dependence on the $K$ vector drops out and we obtain the Kerr stress-energy tensor~\cite{Vines:2017hyw} contracted against the graviton polarization tensor,\footnote{Our three-point amplitude $\covM^{\text{3pt}}_{\text{BH}}$ agrees with the stress-energy tensor in Ref.~\cite{Vines:2017hyw}. However, this comparison requires several changes of conventions, including the difference in the metric signature and Levi-Civita tensor. Moreover, we also need to cross the matter particle from outgoing to incoming. This effectively amounts to flipping $p_1\rightarrow -p_1$ while keeping all the other variables invariant. This leaves \cref{eq:M3_full} formally invariant.}
\begin{align}\label{eq:M3_Kerr}
\covM^{\text{3pt}}_{\text{BH}} &= -(\pol_3\cdot p_1)^2\sum_{n=0}^{\infty}\frac{1}{(2n)!}\bigg[\frac{\kk_3\cdot S}{m}\bigg]^{2n} + \frac{(\pol_3\cdot p_1)\df_3(p_1,S)}{m}\sum_{n=0}^{\infty}\frac{1}{(2n+1)!}\bigg[\frac{\kk_3\cdot S}{m}\bigg]^{2n}\, .
\end{align}
Our goal is to identify the three-point interaction Lagrangian that reproduces Eq.~\eqref{eq:M3_full} through $\order(M^4_{ab})$. 
Through this order, the three-point black-hole amplitude $\mathcal{M}^{\text{3pt}}_{\text{BH}}$ can be reproduced from the Feynman diagram in \cref{fig:3pComp} by the interaction Lagrangian
%
%
\begin{align}\label{eq:L3BH}
    \mathcal{L}^{(3)}_{\text{BH}} &= \frac{1}{8} R_{af_1bf_2} \, \phis \{M^{af_1}M^{bf_2}\}\phis - \frac{i}{12m^2}\nabla_{f_3}R_{af_1bf_2}\nabla^a\phis\{M^{f_3}{}_{c}M^{cf_1}M^{bf_2}\}\phis\nonumber\\
    &\quad + \frac{i}{12m^2}\nabla_{f_3}R_{af_1bf_2} \nabla_{c}\phis \{M^{f_3c}M^{af_1}M^{bf_2}\}\phis \nonumber\\
    &\quad + \frac{1}{96m^2} \nabla_{(f_4}\nabla_{f_3)}R_{af_1bf_2} \phis\{M^{f_3c}M_{c}{}^{f_4}M^{af_1}M^{bf_2}\}\phis \nonumber\\
    &\quad + \frac{1}{24m^4}\nabla_{(f_4}\nabla_{f_3)}R_{af_1bf_2} \nabla_{c}\phis \{M^{f_4c}M^{f_3}{}_{d}M^{da}M^{bf_2}\}\nabla^{f_1}\phis \nonumber\\
    &\quad + \frac{1}{24m^4}\nabla_{(f_4}\nabla_{f_3)}R_{af_1bf_2} \nabla_{c}\phis\{M^{f_4c}M^{f_3d}M^{af_1}M^{bf_2}\}\nabla_d\phis + \order(M_{ab}^5)\,,
\end{align}
which is identified as the three-point interaction for Kerr black holes in our field-theory formalism.
The above interactions introduce $K$ dependence in the three-point vertex, while the coefficients are chosen such that the $K$ dependence from the vertex is exactly canceled by the $K$-dependent exponential factor in \cref{eq:M_def}. We note that this exponential factor originates in the polarization product~\eqref{epdotep}. 

The first (kinematic) term in \cref{eq:full_L} fixes $\mathcal{C}_{0,0}=\mathcal{C}_{1,0}=1$ and $\mathcal{C}_{0,1}=0$, such that they also hold for the most generic \gco  case. We note that the $\mathcal{C}_{0,1}$ term is a mass dipole. Therefore, the \gco considered here is consistent with the absence of long-range dipole-mediated interactions in general relativity. We choose to parameterize the rest of $\mathcal{C}_{n,l}$ up to $\order(M_{ab}^4)$ as follows:
\begin{align}
    & \mathcal{C}_{2,0} = 1 + C_2 & & \mathcal{C}_{1,1} = D_2  & & \mathcal{C}_{0,2} = E_2  & &  & &\nonumber\\
    & \mathcal{C}_{3,0} = 1 + C_3 
    & & \mathcal{C}_{2,1} = D_3-C_2 & & \mathcal{C}_{1,2} = E_3  & & \mathcal{C}_{0,3} = F_3 & & \nonumber\\
    & \mathcal{C}_{4,0} = 1 + C_4 & & \mathcal{C}_{3,1} = D_4-C_3 & & \mathcal{C}_{2,2} = E_4-C_2 & & \mathcal{C}_{1,3} = F_4 & & \mathcal{C}_{0,4} = G_4\,.
\end{align}
The amplitude~\eqref{eq:M3_full} is then recovered by the interactions,
\begin{align}\label{eq:L3gen}
\mathcal{L}^{(3)}_{\text{gen}} &= -  \frac{C_{2}}{2m^2} R_{a f_1 b f_2} \nabla^a\phis \PLS^{(f_1}\PLS^{f_2)}\nabla^b\phis + \frac{D_2}{2m^2} R_{abcd} \nabla_i\phis \{M^{ai}M^{cd}\} \nabla^b\phis \nonumber\\
&\quad + \frac{E_2-2D_2}{2m^4} R_{abcd} \nabla^{(a}\nabla^{e)}\phis \{M^{b}{}_{e}M^{d}{}_{f}\}\nabla^{(c}\nabla^{f)}\phis \nonumber\\
&\quad+ \frac{C_3}{6m^3} \nabla_{f_3}\widetilde{R}_{af_1bf_2} \nabla^a\phis \PLS^{(f_1}\PLS^{f_2}\PLS^{f_3)} \nabla^b\phis + \frac{i D_3}{8m^2} \nabla_{f_3}R_{af_1bf_2} \nabla_e\phis\{M^{f_3e}M^{af_1}M^{bf_2}\}\phis \nonumber\\
&\quad + \frac{i(2D_3-2D_2-E_3)}{4m^4} \nabla_{f_3}R_{af_1bf_2} \nabla^{(e}\nabla^{f_1)}\phis \{M_{e}^{f_3}M^{bf_2}M^{ag}\} \nabla_g\phis \nonumber\\
&\quad + \frac{i(6D_2{-}3E_2{+}3E_3{-}3D_3{-}F_3)}{6m^6} \nabla_{f_3}R_{af_1bf_2} \nabla_{(c}\nabla_{d)}\phis \{M^{cf_3}M^{da}M^{b}{}_{e}\}\nabla^{(e}\nabla^{f_1}\nabla^{f_2)}\phis 
\nonumber\\
&\quad +  \frac{C_4}{24m^4} \nabla_{(f_4}\nabla_{f_3)} R_{af_1bf_2} \nabla^a\phis \PLS^{(f_1}\PLS^{f_2}\PLS^{f_3}\PLS^{f_4)} \nabla^b\phis \nonumber\\
&\quad + \frac{D_4}{12m^4}\nabla_{(f_4}\nabla_{f_3)}R_{af_1bf_2} \nabla_c\phis \{M^{f_4c}M^{f_3}{}_{d}M^{da}M^{bf_2}\}\nabla^{f_1}\phis \nonumber\\
&\quad + \frac{3E_4-2D_4-6C_2+6D_3}{48m^4} \nabla_{(f_4}\nabla_{f_3)}R_{af_1bf_2} \nabla_c\phis\{M^{f_4c}M^{f_3d}M^{af_1}M^{bf_2}\}\nabla_d\phis \nonumber\\
&\quad + \frac{F_4+3D_2+3E_3}{48m^6} \nabla_{(f_4}\nabla_{f_3)}R_{af_1bf_2} \nabla^{(c}\nabla^{f_1)}\phis \{M^{f_3}{}_{c}M^{f_4d}M^{ae}M^{bf_2}\} \nabla_{(e}\nabla_{d)}\phis \nonumber\\
&\quad + \frac{G_4+6E_2+4F_3}{48m^8} \nabla_{(f_4}\nabla_{f_3)}R_{af_1bf_2} \nabla^{(c}\nabla^{d}\nabla^{f_1)}\phis\{M^{f_3}{}_{c}M^{f_4}{}_{d}M^{a}_{e}M^{b}{}_{g}\} \nabla^{(e}\nabla^{g}\nabla^{f_2)}\phis \nonumber\\
&\quad + \order(M_{ab}^5)\,,
\end{align}
where $\widetilde{R}_{abcd}=(1/2)\varepsilon_{abef}R^{ef}{}_{cd}$ is the dual Riemann tensor. The interactions proportional to $C_2$, $C_3$, and $C_4$ are given in terms of the Pauli-Lubanski spin operator $\PLS$ and thus only contribute to spin-dependent terms in the amplitude. They are spin-induced multipole moments and have a direct correspondence with the non-minimal interactions of \ccos given in, for example, Ref.~\cite{Levi:2015msa}.
On the other hand, the Lorentz generator $M_{ab}$ has both $S$ and $K$ contribution in the amplitudes, while the combination $M_{ab}\nabla^{b}\phis$ only gives a $K$ contribution.\footnote{Expressed in terms of $M_{ab}$, the interactions give simpler Feynman rules. However, as $M_{ab}$ contributes to terms containing both $\qftS$ and $\qftK$ in the amplitudes, we often need a linear combination of $M_{ab}$ structures to get a single $S^{n}K^{l}$ term in \cref{eq:M3_full}.}
Notably, if the Wilson coefficients take the special values,
\begin{align}\label{eq:NS_cond}
    D_3=E_4=C_2\,, \qquad D_4=C_3\,, \qquad D_2=E_2=E_3=F_3=F_4=G_4=0\,, 
\end{align}
we have $\mathcal{C}_{n,l\neq 0}=0$ and the $K$ vector completely drops out of the three-point amplitude. The only remaining Wilson coefficients are $C_n$. We thus recover the amplitude for \ccos, with an imposed SSC~\cite{Levi:2015msa},
\begin{align}\label{eq:NS}
\covM^{\text{3pt}}_{\text{\cco}} &= -(\pol_3\cdot p_1)^2\sum_{n=0}^{\infty}\frac{1+C_{2n}}{(2n)!}\bigg[\frac{\kk_3\cdot \qftS}{m}\bigg]^{2n} + \frac{(\pol_3\cdot p_1)\df_3(p_1,\qftS)}{m}\sum_{n=0}^{\infty}\frac{1+C_{2n+1}}{(2n+1)!}\bigg[\frac{\kk_3\cdot \qftS}{m}\bigg]^{2n}\,.
\end{align}
To summarize, $\mathcal{L}_{\text{BH}}^{(3)}+\mathcal{L}_{\text{gen}}^{(3)}$ gives the non-minimal coupling between Einstein gravity and a compact body whose multipole moments are generated by the $S$  and $K$ vectors. Naturally, this Lagrangian allows for a larger number of Wilson coefficients than that describing \ccos, whose multipole moments are generated by spin only.
\Cref{eq:M3_full,eq:NS,eq:M3_Kerr} are also the stress tensors of an extended compact object, Kerr black hole and conventional compact object contracted with a graviton polarization tensor, respectively. Much like the stress tensor of a moving scalar particle sources the metric of a boosted Schwarzschild black hole~\cite{Duff:1973zz, Jakobsen:2020ksu, Mougiakakos:2024nku}, these stress tensors also source the space-time metrics of the corresponding objects.\footnote{The stress tensor of a Kerr black hole was, in fact, constructed~\cite{Vines:2017hyw} to have this property.} Thus, as a first indication that the $K$ vector has physical consequences, we note that, for a particle at rest, the stress tensor \eqref{eq:M3_full} sources a $K$-dependent stationary metric. 
From a different perspective, it is known that, in certain supergravity theories, classical solutions with the interpretation of black hole microstates exhibit a mass dipole (see e.g.~Refs.~\cite{Bena:2020uup, Bah:2021jno} or the review~\cite{Bena:2022ldq}).
While on its own, this mass dipole is a gauge degree of freedom, it contributes nontrivially to the higher multipole moments of these solutions. 
It would be interesting to extend our analysis in \cref{sec: metricWK} to such horizonless solutions and study their relation to our $K$-dependent stress tensor.



\subsection{Four-point interactions}

With the non-minimal interaction $\mathcal{L}_{\text{BH}}^{(3)}+\mathcal{L}_{\text{gen}}^{(3)}$ understood, one can continue and compute four-point Compton amplitudes shown in \cref{fig:4pComp}. Using Feynman diagrams, we first compute the full quantum amplitudes, which we later use to construct one-loop two-body amplitudes via generalized unitarity~\cite{Bern:1994zx, Bern:1994cg, Kosmopoulos:2020pcd}. We then take the classical limit by scaling the massless momenta and spin variables as
\begin{align}\label{eq:soft_scale_Comp}
    k_3\rightarrow \lambda k_3\,,\qquad k_4\rightarrow \lambda k_4\,,\qquad S\rightarrow \lambda^{-1} \qftS\,,\qquad K\rightarrow \lambda^{-1}\qftK \,.
\end{align}
The Compton amplitudes under the $\lambda$ expansion scale as $\lambda^0$ at the leading order, which corresponds to the classical contribution. 
In four dimensions, the opposite $(-+)$ and the same $(++)$ graviton helicity amplitudes encode the complete physical information. 
We choose the following polarization vectors for on-shell gravitons,
\begin{align}
& \text{opposite helicity:} & & \pol_3^{-}{}^\mu = \frac{\langle 3|\sigma^{\mu}|4]}{\sqrt{2}\,[34]}  & & \pol_4^{+}{}^\mu = \frac{\langle 3|\sigma^{\mu}|4]}{\sqrt{2}\,\langle 34\rangle}\,, \nonumber\\
& \text{same helicity:} & & \pol_3^{+}{}^\mu = \frac{[4|p_1\sigma^{\mu}|3]}{\sqrt{2}\,[4|p_1|3\rangle} & & \pol_4^{+}{}^\mu = \frac{[3|p_1\sigma^{\mu}|4]}{\sqrt{2}\,[3|p_1|4\rangle}\, .
\end{align}
Following Ref.~\cite{Haddad:2023ylx}, for convenience, we also define the following variables,
\begin{align}
    & \wpm^\mu = \frac{\langle 3|\sigma^{\mu}|4]}{2} \,,& & \ypm = \langle 3|p_1|4]  \,,
 & & \aa^{\mu} = \frac{\qftS^{\mu}}{m} \,,
    \nonumber\\[4pt]
    & \hat{k}_3 = \kk_3 - \frac{2p_1\cdot \kk_3}{y_{-+}}\wpm\,,  & & \hat{k}_4 = \kk_4 + \frac{2p_1\cdot \kk_3}{y_{-+}}\wpm\,,
    & & \kK^{\mu} = \frac{\qftK^{\mu}} {m} \,, 
\nonumber\\[4pt]
    &  \wpp^\mu = \frac{[4|p_1\sigma^{\mu}|3]}{2m}\,, & & \ypp = -m [34] \,, \label{EqConvenientVariables}
\end{align}
where we also include the $K$-dependence.

Given $\mathcal{L}_{\text{BH}}^{(3)}+\mathcal{L}_{\text{gen}}^{(3)}$, the four-point classical Compton amplitudes up to $\order(\gS^{3})$ are uniquely fixed because classically relevant four-point contact interactions only exist for $\order(M_{ab}^{4})$ and beyond. Therefore, the decoupling of $K$ under the special values~\eqref{eq:NS_cond} should hold automatically at $\order(\gS^3)$ for the four-point Compton amplitudes. This is a self-consistency check. At $\order(M_{ab}^4)$, we first enumerate all the independent contact interactions, including both spin and $K$ dependence. We then find the special values for the Wilson coefficients that decouple the $K$ degrees of freedom and reduce to conventional compact objects, including Kerr black holes. 

We start with Kerr black holes. It turns out that by using the kinematic term and $\mathcal{L}_{\text{BH}}^{(3)}$ only, we exactly reproduce the known opposite-helicity amplitude~\cite{Aoude:2020onz,Chen:2021kxt} up to $\order(\aa^4)$, 
\begin{align}
    \covM^{-+}_{\text{BH}} &= - \frac{(\ypm)^4}{32(p_1\cdot \kk_3)^2(\kk_3\cdot \kk_4)}\sum_{n=0}^{4} \frac{(\hat{k}_4\cdot \aa-\hat{k}_3\cdot \aa)^{n}}{n!} + \order(\aa^5) \,.
\end{align}
On the other hand, the same helicity amplitudes for Kerr black holes are expected to have an all-orders-in-spin closed formula~\cite{Aoude:2020onz,Chen:2021kxt}, 
\begin{align}\label{eq:BH_pp}
    \covM^{++}_{\text{BH}} &= - \frac{(\ypp)^4}{32(p_1\cdot \kk_3)^2(\kk_3\cdot \kk_4)}\sum_{n=1}^{\infty} \frac{(\kk_3\cdot\aa + \kk_4\cdot\aa)^{n}}{n!} \,,
\end{align}
as was recently verified through 8\textsuperscript{th} order in the spin from a solution to the Teukolsky equation~\cite{Bautista:2023sdf}.
Using $\mathcal{L}_{\text{BH}}^{(3)}$, we exactly reproduce this result up to $\order(\aa^3)$.
At $\order(\aa^4)$, the amplitude produced by $\mathcal{L}_{\text{BH}}^{(3)}$ differ from \cref{eq:BH_pp} only by contact terms. They can be canceled by the following contact interactions, which only contribute to the same-helicity sector at $\order(\aa^4)$,
\begin{align}\label{eq:L4BH}
\mathcal{L}_{\text{BH}}^{(4)} &= \frac{1}{384m^2}R_{abcd}R^{abcd} \phis\{M^{ef}M_{ef}M^{gh}M_{gh}\}\phis \nonumber\\
&\quad - \frac{1}{64m^2} R_{a f_1 cd}R^{cd}{}_{bf_2} \phis\{M^{af_1}M^{bf_2}M^{gh}M_{gh}\}\phis \nonumber\\
&\quad + \frac{1}{96m^4}  R_{a f_1 cd}R^{cd}{}_{bf_2} \nabla^{f_2}\phis\Big[\{M^{af_1}M^{be}M^{gh}M_{gh}\} - 5\{M^{ab}M^{f_1e}M^{gh}M_{gh}\} \nonumber\\
& \qquad\qquad\qquad\qquad\qquad\qquad\quad +2\{M^{af_1}M^{bg}M_{gh}M^{he}\}\Big]\nabla_e\phis \nonumber\\
&\quad - \frac{1}{384m^4} R_{abcd}R^{abcd} \phis \Big[ \{M^{ef_1}M_{ef_1}M^{gf_2}M_{f_2}{}^{h}\} -2 \{M^{ge}M_{ef_1}M^{f_1f_2}M^{f_2h}\}\Big]\nabla_g\nabla_h\phis \nonumber\\
&\quad - \frac{5}{48m^6} R_{a f_1 cd}R^{cd}{}_{bf_2} \nabla^{f_1}\nabla^{f_2}\phis \{M^{be}M_{eg}M^{gh}M^{ai}\} \nabla_h\nabla_i\phis  \nonumber\\
&\quad - \frac{5}{96m^6} R_{a f_1 cd}R^{cd}{}_{bf_2} \nabla^{f_1}\nabla^{f_2}\phis \{M^{ai}M^{bh}M^{eg}M_{eg}\} \nabla_h\nabla_i\phis  \nonumber\\
&\quad + \frac{11}{192m^4} R_{a f_1 cd}R^{cd}{}_{bf_2} \phis \{ M^{af_1}M^{bf_2}M^{eg}M_{g}{}^{h}\} \nabla_e\nabla_h\phis \nonumber\\
&\quad - \frac{5}{48m^6} R_{a f_1 cd}R^{cd}{}_{bf_2} \phis \Big[ \{M^{ab}M^{f_2h}M^{f_3e}M_{f_3}{}^{g}\} + \{ M^{bf_2}M^{ah}M^{f_3e}M_{f_3}{}^{g}\} \nonumber\\
&\qquad\qquad\qquad\qquad\qquad\quad + \{M^{f_2f_3}M_{f_3}{}^{h}M^{ae}M^{bg}\} \Big]\nabla^{f_1}\nabla_{e}\nabla_{g}\nabla_{h}\phis \,.
\end{align}
That is, within our field-theory formalism, the following Lagrangian,
\begin{align}
\mathcal{L} = - \frac{1}{2} \phis (\nabla^2+m^2) \phis + \mathcal{L}^{(3)}_{\text{BH}} + \mathcal{L}^{(4)}_{\text{BH}} \,,
\end{align}
is sufficient to describe Kerr black holes in a scattering process up to $\order(\aa^4)$. Consequently, because the Compton amplitudes are the basic building blocks, it is also sufficient for describing the scattering of two Kerr black holes at one-loop level through $\order(\aa^4)$.

Next, we consider the generic compact object described by $\mathcal{L}_{\text{BH}}^{(3)}+\mathcal{L}_{\text{gen}}^{(3)}$. Up to $\order(\gS^3)$, the opposite-helicity Compton amplitude is given by
\begin{align}\label{eq:Mgco_S3pm}
    & \covM^{-+}_{\text{\gco}} = \covM^{-+}_{\text{\cco}} - \frac{(\ypm)^4}{32(p_1\cdot k_3)^2(k_3\cdot k_4)} 
    \left\{ - D_2 \left[(\aa\cdot\hat{k}_3)(\kK\cdot\hat{k}_3)-(\aa\cdot\hat{k}_4)(\kK\cdot\hat{k}_4)\right]\vphantom{\frac{\aa\cdot\wpm}{\wpm}} \right. \nonumber\\
    &\quad + \frac{E_2}{2}\left[(\kK\cdot\hat{k}_3)^2 + (\kK\cdot\hat{k}_4)^2\right] + 2D_2^2 (p_1\cdot k_3)(\aa\cdot\hat{k}_3)(\aa\cdot\hat{k}_4)\frac{\aa\cdot\wpm}{\ypm} \nonumber\\
    &\quad + \frac{D_3-C_2}{2}\left[(\aa\cdot\hat{k}_3)^2(\kK\cdot\hat{k}_3)+(\aa\cdot\hat{k}_4)^2(\kK\cdot\hat{k}_4)\right] - D_2(\aa\cdot\hat{k}_3)(\aa\cdot\hat{k}_4)(\kK\cdot\hat{k}_3+\kK\cdot\hat{k}_4) \nonumber\\
    &\quad + 2D_2^2 (p_1\cdot k_3)\left[\frac{\aa\cdot\wpm}{\ypm}(\kK\cdot\hat{k}_3)(\kK\cdot\hat{k}_4) + \left[(\aa\cdot\hat{k}_4)(\kK\cdot\hat{k}_3)+(\aa\cdot\hat{k}_3)(\kK\cdot\hat{k}_4)\right]\frac{\kK\cdot\wpm}{\ypm}\right] \nonumber\\
    &\quad + 2 (C_2-E_2) D_2 (p_1\cdot k_3)\frac{\aa\cdot\wpm}{\ypm}\left[(\aa\cdot\hat{k}_4)(\kK\cdot\hat{k}_3)-(\aa\cdot\hat{k}_3)(\kK\cdot\hat{k}_4)\right] \nonumber\\
    &\quad - 2 C_2 E_2 (p_1\cdot k_3)\frac{\kK\cdot\wpm}{\ypm}\left[(\aa\cdot\hat{k}_4)(\kK\cdot\hat{k}_3)-(\aa\cdot\hat{k}_3)(\kK\cdot\hat{k}_4)\right]\nonumber\\
    &\quad + \frac{E_2}{2}\left[(\aa\cdot\hat{k}_4)(\kK\cdot\hat{k}_3)^2-(\aa\cdot\hat{k}_3)(\kK\cdot\hat{k}_4)^2\right] - 2E_2^2 (p_1\cdot k_3) \frac{\aa\cdot\wpm}{\wpm}(\kK\cdot\hat{k}_3)(\kK\cdot\hat{k}_4) \\
    &\quad - \left.\frac{E_3}{2}\left[(\aa\cdot\hat{k}_3)(\kK\cdot\hat{k}_3)^2-(\aa\cdot\hat{k}_4)(\kK\cdot\hat{k}_4)^2\right] + \frac{F_3}{6}\left[(\kK\cdot\hat{k}_3)^3+(\kK\cdot\hat{k}_4)^3\right]\right\} + \order(\gS^4) \,,\nonumber
\end{align}
where $\covM^{-+}_{\text{\cco}}$ is the Compton amplitude for the conventional compact body with only the spin degrees of freedom. It has the following closed formula up to $\order(\aa^4)$~\cite{Haddad:2023ylx},
\begin{align}\label{eq:MccoS4PM}
    \covM^{-+}_{\text{\cco}} &= - \frac{(\ypm)^4}{32(p_1\cdot k_3)^2(k_3\cdot k_4)} \left[\sum_{0\leqslant n+l\leqslant 4}\frac{(1+C_n)(1+C_l)}{n!\;l!}(-\hat{k}_3\cdot\aa)^{n}(\hat{k}_4\cdot\aa)^{l} \right. \\
    &\quad + \left.\frac{2(p_1\cdot k_3)(\aa\cdot w_{-+})}{y_{-+}}\!\sum_{0\leqslant n+l\leqslant 3}\!\frac{(C_n-C_{n+1})(C_l-C_{l+1})}{n! \; l!}(-\hat{k}_3\cdot\aa)^{n}(\hat{k}_4\cdot\aa)^{l}\right] \!\! + \order(\aa^5)\,, \nonumber
\end{align}
where $C_0=C_1=0$.
Indeed, when the Wilson coefficients take the special values as in \cref{eq:NS_cond}, $\covM^{-+}_{\text{\gco}}$ reduces to $\covM^{-+}_{\text{\cco}}$. The same is true for the same-helicity sector. Up to $\order(\gS^3)$, we have
\begin{align}\label{eq:Mgco_S3pp}
\covM^{++}_{\text{\gco}} = \covM^{++}_{\text{\cco}} + \covM_{\text{extra}}^{++}\,,
\end{align}
where $\covM_{\text{\cco}}^{++}$ is given by
\begin{align}
    \covM^{++}_{\text{\cco}} &= - \frac{(\ypm)^4}{32(p_1\cdot k_3)^2(k_3\cdot k_4)}  \left\{ \frac{C_2}{2}\left[(\aa\cdot k_{3}+\aa\cdot k_4)^2-(k_3\cdot k_4)\left(\aa^2+4m^2\frac{(\aa\cdot\wpp)^2}{\ypp^2}\right)\right] \right. \nonumber\\
    &\quad + \frac{C_2 (k_3\cdot k_4)}{4}\left(\aa\cdot k_{3}+\aa\cdot k_4+12(p_1\cdot k_3)\frac{\aa\cdot\wpp}{\ypp}\right)\left(\aa^2+4m^2\frac{(\aa\cdot\wpp)^2}{\ypp^2}\right) \nonumber\\
    &\quad + C_2^2 (k_3\cdot k_4)(p_1\cdot k_3)\frac{\aa\cdot\wpp}{\ypp}\left(\aa^2+4m^2\frac{(\aa\cdot\wpp)^2}{\ypp^2}\right) + \frac{C_3}{6}\left[(\aa\cdot k_3+\aa\cdot k_4)^3\vphantom{\frac{\aa\cdot\wpp}{\ypp}}\right.\nonumber\\
    &\quad - \left.\left. \frac{3k_3\cdot k_4}{2}\left(\aa\cdot k_{3}+\aa\cdot k_4+4(p_1\cdot k_3)\frac{\aa\cdot\wpp}{\ypp}\right)\left(\aa^2+4m^2\frac{(\aa\cdot\wpp)^2}{\ypp^2}\right)\right]\right\} \nonumber\\
    &\quad + \covM^{++}_{\text{BH}} + \order(\aa^4)\,.
\end{align}
This expression agrees with Eq.~(B.4) of Ref.~\cite{Haddad:2023ylx}.
On the other hand, $\covM_{\text{extra}}^{++}$ contains the $K$ dependence and vanishes under the condition~\eqref{eq:NS_cond}. The full expression of \cref{eq:Mgco_S3pp} in spinor variables is included in the ancillary file {\tt CompGCO\_M3.m}.

Next, we examine generic contact interactions at $\order(M_{ab}^4)$ that include both $S$ and $K$ dependence. 
We organize them based on the powers of $S$ and $K$,
\begin{align}
    \underset{\eqref{eq:LS4_cont}}{\mathcal{L}^{(4)}_{S^4}} + \underset{\eqref{eq:LS3K_cont}}{\mathcal{L}^{(4)}_{S^3K}} + \underset{\eqref{eq:LS2K2_cont}}{\mathcal{L}^{(4)}_{S^2K^2}} + \underset{\eqref{eq:LSK3_cont}}{\mathcal{L}^{(4)}_{SK^3}} + \underset{\eqref{eq:LK4_cont}}{\mathcal{L}^{(4)}_{K^4}} \,.
\end{align}
The contact interactions that only have spin contributions are identical to those for conventional compact objects. There are in all six independent operators, and we write them in a form that resembles the worldline operators in Ref.~\cite{Scheopner:2023rzp},
\begin{align}\label{eq:LS4_cont}
\mathcal{L}^{(4)}_{S^4} &= \frac{C_{4a}}{48m^6}(R_{af_1bf_2}R_{cf_3df_4}+\widetilde{R}_{af_1bf_2}\widetilde{R}_{cf_3df_4})\nabla^a\nabla^c\phis \PLS^{(f_1}\PLS^{f_2}\PLS^{f_3}\PLS^{f_4)} \nabla^b\nabla^d\phis \nonumber\\
&\quad + \frac{C_{4b}}{48m^6}(R_{af_1bf_2}R_{cf_3df_4}-\widetilde{R}_{af_1bf_2}\widetilde{R}_{cf_3df_4})\nabla^a\nabla^c\phis \PLS^{(f_1}\PLS^{f_2}\PLS^{f_3}\PLS^{f_4)} \nabla^b\nabla^d\phis \nonumber\\
&\quad - \frac{C_{4c}}{48m^6}(R_{af_1bf_2}R^{f_2}{}_{cdf_4}+\widetilde{R}_{af_1bf_2}\widetilde{R}^{f_2}{}_{cdf_4})\nabla^a\nabla^c\phis \PLS^{(f_1}\PLS^{f_4)}\PLS_{i}\PLS^{i} \nabla^b\nabla^d\phis \nonumber\\
&\quad - \frac{C_{4d}}{48m^6}(R_{af_1bf_2}R^{f_2}{}_{cdf_4}-\widetilde{R}_{af_1bf_2}\widetilde{R}^{f_2}{}_{cdf_4})\nabla^a\nabla^c\phis \PLS^{(f_1}\PLS^{f_4)}\PLS_{i}\PLS^{i} \nabla^b\nabla^d\phis \nonumber\\
&\quad + \frac{C_{4e}}{48m^6}(R_{af_1bf_2}R^{f_1}{}_{c}{}^{f_2}{}_{d}+\widetilde{R}_{af_1bf_2}\widetilde{R}^{f_1}{}_{c}{}^{f_2}{}_{d})\nabla^a\nabla^c\phis \PLS_{k}\PLS^{k}\PLS_{i}\PLS^{i} \nabla^b\nabla^d\phis \nonumber\\
&\quad + \frac{C_{4f}}{48m^6}(R_{af_1bf_2}R^{f_1}{}_{c}{}^{f_2}{}_{d}-\widetilde{R}_{af_1bf_2}\widetilde{R}^{f_1}{}_{c}{}^{f_2}{}_{d})\nabla^a\nabla^c\phis \PLS_{k}\PLS^{k}\PLS_{i}\PLS^{i} \nabla^b\nabla^d\phis \,.
\end{align}
They lead to the following contact structures in on-shell amplitudes,
\begin{align}\label{eq:S4_amp_cont}
    S^4\text{ contact terms:}\qquad \frac{(y_{\pm +})^4 \aa^4}{m^2}\, , \hskip 1 cm  (y_{\pm +})^2(\aa\cdot w_{\pm +})^2 \aa^2\,, \hskip 1 cm m^2 (\aa\cdot w_{\pm +})^4 \,.
\end{align}
The above counting and on-shell structures agree with those given in Ref.~\cite{Haddad:2023ylx}. In this paper, we are interested in purely conservative dynamics. Thus, dissipative contact terms that are proportional to odd powers of $|\aa|$ are not included.

We now perform a similar on-shell analysis of the contact interactions that contribute to the $K$ dependence.
We only keep the classically relevant terms, namely, those that scale as $\lambda^{0}$ under \cref{eq:soft_scale_Comp}. Assuming conservative dynamics, we find $6$ independent structures in the $S^3K$ sector and $12$ in the $S^2K^2$ sector,
\begin{subequations}\label{eq:M4_amp_cont}
\begin{align}
    \label{eq:S3K_amp_cont}
    & S^3K\text{ contact terms:} & & \frac{(y_{-+})^3\aa^2(\varepsilon_{\mu\nu\rho\sigma}\kK^{\mu}w^{\nu}_{-+}p^{\rho}_1\aa^{\sigma})}{m^2}\,,  & & y_{-+}(\aa\cdot w_{-+})^2(\varepsilon_{\mu\nu\rho\sigma}\kK^{\mu}w^{\nu}_{-+}p^{\rho}_1\aa^{\sigma})\,, \nonumber\\[3pt]
    & & & \frac{(y_{++})^4\aa^2(\aa\cdot\kK)}{m^2}\,, & & (y_{++})^2\aa^2(\aa\cdot w_{++})(\kK\cdot w_{++}) \,,\nonumber\\
    & & & (y_{++})^2(\aa\cdot w_{++})^2(\aa\cdot\kK)\,,  & & m^2(\aa\cdot w_{++})^3(\kK\cdot w_{++}) \,,\\[3pt]
    & S^2K^2\text{ contact terms:} & & \frac{(y_{\pm +})^4\aa^2\kK^2}{m^2}\,,  & & (y_{\pm +})^2(\aa\cdot\kK)(\aa\cdot w_{\pm +})(\kK\cdot w_{\pm +}) \,, \nonumber\\[3pt]
    & & & \frac{(y_{\pm +})^4(\aa\cdot\kK)^2}{m^2}\,, & & (y_{\pm +})^2\kK^2(\aa\cdot w_{\pm +})^2 \,,\nonumber\\[3pt]
    & & & (y_{\pm +})^2\aa^2(\kK\cdot w_{\pm +})^2\,, & & m^2(\aa\cdot w_{\pm +})^2(\kK\cdot w_{\pm +})^2\,, \\[3pt]
    & SK^3\text{ contact terms:} & & \aa\leftrightarrow\kK\text{ in \cref{eq:S3K_amp_cont}} \,,\\[3pt]
    & K^4\text{ contact terms:} & & \aa\rightarrow\kK\text{ in \cref{eq:S4_amp_cont}}\,.
\end{align}
\end{subequations}
Note that the $SK^3$ and $K^4$ contact structures can be obtained by a relabeling between the $S$ and $K$ vectors. As before, we only consider conservative dynamics. The independent gauge-invariant contact structures can be mapped into corresponding $R^2$ contact interactions in the Lagrangian,
\begin{align}\label{eq:LM4_cont}
    \mathcal{L}_{S^3K}^{(4)} &= \Big[\text{\;\cref{eq:LS3K_cont}\;}\Big] & & \text{Wilson coefficients:}\quad \{D_{4a},D_{4b},\ldots,D_{4f}\} \,,\nonumber\\
    \mathcal{L}_{S^2K^2}^{(4)} &= \Big[\text{\;\cref{eq:LS2K2_cont}\;}\Big] & & \text{Wilson coefficients:}\quad \{E_{4a},E_{4b},E_{4c},E_{4d},\ldots,E_{4l}\}\,, \nonumber\\
    \mathcal{L}_{SK^3}^{(4)} &= \Big[\text{\;\cref{eq:LSK3_cont}\;}\Big] & & \text{Wilson coefficients:}\quad \{F_{4a},F_{4b},\ldots,F_{4f}\} \,,\nonumber\\
    \mathcal{L}_{K^4}^{(4)} &= \Big[\text{\;\cref{eq:LK4_cont}\;}\Big] & & \text{Wilson coefficients:}\quad \{G_{4a},G_{4b},\ldots,G_{4f}\}\,,
\end{align}
where we collect the explicit formulas in Appendix~\ref{sec:LcontactK}. The number of independent on-shell contact terms given in \cref{eq:M4_amp_cont} matches the number of independent Wilson coefficients. This completes the Lagrangian~\eqref{eq:full_L} for our compact spinning body with both $S$ and $K$ degrees of freedom.

We also need to derive the specific values of the Wilson coefficients listed in \cref{eq:LM4_cont} that eliminate the $K$ dependence from the four-point Compton amplitudes at $\order(\gS^4)$.
Using $\mathcal{L}_{\text{BH}}^{(3)}+\mathcal{L}_{\text{gen}}^{(3)}$ and the condition~\eqref{eq:NS_cond}, we find that the spin part agrees with the known results, while the $K$ dependent part is local. More specifically, our $\order(\aa^4)$ Compton amplitude agrees with the Compton amplitude used in Ref.~\cite{Bern:2022kto}.\footnote{The comparison is carried out when all the extra Wilson coefficients in Ref.~\cite{Bern:2022kto} are removed.} In terms of the spinor helicity variables, the opposite-helicity Compton amplitudes agree with the $\order(\aa^4)$ part of \cref{eq:MccoS4PM}, while the same-helicity amplitude agrees with Eq.~(3.17) of Ref.~\cite{Haddad:2023ylx}.\footnote{The comparison with Ref.~\cite{Haddad:2023ylx} requires a change of convention $p_1\rightarrow -p_1$, $k_{3,4}\rightarrow q_{3,4}$, $\aa\rightarrow\mathfrak{a}$ and $C_{i}\rightarrow C_{S^{i}}-1$.} All the agreements here are up to the spin-dependent contact terms given in \cref{eq:S4_amp_cont}.

On the other hand, it is remarkable that the $K$ dependence computed from $\mathcal{L}_{\text{BH}}^{(3)}+\mathcal{L}_{\text{gen}}^{(3)}$ and \cref{eq:NS_cond} is local,\footnote{The exponential factor $\exp\left[\frac{q\cdot K}{m}\right]$ in $\polM_1\cdot\polM_2$ is crucial for the final $K$ dependence to be local. For example, there are two contributions to the $\order(S^3K)$ amplitude: one from $\widetilde{\mathcal{M}}(S^3,K)$ and the other from $\frac{(k_3+k_4)\cdot K}{m}\widetilde{\mathcal{M}}(S^3)$. When \cref{eq:NS_cond} is imposed, the non-local terms in the $\order(S^3 K)$ amplitude cancel among the two contributions and, at most, leave behind a local contact term.} which can then be canceled by the $R^2$ contact interactions given in \cref{eq:LM4_cont}. This requirement fixes the Wilson coefficients in \cref{eq:LM4_cont} to the following special values,
\begin{align}\label{eq:NS_cont_cond}
    \renewcommand{\arraystretch}{1.3}
    \begin{array}{cccccc}
    D_{4a} = D_{4b} = C_2  & D_{4c} = C_2 + \frac{C_3}{4} & \;\; & D_{4d} = C_2 - \frac{C_3}{4} & \;\; & D_{4e} = D_{4f} = C_3 \\
    \multicolumn{2}{c}{E_{4a} = E_{4c} = E_{4e} = E_{4g} = E_{4i} = E_{4k} = C_2} & & E_{4b} = C_2 + \frac{C_3}{2} & &E_{4d} = C_2 + \frac{3C_3}{2} \\
    E_{4f} = C_{2} - \frac{C_3}{2}  & E_{4h} = C_2 + \frac{5C_3}{2} & & E_{4j} = C_2 - \frac{7C_3}{2}\,, & & E_{4l} = C_2 - \frac{3C_3}{2} \\
    F_{4a} = C_2 & F_{4b} = F_{4f} = 0 \,,& & F_{4c} = C_2 + \frac{C_3}{4}  & & F_{4d} = F_{4e} = C_3 \\
    \multicolumn{6}{c}{G_{4a} = G_{4b} = G_{4c} = G_{4d} = G_{4e} = G_{4f} = 0\,.}
    \end{array}
\end{align}
Imposing the condition~\eqref{eq:NS_cond} and~\eqref{eq:NS_cont_cond} onto the Lagrangian~\eqref{eq:full_L} leads to a field-theory description of conventional compact objects. The $K$ dependence, although existing in the Lagrangian and intermediate steps, drops out of the on-shell amplitudes and observables. The compact object is then characterized by the Wilson coefficients $C_{2,3,4}$ and those for contact interactions in \cref{eq:LS4_cont}.
For such cases, we obtain the field-theory model for conventional compact objects (and, in particular, Kerr black holes when these Wilson coefficients vanish),  equivalent to imposing the SSC. 
This indicates that our field-theory model with additional degrees of freedom is a consistent generalization of the conventional spin dynamics.
We emphasize that the decoupling of $K$ is not realized by introducing additional constraints, as accomplished by the SSC, but rather, it appears as a redundancy of the theory for the special choice of Wilson coefficients.
The generic $\order(\gS^4)$ Compton amplitudes in terms of spinor variables,
\begin{align}
    \covM^{-+}_{\text{\gco}}\Big|_{\order(\gS^4)} \quad\text{and}\quad \covM^{++}_{\text{\gco}}\Big|_{\order(\gS^4)}
\end{align}
are included in the ancillary file {\tt CompGCO\_M4.m}.
The specialization to conventional compact bodies and Kerr black holes can also be found therein.

We note that the Lagrangian~\eqref{eq:L_NS} describing conventional compact objects still contains interactions that depend on $M_{ab}\nabla^b\phis\sim K_a$, while the decoupling of $K$ only happens at the level of on-shell amplitudes and observables. This feature also exists in the recent worldline QFT formalism that models the spin tensor as a pair of bosonic oscillators~\cite{Haddad:2024ebn}, which also proposes conditions that are similar in spirit to \cref{eq:NS_cond,eq:NS_cont_cond} for the $K$-dependent operators.\footnote{The vector $K$ in our work is called $Z$ in Ref.~\cite{dAmbrosi:2015ndl, vanHolten:2015vfa, dAmbrosi:2015wqz, Haddad:2024ebn}.} On the other hand, operators that contribute to $\order(K^{n\geqslant 2})$ are necessary in our formalism, while only linear-in-$K$ operators are present in Ref.~\cite{Haddad:2024ebn}. This difference originates from whether $K$ is treated as a dynamical degree of freedom. Given the subtle technical nature of this point, it is best to compare at the level of observables rather than intermediate steps; further comments are made in \cref{sec:wl}.


\section{Worldline}
\label{sec:wl}


In this section, we describe a worldline gravitational theory without imposing a spin supplementary condition (SSC), contrasting it with the formulation where an SSC is imposed. Ref.~\cite{Bern:2023ity} provides a detailed analysis of the analogous electrodynamic case. The gravitational case was briefly summarized in Ref.~\cite{Alaverdian:2024spu}, showing similar results. Those papers demonstrate that through $\order(G^2 \gS^2)$, the two approaches yield physically distinct results for generic choices of Wilson coefficients—specifically, differences in scattering angles and waveforms. Interestingly, for a subset of Wilson coefficients, including those describing Kerr black holes, both theories produce identical physical observables. Here, we present further details on the gravitational case, including worldline impulses and Compton amplitudes, and describe the connection to dynamical multipole moments. 

A similar worldline action was constructed in Ref.~\cite{Haddad:2024ebn} to leading order in the $\covclK$ vector (referred to there as the $Z$ vector). In this reference, the goal was to impose the SSC by demanding the vanishing of the change of the $\covclK$ vector, for which terms linear in $\covclK$ are sufficient. 
Another worldline theory that includes $p_a \gS^{ab}$ as additional degrees of freedom has been proposed in Ref.~\cite{Porto:2008tb, Porto:2008jj}; in this case, the dynamical decoupling was imposed as an equation of motion.
Here, we keep the complete dependence on $\covclK$ because, as we saw in earlier sections and discuss further in subsequent ones, physical systems can exhibit such degrees of freedom, and the corresponding physical observables are distinct.

\subsection{Lagrangian and equations of motion}

We begin with a brief explanation of the worldline formalism that we use here.  Our  setup follows the basic dynamical-mass setup of Ref.~\cite{Steinhoff:2014kwa}; however, we consider the modification that the Lagrange multiplier that usually enforces the SSC is neglected,
\begin{equation}
	S= \int_{-\infty}^\infty \left(
    -p_\mu \dot z^\mu + \frac{1}{2} \gS^{\mu\nu}\Lambda_{A\mu} \frac{D\Lambda^A{}_{\nu}}{D\lambda} + \frac{\xi}{2}(p^2 - \mathrm{M}^2)\right) d\lambda \,.
\label{eq:WLactbig}
\end{equation}
We find that for special choices of Wilson coefficients and initial conditions for which $\covclK^\mu = 0$ (as defined in terms of the spin tensor \eqref{eq:gStensor}), then the $\covclK^\mu = 0$ condition is preserved dynamically, and the equations of motion of this action reduce to the usual equations of motion of a spinning body in general relativity with the minimal degrees of freedom. 

The dynamical variable representing the timelike worldline is $z^\mu(\lambda)$, which we take to locate the body's ``center.''  The variable $\lambda$  parameterizes the worldline and can be interpreted as worldline time. The other variables are the conjugate momentum $p_\mu(\lambda)$, a body tetrad $\Lambda^A{}_\mu(\lambda)$, and the body spin tensor $\gS^{\mu\nu}(\lambda)$.    In addition, $\xi(\lambda)$ is a Lagrange multiplier that enforces the on-shell constraint, and $\mathrm{M}\equiv \mathrm{M}(z,p,\gS)$ is the dynamical mass function of the body, which contains the mass and all its non-minimal couplings to gravity.   The tetrad has the usual properties, 
\begin{equation}
	{\Lambda^\mu}_A{\Lambda^\nu}_B \eta^{AB} = g^{\mu\nu} , \hskip 1.5 cm
        g_{\mu\nu}{\Lambda^\mu}_A{\Lambda^\nu}_B = \eta_{AB} \,.
	\label{eq:WLtetradconditions}
\end{equation}

The variation of the action~\eqref{eq:WLactbig} gives the Mathisson–Papapetrou–Dixon (MPD)~\cite{Dixon:1970mpd,Mathisson:1937zz,Papapetrou:1951pa} equations, 
\begin{align}
    &\dot z^\mu = \hat p^\mu - \frac{1}{\mathrm{M}}\frac{\partial \mathrm{M}}{\partial \hat p_\mu} + \frac{\hat p^\mu}{\mathrm{M}}\hat p_\nu \frac{\partial\mathrm{M}}{\partial \hat p_\nu} \,,\label{EqOfMotionz} \\
    &\dot p_\mu = -\frac{1}{2}R_{\mu\nu\rho\sigma}\dot z^\nu \gS^{\rho\sigma} + \nabla_\mu \mathrm{M} \,,\label{EqOfMotionp} \\
    &\dot \gS_{\mu\nu} = p_\mu \dot z_\nu - p_\nu\dot z_\mu - 2 {\gS_\mu}^\rho \frac{\partial \mathrm{M}}{\partial \gS^{\rho\nu}} + 2 {\gS_{\nu}}^\rho\frac{\partial\mathrm{M}}{\partial \gS^{\rho\mu}} - \frac{\partial \mathrm{M}}{\partial\hat p^\mu} \hat p_\nu + \frac{\partial\mathrm{M}}{\partial\hat p^\nu}\hat p_\mu \,,
\label{EqOfMotions}
\end{align}
where ${\hat p}{}^\mu = {p^\mu}/{\sqrt{p^2}}$. 
Through cubic order in the spin tensor, the dynamical mass function is
\begin{align}
    \frac{\mathrm{M}^2}{m^2} =&\ 1 + \frac{1+C_2}{4} R_{\hat p\aa\hat p\aa} - \frac{1+D_2}{2} \widetilde{R}_{\hat p\aa\hat p\covclk} + \frac{1+E_2}{4} R_{\hat p\covclk\hat p\covclk} \nonumber \\
    &\ + \frac{1+C_3}{3}(\aa\cdot\nabla)\widetilde{R}_{\hat p \aa\hat p \aa} + \frac{1+D_3}{4}(\covclk\cdot\nabla)R_{\hat p \aa\hat p \aa}\nonumber \\
    &\ - \frac{1+E_3+2D_2}{4}(\covclk\cdot \nabla)\widetilde{R}_{\hat p \aa \hat p \covclk} +\frac{F_3+3E_2}{12}(\covclk\cdot\nabla)R_{\hat p \covclk \hat p \covclk} \,,
    \label{eq:dynamicalmassfunction}
\end{align}
where $\widetilde{R}_{abcd}=(1/2)\varepsilon_{abef}R^{ef}{}_{cd}$ is the dual Riemann tensor, and $\covclk^\mu \equiv i\kK^\mu$ and $\aa^\mu$ and $\kK^\mu$ are defined precisely as in \cref{EqConvenientVariables}. 
There are a few more terms that can be written down, in particular $(\hat p\cdot\nabla)R_{\hat p \aa \aa \covclk}$,  $(\hat p\cdot\nabla){\widetilde R}_{\hat p \covclk \aa \covclk}$, $(\aa\cdot\nabla)R_{\hat p \aa \hat p \covclk}$, $(\aa\cdot\nabla){\widetilde R}_{\hat p\covclk\hat p \covclk}$, however these terms do not contribute independently to observables and through field redefinitions can be reduced to the terms we have included. The gravitational Compton amplitude of this worldline theory with the coefficients as identified precisely matches the one from the quantum field theory of the previous section. The SSC can be ``turned on'' in the above equations by setting the coefficients to the special values designated in \cref{eq:NS_cond}. These special values are determined by requiring that if the initial conditions are SSC-satisfying, then the SSC is preserved dynamically. With these special values, the worldline becomes identical to the usual minimal gravitational worldline through the same order in spin. Consequently, this modified worldline formalism is strictly more general than the conventional one, as the former contains the latter as a special case when appropriate initial conditions and Wilson-coefficient values are selected. 

\subsection{Worldline Compton amplitude}
\label{sec:WLCompton}

Using the equations of motion, we compute the classical Compton amplitude to order ${\cal O}(G \gS^3)$ for general values of the Wilson coefficients. The classical Compton amplitude is computed by computing the coefficient of the outgoing spherical wave produced by the response of the spinning body to an incoming plane wave ~\cite{Kosower:2018adc, Saketh:2022wap, Scheopner:2023rzp}. We do so following the approach of~\cite{Scheopner:2023rzp}. In particular, we consider a perturbation of Minkowski space in de Donder gauge defined by: 
	\begin{equation}
    \label{eq:metricexpansion}
		{g}^{\mu\nu}\sqrt{-\det g} = \eta^{\mu\nu} + \kappa h^{\mu\nu}, \hskip 1.2 cm \kappa = \sqrt{32\pi G}, \hskip 1.2 cm
		\partial_\nu h^{\mu\nu} = 0 \,.
	\end{equation}
In terms of $h^{\mu\nu}$, Einstein's equations are exactly,
	\begin{align}
		-\partial^2 h^{\mu\nu} = \kappa&\left(-\frac{T^{\mu\nu}}{2}|\det {g}| + h^{\rho\sigma}\partial^2_{\rho\sigma}h^{\mu\nu} -\partial_\sigma h^{\mu\rho}\partial_\rho h^{\nu\sigma} - g_{\alpha\beta}g^{\rho\sigma}\partial_\rho h^{\mu\alpha}\partial_\sigma h^{\nu\beta}\right. \nonumber \\
		&\ \ \ \left.+ g^{\mu\beta}g_{\rho\alpha}\partial_\sigma h^{\nu\alpha}\partial_\beta h^{\rho\sigma} + g^{\nu\beta}g_{\rho\alpha}\partial_\sigma h^{\mu\alpha}\partial_\beta h^{\rho\sigma} - \frac{1}{2}g^{\mu\nu}g_{\alpha\rho}\partial_\sigma h^{\alpha\beta}\partial_\beta h^{\rho\sigma} \right.\nonumber \\
		&\ \ \ \left. \vphantom{\frac{T^{\mu\nu}}{2}}- \frac{1}{8}(2g^{\mu\tau}g^{\nu\omega}-g^{\mu\nu}g^{\tau\omega})(2g_{\alpha\rho}g_{\beta\sigma}-g_{\alpha\beta}g_{\rho\sigma})\partial_\tau h^{\alpha\beta}\partial_\omega h^{\rho\sigma}\right).
	\end{align}
Both $h^{\mu\nu}$ and $T^{\mu\nu}$ have solutions in powers of $\kappa$:
	\begin{align}
		h^{\mu\nu} &= h^{\mu\nu}_{(0)} + \kappa h^{\mu\nu}_{(1)} + \kappa^2 h^{\mu\nu}_{(2)} + \order(\kappa^3) \,, \nonumber \\
		T^{\mu\nu} &= T^{\mu\nu}_{(0)} + \kappa T^{\mu\nu}_{(1)} + \order(\kappa^2)\,.
	\end{align}
For Compton scattering, we  consider the incoming gravitational field to be a plane wave,
    \begin{equation}
        h^{\mu\nu}_{(0)} = \text{Q}\mathcal{E}_1^{\mu\nu}e^{ik_1\cdot x} \,,
    \end{equation}
for some polarization tensor $\mathcal{E}_1^{\mu\nu}$ and book-keeping parameter $\text{Q}$ which controls the strength of the incoming gravitational wave. 
For Compton scattering, we are only concerned with the response of the system to linear order in the strength $\text{Q}$ of the incoming wave, and so it is useful to define $h^{\mu\nu}_\text{stat}$ as the stationary response of the metric perturbation to the unperturbed energy-momentum tensor. In the equation of motion for $h_{(1)}^{\mu\nu}$, all of the contributions from $h^{\mu\nu}_{(0)}$ are of order $\text{Q}^2$ and so: 
	\begin{equation}
		h^{\mu\nu}_{(1)} = h^{\mu\nu}_\text{stat} + \mathcal{O}(\text{Q}^2), \hskip 1.3 cm 
        -\partial^2 h^{\mu\nu}_\text{stat} = -\frac{1}{2}T^{\mu\nu}_{(0)}\,.
	\end{equation}
The leading correction to the metric perturbation can be decomposed into a stationary piece from iterating corrections from $h^{\mu\nu}_\text{stat}$, which is $\text{Q}$ independent, a piece that is linear in $\text{Q}$, and pieces that are of at least $\order(\text{Q}^2)$:
    \begin{equation}
		h^{\mu\nu}_{(2)} = h^{\mu\nu}_{\text{stat},2}+ \text{Q} \delta h^{\mu\nu} 
        + \order(\text{Q}^2)\,.
	\end{equation}
The Compton amplitude is thus contained in the behavior of $\delta h^{\mu\nu}$ near spatial infinity. In particular, for large $r$, $\delta h^{\mu\nu}$ satisfies
    \begin{equation}
        \mathcal{E}_{2\mu\nu}\delta h^{\mu\nu} = \frac{e^{i\omega(r-t)}}{8\pi m r} \mathcal{A} + \order\left(\frac{1}{r^2}\right),
    \end{equation}
where $\mathcal{A}$ is the gravitational Compton amplitude and $\mathcal{E}_{2\mu\nu}$ is the outgoing polarization tensor of the scattered graviton. We have confirmed that this worldline gravitational Compton amplitude precisely matches the classical limit of the Compton amplitude obtained using the quantum field theory of \sect{sec:HigherOrderFT} through cubic order in the spin tensor. 


\subsection{Worldline impulses}
\label{sec:WLImpulses}

For computing observables with these equations of motion, we consider the probe limit of a spinning particle of mass $m$ scattering off a stationary scalar source. We take the probe limit for simplicity; even so, the result is sufficiently complex to demonstrate a rather nontrivial comparison with the field-theory calculations. For the source, we consider a point mass of mass $m_2$ with four-velocity $u_2$. 
The solutions to the equations of motion of the probe in powers of $G$ are of the form:
\begin{align}
    & z^\mu(\lambda) = b^\mu + u_1^\mu \lambda + G \delta z_{(1)}^\mu(\lambda) + G^2 \delta z_{(2)}^\mu(\lambda) + \order(G^3)\,,\nn \\
    & p^\mu(\lambda) = m_1 u_1^\mu + G \delta p_{(1)}^\mu(\lambda) + G^2 \delta p_{(2)}^\mu(\lambda) + \order(G^3)\,, \nn\\
    & \gS^{\mu\nu}(\lambda) = \gS_1^{\mu\nu} + G \delta \gS^{\mu\nu}_{(1)}(\lambda) + G^2 \delta \gS^{\mu\nu}_{(2)} + \order(G^3)\,.
\end{align}
The impact parameter $b^\mu$ is defined to be transverse to the initial momentum, $b\cdot p_1 = 0$. The initial momentum $m_1 u_1^\mu$ defines the initial four-velocity $u_1^\mu$.  All perturbations of $p^\mu$ and $\gS^{\mu\nu}$ asymptotically vanish for $\lambda \to \pm \infty$ while the trajectory perturbations are logarithmically divergent with the worldline time due to the long-range nature of the $\frac{1}{r}$ potential. Due to this logarithmic divergence, in order to treat the $\order(G^2)$ and higher-order solutions correctly, all the perturbations may be set to 0 at an initial cutoff time $\lambda = -T$. Impulse observables are then computed by taking the difference in observables at time $T$ and $-T$ and, at the end, taking the limit $T\to\infty$. Equivalently, the perturbations may be given representations in terms of standard Feynman integrals and computed using dimensional regularization, such as done for the nonspinning case in Ref.~\cite{Kalin:2020mvi}.

The momentum impulse $\Delta p$ and spin kick $\Delta\qftS$ have been computed through $\order(G \gS^3)$ in the worldline formalism as described above. See \cref{eq:LOimpulseToS3,eq:LOimpulseToSsq} for the impulse through $\order(G \gS^3)$. See also the ancillary file {\tt twoBody.m} for the two-body amplitudes that produce the observables for these orders using the eikonal formula discussed in \cref{sec:eikonalFormula}.
For generic values of Wilson coefficients, $\Delta p$, $\Delta\qftS$ and $\Delta\covclK$ depend on the initial values of $\qftS$ and $\covclK$. Note that $\Delta\covclK$ is nonzero even if the initial $\covclK$ vanishes unless the Wilson coefficients take the special values as in \cref{eq:NS_cond}.
When the Wilson coefficients take those values, the $\covclK$ dependence in $\Delta p$ and $\Delta\qftS$ drops out, as we are now considering the \ccos.
With the special values in Eq.~\eqref{eq:NS_cond}, an initial nonzero $\covclK$ leads to a nonzero $\Delta\covclK$, but as we have discussed above, we can absorb the entire $\covclK$ dependence into a shift of worldline.
For this choice of Wilson coefficients, the initial value of $\covclK$ may be thought of a choice of SSC. Hence, the above is in accordance with the interpretation of the SSC as a gauge choice related to the choice of center of the worldline~\cite{Levi:2015msa, Steinhoff:2015ksa, Vines:2016unv}.


\subsection{Relation of dynamical multipole moments}
\label{sec:MultipoleRelations}

The generic MPD equations describe the motion of a compact object in general relativity with arbitrary internal dynamics. They, however, do not close as a system of equations of motion because they depend on Dixon's tower of multipole moments whose dynamics, beginning with that of the quadrupole, is generically unconstrained~\cite{Dixon:1970zza}.
Any generally covariant theory of compact objects is equivalent to the MPD equations with some choice of dynamics for the higher multipole moments. 
In the traditional minimal worldline approach with spin, only the variables whose dynamics is supplied by the MPD formalism are kept: the worldline trajectory, the total linear momentum, and the spin vector~\cite{Steinhoff:2015ksa, Levi:2015msa}. The compact object’s independent degrees of freedom are truncated so that the quadrupole and all higher multipole moments are functions of these variables.
Our description self-consistently includes an extra vector degree of freedom beyond that minimal set, so that Dixon's tower of moments can in principle be expressed as functions of $p, S, \covclK$, rather than than just $p, S$. 
The dynamics of this degree of freedom is fixed by the requirement of Lorentz invariance.

It is interesting to compare the worldline theory we just discussed with previous descriptions of compact bodies with internal degrees of freedom. 
These approaches, in terms of dynamical multipole moments, employ localized degrees of freedom on the worldline to describe absorptive properties, e.g., as in Refs.~\cite{Goldberger:2005cd, Porto:2005ac, Porto:2007qi, Goldberger:2020fot}. For example, the action for dynamical quadrupoles is
\begin{align}
    S_{\text{fin}} = \int d\lambda \left(Q_{E}^{ij}\frac{E_{ij}}{\sqrt{u^2}} + Q_{B}^{ij}\frac{B_{ij}}{\sqrt{u^2}}\right)+\ldots\,,
\end{align}
where $Q_{E}$ and $Q_{B}$ are the parity even and odd quadrupole moments, respectively, and $E_{ij}$ and $B_{ij}$ are the electric and magnetic components of the Riemann tensor, $E_{\mu\nu} = R_{\mu\alpha\nu\beta}u^\alpha u^\beta$ and $B_{\mu\nu} = \frac{1}{2}\varepsilon_{\alpha\beta\gamma\mu} R^{\alpha\beta}{}_{\delta\nu} u^\gamma u^\delta$.\footnote{Here, we write the action in a local co-moving frame with $e_{0}^{\mu}=u^{\mu}$ such that only spatial components are relevant. We also omit the contribution from higher multipoles.}
A similar action, extended with further rotation-induced interactions between modes of the quadrupole, was used to study the interplay between spin and tidal deformations in, e.g., Ref.~\cite{Steinhoff:2021dsn}. 

The properties of compact bodies are determined by correlation functions of the dynamical multipoles, such as the quadrupoles $Q_{E}$ and $Q_{B}$ above, which parameterize both conservative and absorptive dynamics. 
In Refs.~\cite{Goldberger:2005cd, Porto:2007qi, Goldberger:2020fot}, the absorptive properties of black-hole quadrupoles were extracted by matching with the long-wavelength graviton absorption cross section~\cite{Starobinskil:1974nkd, Page:1976df}. 
The (quantum) field theory formalism incorporates such effects in a similar manner, by including additional degrees of freedom whose correlation functions are constructed to match gravitational calculations as in, e.g., 
Refs.~\cite{Jones:2023ugm, Chen:2023qzo, Aoude:2023fdm, Aoude:2024jxd, Bautista:2024emt}. 
In these works, the correlation functions of the dynamical multipoles are a necessary additional ingredient of the analysis.
Here, however, we are interested in the conservative dynamics of a \gco, which can be signaled by, e.g., the presence of massless poles in these correlation functions or by the presence of contributions of zero-frequency excitations. 
The restriction to conservative dynamics allows us to treat the multipoles of a \gco in the same way as the spin-induced multipoles of a \cco, without requiring knowledge of additional correlation functions.

If the position of the body's center $z^\mu$ and the body's spin $\qftSrf$ are the only light degrees of freedom, then there is a unique spin-induced quadrupole moment that respects parity~\cite{Porto:2005ac},
\begin{align}
    Q_{E}^{ij} = \frac{C_2}{2m} S^{i} S^{j} \, ,
\end{align}
corresponding to the first curvature-dependent term in the dynamical mass function~\eqref{eq:dynamicalmassfunction}.
As we discuss in Secs.~\ref{sec:TwoBodyAmplitude} and \ref{sec:Hamiltonian},  $\clK$ evolves conservatively (at least on the time scale of the scattering process) and thus corresponds to degrees of freedom that are gapless.\footnote{Intuitively, they capture internal energy-preserving rearrangements of the mass distribution of a compact body as a consequence of interactions.} 
Thus, they can appear in the expressions of the dynamical multipoles. For example, $\clK$ induces additional couplings that are linear in the Riemann tensor at the quadrupolar level,
\begin{align}
    Q_{E}^{ij} = \frac{C_2}{2m}S^{i} S^{j} + \frac{E_2}{2m}\mathrm{K}^{i}\mathrm{K}^{j}\,,\qquad Q_{B}^{ij} = \frac{D_2}{2m} S^{(i}\mathrm{K}^{j)} \, ,
\end{align}
and correspond to the second and third curvature-dependent terms in the dynamical mass function~\eqref{eq:dynamicalmassfunction}. 
The subsequent linear in $R$ terms in that equation are spin and $K$-induced
dynamical octupole moments. We compared their dynamics as well as that of the
hexadecapole $Q_{E(B)}^{ijkl}$ in the presence of both $\qftSrf$ and $\clK$ and found that it matches that of the field-theory construction. 

In general, there is a nonvanishing $\Delta \clK$ between the initial and final values. This distinguishes $\clK$ from, e.g., elastic tidal deformations, 
which relax to the equilibrium state once the external tidal field disappears or normal-mode oscillations, which are typically gapped. See Ref.~\cite{Steinhoff:2016rfi,Gupta:2020lnv} for examples of tidally induced multipole moments.

In our formalism, the dynamics of these multipoles are determined by Lorentz invariance.
In particular, $\clK$ and $\bm S$ contribute to the equations of motion of each other because their Poisson bracket, $\{\clK, \bm S\}\propto \clK$ and $\{\clK, \clK\}\propto \bm S$,  are inherited from the commutation relations of Lorentz generators. 
The formalism of Refs.~\cite{Goldberger:2005cd, Porto:2007qi, Goldberger:2020fot} can accommodate such dynamical multipoles, and the light degrees of freedom described by $\clK$ would be encoded by properties of their correlation functions. The absence of unusual properties of their two-point functions for the Kerr black hole~\cite{Starobinskil:1974nkd} is intuitively in agreement with our observation that for \cco\ -- and in particular for the Kerr black hole -- $\clK$ decouples.


\section{Scattering waveform at leading order}
\label{sec:waveform}

To determine whether a \cco without $K$ is physically distinct from the \gco with $K$, it is essential to compare their physical observables.
The waveform emitted in a scattering event is an especially good observable because it can, in principle, be directly measured by gravitational-wave detectors. 
We specifically investigate whether a waveform signal that can be accurately fitted by a $K=0$ waveform can still be fitted when $K$ is turned on and the Wilson coefficients are appropriately readjusted in an attempt to absorb any differences. Our analysis carried out to the relevant order demonstrates that no such readjustments can eliminate the observable effects of $K$.

\subsection{Leading-order waveform and its soft expansion}
\label{sec:LOwf_and_soft_expansion}

The connection between scattering waveforms, which describe the metric at infinity, ${h}^\infty_{\mu\nu}$, and scattering amplitudes is discussed in Ref.~\cite{Cristofoli:2021vyo}. We write the metric at infinity as \footnote{This identification of metric fluctuations about Minkowski space is slightly different from that used in Eq.~\eqref{eq:metricexpansion}. The trace terms coming from the expansion of the determinant are irrelevant because $h^\infty_{\mu\nu}$ is traceless. The sign coming from the expansion of the inverse metric must be accounted for when comparing the waveform discussed here with waveforms computed with the metric parametrization in the previous section.} 
\begin{align}
\label{eq:metricQFT}
g_{\mu\nu} = \eta_{\mu\nu} + \frac{\kappa}{4\pi |\bm{x}|} \, {h}^\infty_{\mu\nu} \,.
\end{align}
At leading order, this metric perturbation takes the form
\begin{align}
{h}^\infty_{\mu\nu} = \kappa\, \mathsf{M}\left( \frac{\kappa^2 \mathsf{M}}{\sqrt{-b^2}}\,\hat{h}^{(1)}_{\mu\nu}+
\order(\kappa^4)\right),
\end{align}
with $\mathsf{M}=m_1+m_2$. This leads to the expression for the waveform in the time domain,
\begin{align}
    W^\text{LO}(\tau) &\equiv \pol^{\mu\nu}h_{\mu\nu}^\infty\Big|_{\order(\kappa^3)} = \int_{0}^{+\infty}\frac{d\omega}{2\pi} \, \Big[ \Theta(\omega){\cal W}^\text{LO}(\omega) \, e^{-i\omega \tau } + \text{h.c.} \Big] \, ,
\label{TDwaveform}    
\end{align}
where $\tau = t - |\bm x|$ is the retarded time. The frequency domain waveform is given by
\begin{align}
\label{eq:MtoWEFT}
    {\cal W}^\text{LO}  (\omega)
    &= -2 \int\mu(k,q_1,q_2)  e^{-iq_1 \cdot b}{\cal M}^\text{tree}_5 \,,
\end{align}
where ${\cal M}^\text{tree}_5 \equiv {\cal M}^\text{tree}_5(p_{1},p_{2}, q_1, k, \pol, \qftS_{i}, \qftK_{i})$ represents the five-point amplitude involving an outgoing graviton with momentum $k$ and polarization tensor $\pol$, along with the two matter particles having incoming momenta $p_1$ and $p_2$, and outgoing momenta $p_4$ and $p_3$.
The momentum loss $q_i$ of the $i$-th matter particle is given by
$q_1 = -p_1-p_4$ and $q_2 = -p_2-p_3$ in an all-outgoing convention, so that momentum conservation requires that 
\begin{align}
\label{eq:Momentumk}
q_1+q_2 = k \, .
\end{align}
We use this relation to remove the $q_2$ dependence from the amplitude. The matter particles may carry spin and $K$ degrees of freedom, represented by $\qftS_{i}$ and $\qftK_{i}$ for the $i$-th particle.
The integration measure is
\begin{align}\label{eq:cl_measure}
    \mu(k,q_1,q_2) = \frac{d^4 q_1}{(2\pi)^4}\frac{d^4 q_2}{(2\pi)^4} & \,\hdelta(2{\bar p}_1\cdot q_1)\,\hdelta(2{\bar p}_2\cdot q_2)\,\hdelta^{4}(q_1+q_2-k)\,, 
\end{align}
with ${\bar p}_i = p_i +q_i/2$, $\hdelta(x) = 2\pi\delta (x)$ and $\hdelta^{d}(x) = (2\pi)^d \delta^d(x)$.
The on-shell condition for $\bar p_i$ is ${\bar p}_i^2 = {\bar m}_i^2$ where ${\bar m}_i^2 = m_i^2 - q_i^2/4$. 
The measure factor suggests that it is more natural to express the amplitude used in \cref{eq:MtoWEFT} in terms of the $\bar p$ variables. Similarly, in general the $S^\mu$ and $K^\mu$ should be evaluated at momenta $S_i^\mu = S^\mu(\bar p_i)$ and $K_i^\mu = K^\mu(\bar p_i)$.
At tree level, the difference between $\bar p$ and $p$ is not relevant. In the rest of this section, we do not distinguish $\bar{p}$ and $p$ in the classical amplitudes and observables.

\begin{figure}
\centering
\pgfmathsetmacro{\w}{1.4}
\pgfmathsetmacro{\h}{1.3}
\pgfmathsetmacro{\x}{0.3}
\pgfmathsetmacro{\r}{0.17}
\raisebox{0.9cm}{$\mathcal{M}_5 \sim $}
\begin{tikzpicture}[scale=1.1]
    \draw [massive] (0,0) node [left]{\small $1$} -- (\w,0) node [right]{\small $4$};
    \draw [massive] (0,\h) node [left]{\small $2$} -- (\w,\h) node [right]{\small $3$};
    \draw [graviton] (\w/2,0) -- ++ (0,\h);
    \draw [graviton] (\w/2,\h) -- (\w+0.1,\h/2 + 0.1) node[below=0]{\small $k$};
    \draw [dashed] (\x,\h/2) -- ({\w-\x},\h/2);
    \node at (\x-\r-0.05,\h/2) {$q_1$};
    \filldraw [fill=gray!50] (\w/2,0) circle (\r);
    \filldraw [fill=gray!50] (\w/2,\h) circle (\r);
\end{tikzpicture}
\raisebox{0.9cm}{$\;+$}
\begin{tikzpicture}[scale=1.1]
    \draw [massive] (0,0) node [left]{\small $1$} -- (\w,0) node [right]{\small $4$};
    \draw [massive] (0,\h) node [left]{\small $2$} -- (\w,\h) node [right]{\small $3$};
    \draw [graviton] (\w/2,0) -- ++ (0,\h);
    \draw [graviton] (\w/2,0) -- (\w+0.1,\h/2 - 0.1) node[above]{\small $k$};
    \draw [dashed] (\x,\h/2) -- ({\w-\x},\h/2);
    \node at (\x-\r-0.05,\h/2) {$q_2$};
    \filldraw [fill=gray!50] (\w/2,0) circle (\r);
    \filldraw [fill=gray!50] (\w/2,\h) circle (\r);
\end{tikzpicture}
\raisebox{0.9cm}{$\;-$}
\begin{tikzpicture}[scale=1.1]
    \draw [massive] (0,0) node[left]{\small $1$}-- (\w,0) node [right]{\small $4$};
    \draw [massive] (0,\h) node[left]{\small $2$} -- (\w,\h) node [right]{\small $3$};
    \draw [graviton] (\w/2,0) -- ++ (0,\h);
    \draw [graviton] (\w/2,\h/2) -- (\w,\h/2) node[right]{\small $\!k$};
    \draw [dashed] (\x,\h/4) -- ({\w-\x},\h/4);
    \node at (\x-\r-0.05,3*\h/4) {$q_2$};
    \draw [dashed] (\x,3*\h/4) -- ({\w-\x},3*\h/4);
    \node at (\x-\r-0.05,\h/4) {$q_1$};
    \filldraw [fill=gray!50] (\w/2,0) circle (\r);
    \filldraw [fill=gray!50] (\w/2,\h) circle (\r);
    \filldraw [fill=gray!50] (\w/2,\h/2) circle (\r);
\end{tikzpicture}\,
\caption{Diagrammatic representation of Eq.~\eqref{eq:sewing}. The dashed lines across the graviton propagators indicate on-shell conditions and summation over physical states.
}
\label{fig:M5}
\end{figure}

As in the SSC-constrained case~\cite{DeAngelis:2023lvf, Brandhuber:2023hhl}, it turns out that at leading order in Newton's constant, only certain parts of the classical five-point amplitude contribute to waveforms.  Since the calculation of the classical tree-level five-point amplitude is sufficiently simple, we outline it here.
This calculation can be carried out directly using the Feynman rules derived from the Lagrangians~\eqref{eq:Lmin} and~\eqref{eq:ExtraInteractions}, or alternatively, as we do here by sewing together three-point and Compton amplitudes as shown in \cref{fig:M5}, namely,
\begin{align}
\label{eq:sewing}
{\cal M}_5 & =  \sum_{\rm states} \biggl( {\cal M}_3\frac{1}{q_1^2}{\cal M}_4
+ {\cal M}_4 \, \frac{1}{q_2^2} \, {\cal M}_3
- {\cal M}_3\, \frac{1}{q_1^2} \, {\cal M}_3\, \frac{1}{q_2^2} \, {\cal M}_3
  + \hbox{contact terms} \biggr)\,,
\end{align}
where the sum runs over the physical graviton states, and the third term removes the overcount of the double-pole contribution.  As usual, the state sum results in a graviton physical state projector.
The sewing procedure does not correctly capture contact terms involving intersecting matter lines, but such terms are irrelevant in the classical limit, where only long-range forces are considered. We, therefore, omit such contact terms in the following discussion.

Before analyzing the complete leading-order waveform, it is helpful to first gain insight by examining the gravitational-wave memory, which is characterized by the metric failing to return to its initial value after the wave has passed. This quantity is useful as it provides a simple criterion for identifying whether two waveforms are inequivalent.
To this end, we revisit the analysis of Ref.~\cite{Saha:2019tub} using our worldline theory with both $S$ and $K$ vectors present, whose observables match those derived from quantum field theory.
At the leading order in the outgoing graviton frequency, the solution to the point-particle equations of motion depends only on their momenta and does not exhibit an explicit dependence on their intrinsic properties, such as their Lorentz-group representation. It, therefore, follows that the gravitational-wave memory in the presence of both the $S$ and $K$ vectors should have the same general structure as for spinless particles~\cite{Saha:2019tub, Sahoo:2021ctw}, 
\begin{align}
{\cal W}(\omega)=\epsilon^{\mu \nu}h^{\infty}_{\mu\nu}(\omega)  
= \frac{i\,\mathcal{A}}{\omega} + \order(\omega^0)\,,
\end{align}
with 
\begin{align}
\mathcal{A}
= - \frac{(\epsilon\cdot p_3)^2}{\hat k\cdot p_3}
- \frac{(\epsilon\cdot p_4)^2}{\hat k\cdot p_4} + \frac{(\epsilon\cdot p_1)^2}{\hat k\cdot p_1}+ \frac{(\epsilon\cdot p_2)^2}{\hat k\cdot p_2} \,,
\label{eq:memory}
\end{align}
where $\hat k = k/\omega$, $p_4 = p_1+\Delta p$, $p_3 = p_2-\Delta p$ with $\Delta p$ being the impulse including the $S$ and $K$-dependent contributions. For convenience, in \cref{eq:memory} we express the graviton polarization tensor $\pol^{\mu\nu}$ as a product of two identical polarization vectors, $\epsilon^{\mu}\epsilon^{\nu}$, each satisfying $\epsilon^2 = k \cdot \epsilon = 0$. In this way, the graviton polarization tensor inherits the properties of standard circular vector polarizations.
The observations above align with conjectures formulated in Ref.~\cite{Saha:2019tub}.

We verify this at leading order~\cite{Alaverdian:2024spu}, by directly extracting the gravitational-wave memory. It is 
\begin{align}
\pol^{\mu\nu}\Delta h_{\mu\nu}^\infty \Big|^\text{LO}&= \int_{-\infty}^
{+\infty} \pol^{\mu\nu} \partial_t h_{\mu\nu}^\infty\Big|^\text{LO} \nn
\\
&=
\lim_{\omega\rightarrow 0^+}(-i\omega)\Theta(\omega){\cal W}{}^\text{LO}  (\omega) 
+
\lim_{\omega\rightarrow 0^-}(-i\omega)\Theta(-\omega){\cal W}^*{}^\text{LO}  (|\omega|) \,,
\end{align}
which is given by the soft limit of the five-point amplitude~\eqref{eq:sewing} via \cref{eq:MtoWEFT}. Carrying out the Fourier transform to impact-parameter space reproduces Eq.~\eqref{eq:memory} expanded to leading order in Newton's constant and with the impulse $\Delta p$ given by \cref{eq:LOimpulseToS3}.

It is interesting to discuss the influence of the $K$ vector on the spin memory discussed in Ref.~\cite{Pasterski:2015tva}, which in turn is related to the properties of amplitudes at subleading order in the soft expansion~\cite{Low:1954kd, Cachazo:2014fwa}.  Ref.~\cite{Saha:2019tub} conjectured that the first subleading term in the soft expansion of the waveform for spinning particles is proportional to 
\begin{align}
{\cal W}(\omega)
&= \frac{i\,\mathcal{A}}{\omega} + \mathcal{B}\log \omega + \order(\omega^1)\,,
\\
\mathcal{B} &\propto \sum_{i} \frac{\hat k_\rho J_i^{\rho(\mu} p_i^{\nu)}}{p_i\cdot \hat k}
\,, \qquad
J_i^{\rho\mu} = X_i^\rho p_i^\mu - X_i^\mu p_i^\rho + S^{\rho\mu}(p_i) \,,
\end{align}
where $J_i^{\rho\mu}$ is the {\em total} angular momentum operator of the $i$-th particle located at $X_i$ and we discarded $\omega$-independent terms which yield local terms in time. 
While in principle, the spin tensor $\gS^{\rho\mu}(p)$, not satisfying the SSC, could have appeared in the total angular momentum, the absence of a dipole interaction for the $K$ vector suggests that only its SSC-satisfying part can appear in the subleading soft term. The
$K$ vector however enters in the $\mathcal{B}$ coefficient through the impulse $\Delta p$, as it does in the $\mathcal{A}$ coefficient.

\subsection{Analytic time-domain LO waveform}

To compute the analytic waveform in the time domain, it is convenient to consider together the $q_i$ and $\omega$ integrals in \cref{eq:MtoWEFT,TDwaveform}. As discussed in Ref.~\cite{Herderschee:2023fxh} for spinless particles and Refs.~\cite{DeAngelis:2023lvf, Brandhuber:2023hhl} for spinning ones, we  isolate the $\omega$ dependence and evaluate the $\omega$ integral first. The classical amplitude (and more generally, the KMOC matrix elements) satisfies the following scaling property,
\begin{align}\label{eq:M5scaling}
& \covM^\text{tree}_5(p_1, p_2, \lambda q_1, \lambda k, \pol, \lambda^{-1}\qftS_i, \lambda^{-1}\qftK_i) = \lambda^{-2} \covM^\text{tree}_5(p_1, p_2, q_1, k, \pol, \qftS_i, \qftK_i) \, .
\end{align}
Since $\covM^{\text{tree}}_5$ is a polynomial in $S_i$ and $K_i$, we can schematically write it as
\begin{align}
\covM^{\text{tree}}_{5}(p_1,p_2,q_1,\kk,\pol,\qftS_i,\qftK_i) = \sum_{n=0}^{\infty} \covM^{\text{tree}}_{5,(n)}(p_1,p_2,q_1,\kk,\pol,\qftS_i,\qftK_i)\,,
\end{align}
where $\covM^{\text{tree}}_{5,(n)}$ is a homogeneous degree-$n$ polynomial in $\qftS_i$ and $\qftK_i$.
\Cref{eq:M5scaling} implies that under the scaling $q_i\rightarrow\lambda q_i$ and $k\rightarrow\lambda k$ the partial amplitude $\covM^{\text{tree}}_{5,(n)}$ scales as
\begin{align}
\covM^{\text{tree}}_{5,(n)}(p_1,p_2,\lambda q_1,\lambda k,\pol,\qftS_i,\qftK_i) = \lambda^{-2+n}\covM^{\text{tree}}_{5,(n)}(p_1,p_2,q_1,k,\pol,\qftS_i,\qftK_i)\,.
\end{align}
Thus, we may identify $\lambda=\omega$ and change the integration variable $q_1\rightarrow\omega q_1$ in \cref{eq:MtoWEFT,TDwaveform}. 

While, for the sake of consistency, we express the waveform in terms of the $\qftK$ vector, its hermiticity properties must be carefully accounted for because the relation between the 5-point amplitude and the waveform involves the conjugation of the latter, cf. Eq.~\eqref{TDwaveform}.
To this end, we analytically continue (or identify) $ i \qftK_i\mapsto \covclK_i$ with both $\qftK_i$ and $\covclK_i$ real, which renders the amplitude real or imaginary (for real polarization tensor $\pol^{\mu\nu}$) for even or odd powers of $\qftS$ and $\covclK$ respectively, carry out the $\omega$ integral, and then analytically continue back. 
This leads to a uniform general expression for the partial waveform that is degree-$n$ in $\qftS_i$ and $\qftK_i$, which are thus given by
\begin{align}
W^\text{LO}_{(n)}(\tau) &= -2
\int_{-\infty}^{+\infty} \frac{d\omega}{2\pi} \mu(\hat k, q_1, q_2)\, e^{i\omega (\tau -q_1\cdot b)}\, \omega^{n} 
\covM^\text{tree}_{5,(n)} (p_1, p_2, q_1, \hat k, \pol, \qftS_i,\qftK_i) \nonumber\\
&= - 2\, (-i)^n \partial_\tau^n
\int \mu(\hat k, q_1, q_2)\, \delta(\tau-q_1\cdot b) \,
\covM^\text{tree}_{5,(n)} (p_1, p_2, q_1, \hat k, \pol, \qftS_i,\qftK_i)\,,
\label{eq:WLOafteromegaint}
\end{align}
where $\hat{k}=k/\omega=(1,\bs{n})$ with $\bs{n}^2=1$. Thus, prior to the evaluation of the integral over the momentum transfer, the Fourier transform to the time domain yields a delta function. This extends to the presence of the $K$ vector, the analysis of the all-order-in-spin tree amplitudes in Refs.~\cite{DeAngelis:2023lvf, Brandhuber:2023hhl, Aoude:2023dui} and, as stated above, requires that the hermiticity properties of the $K$ vector be carefully taken into account.
Finally, the full LO waveform is given by
\begin{align}
    W^{\text{LO}}(\tau) = \sum_{n=0}^{\infty} W^{\text{LO}}_{(n)}(\tau)\,.
\end{align}

To evaluate the remaining integrals, we first use the momentum-conserving delta function in the definition of $\mu(\hat k, q_1, q_2) $ to solve for $q_2$. For the remaining $q_1$ integral, we fix $d=4$ and decompose $q_1$ in a basis of four vectors as in Ref.~\cite{Cristofoli:2021vyo}, i.e.
\begin{align}\label{eq:qbasis}
    & q_1^{\mu}=z_1 {\bar u}_1^{\mu} + z_2 {\bar u}_2^{\mu} + z_b {\tilde b}^{\mu} + z_v {\tilde v}^{\mu}\,,   & &
    v^{\mu}\equiv 4\varepsilon^{\mu\nu\rho\sigma}{\bar u}_{1\nu}{\bar u}_{2\rho}b_{\sigma} \,, 
    & &
    {\bar u}_i = {\bar p}_i/{\bar m}_i\,,\nn
    \\
    & {\tilde b^\mu}= b^\mu/\sqrt{-b^2} \,,
    \qquad
    {\tilde v^\mu} = v^\mu/\sqrt{-v^2} \,,
    & &
    v^2 = 16 b^2 (y^2-1)\,,
    & &
    y = {\bar u}_1\cdot {\bar u}_2\,,
\end{align}
and integrate over the coefficients $\{z_1,z_2,z_b,z_v\}$.  The Jacobian of this change of variables is $d^4 q_1 = \sqrt{y^2-1}\,{dz_1 dz_2 dz_v dz_b}$.
The integrals over $z_1$ and $z_2$ can be easily evaluated using the 
two one-dimensional delta functions in the definition of $\mu(\hat k, q_1, q_2)$. They set 
\begin{align}
z_1 =\frac{{\hat w}_2 y}{y^2-1}
\,, \qquad
z_2 = -\frac{{\hat w}_2}{y^2-1}\,,
\qquad
{\hat w}_i = {\hat k}\cdot {\bar u}_i \,, \ \ i=1,2 \, ,
\end{align}
and introduce the measure factor $1/(4 {\bar m}_1 {\bar m}_2 (y^2-1))$, resulting in a net Jacobian 
\begin{align}
\int d^4q \rightarrow \mathcal{J} \int dz_v dz_b  \ \, ,\qquad \mathcal{J} = \frac{1}{4 {\bar m}_1 {\bar m}_2 \sqrt{y^2-1}} \, .
\end{align}

Of the remaining integrals over $z_b$ and $z_v$, the former can be evaluated straightforwardly using the delta function in Eq.~\eqref{eq:WLOafteromegaint}, which relates it to the retarded time, 
\be
z^*_b = -\frac{\tau}{\sqrt{-b^2}} \, .
\ee
The remaining $z_v$ integral can be evaluated through Cauchy's residue theorem, Refs.~\cite{DeAngelis:2023lvf, Brandhuber:2023hhl}
by closing the contour in, e.g., the upper-half plane 
after its convergence at infinity is improved by making use of the symmetry of the integration domain to replace the integrand, say $f(z_v)$, by 
\be
f(z_v) \rightarrow \frac{1}{2}\Big[ f(z_v) + f(-z_v) \Big] \, . 
\ee
This is equivalent to the principal-value prescription of Ref.~\cite{Brandhuber:2023hhl} and with the prescription of taking the difference between the upper-half plane and the lower-half plane contours as in Ref.~\cite{DeAngelis:2023lvf}.
Moreover, as there, it suffices to consider the contributions of the physical poles at $q_1^2=0$ and $(q_1-k)^2 = 0$.

Thus, from the three terms in Eq.~\eqref{eq:sewing}, only the first term contributes to the residue at $q_1^2 = 0$, and only the second term contributes to the residue at $(q_1-k)^2 = 0$; in both cases,  the contributions of the other two terms cancel each other out.  We are left with  
\begin{align}
\label{eq:I1}
I_1(S_i, K_i) &= \frac{1}{2}\left.\sum_h{\cal M}_3(p_1, q_1^h)
{\cal M}_4(p_2, -q_1^h, k) \right|_{z_v= z_{v, 1}}
  \!\!\!\!\!\!\!\!\!
  {\rm Res}_{z_v = z_{v, 1}}
\frac{1}{q_1^2}
~~  + (z_{v, 1}\rightarrow -z_{v, 1}) \,,
\\
\label{eq:I2}
I_2(S_i, K_i) &= \frac{1}{2}\left.\sum_h{{\cal M}_3(p_2, q_2^h)\cal M}_4(p_1, -q_2^h, k)\right|_{z_v= z_{v, 2}} 
  \!\!\!\!\!\!\!\!\!
  {\rm Res}_{z_v = z_{v, 2}}
\frac{1}{q_2^2} ~~ + (z_{v, 2}\rightarrow -z_{v, 2}) \,,
\end{align}
where the summation is performed over the on-shell polarization states $h$ of the corresponding internal graviton and
\begin{align}
z_{v, 1} &= i \sqrt{z_b^2+\frac{\hat w^2_2}{y^2-1}}\,,
\qquad
z_{v, 2} = -k\cdot \tilde v + i \sqrt{(z_b+{\tilde b}\cdot k)^2+\frac{\hat w_1^2}{y^2-1}} \, .
\end{align}
When solving $(k-q_1)^2=0$, we used Levi-Civita identities to 
rewrite $(k\cdot \tilde v )^2$ in terms of other dot products.
Putting together the various Jacobians, the leading order time-domain waveform is  
given by
\begin{align}
W^\text{LO}(\tau) &=  
\frac{\mathcal{J}}{i\pi\sqrt{-b^2}}\int_{-\infty}^{+\infty} dz_b \Big[I_1(S_{i}, K_{i})+I_2(S_{i}, K_{i})\Big]\Bigg|_{S\rightarrow S(-i\partial_\tau)}^{K\rightarrow K(-i\partial_\tau)}\delta\Big(z_b+\frac{\tau}{\sqrt{-b^2}}\Big) \,. 
\label{eq:LOwaveformFinal}
\end{align}
As already mentioned, the last integral, over $z_b$, is evaluated using the explicit delta function, and the derivatives with respect to the retarded time $\tau$ can be evaluated afterward. Effectively, one is to differentiate $n$ times with respect to $\tau$ the coefficient of each monomial of degree $n$ in $S$ and $K$.

\Cref{eq:I1,eq:I2,eq:LOwaveformFinal} imply that properties of the Compton amplitude translate directly into properties of the waveform. For example, as we discussed in Ref.~\cite{Alaverdian:2024spu}, 
in the Compton amplitude, the terms bilinear in the spin vector originating from the $C_2$ Wilson coefficient are the same (up to $\qftS\rightarrow \qftK$) as the terms bilinear in $K$ originating from the $E_2$ Wilson coefficient, i.e.
\begin{align}
W^\text{LO}_{(2)}(\tau)\Big|^{\qftS\rightarrow \qftK}_{C_2}=W^\text{LO}_{(2)}(\tau)\Big|_{E_2} \, .
\label{eq:propC2E2}
\end{align}
Similarly, the classical five-point amplitude exhibits the same property as a consequence of its close relation \eqref{eq:sewing} with the three-point and Compton amplitudes.

\begin{figure}[tb]
\includegraphics[width=2in]{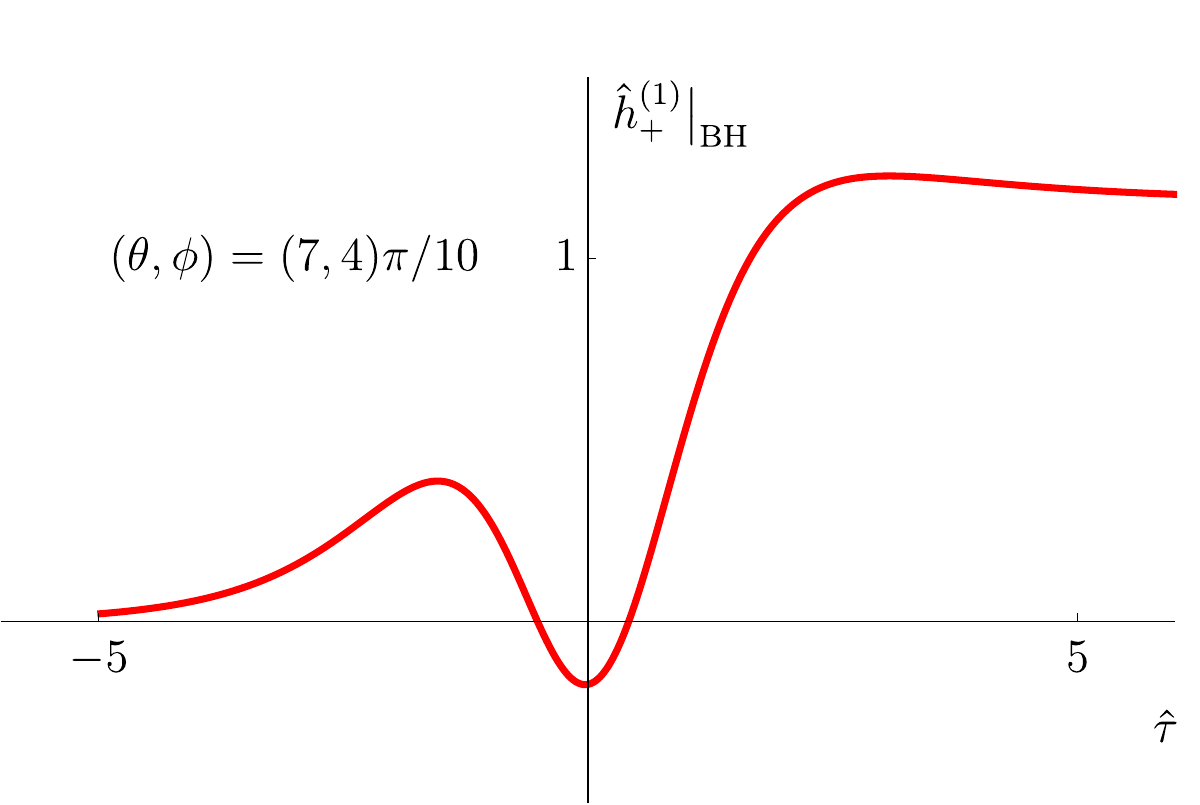}~\raisebox{-0.3cm}{\includegraphics[width=2in]{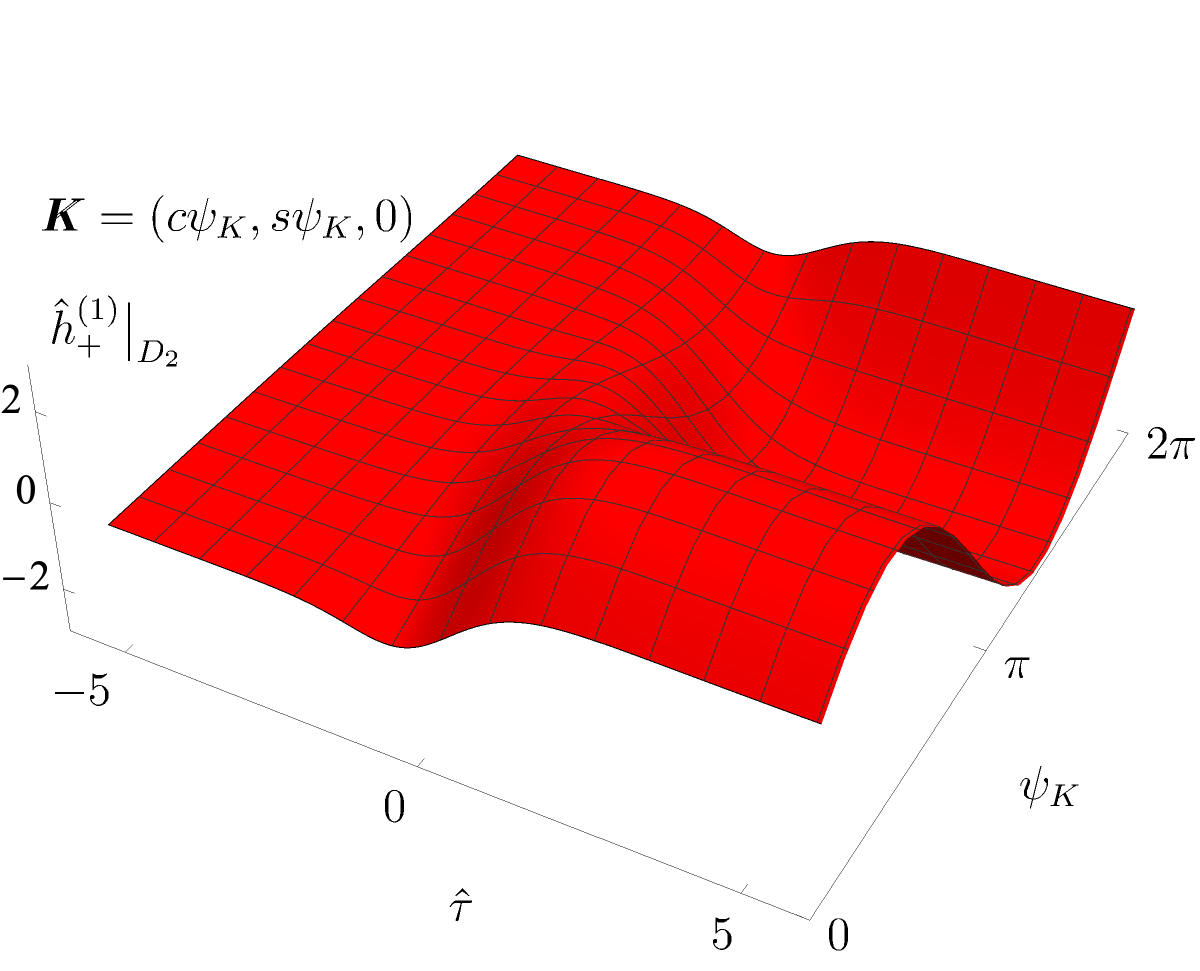}~\includegraphics[width=2in]{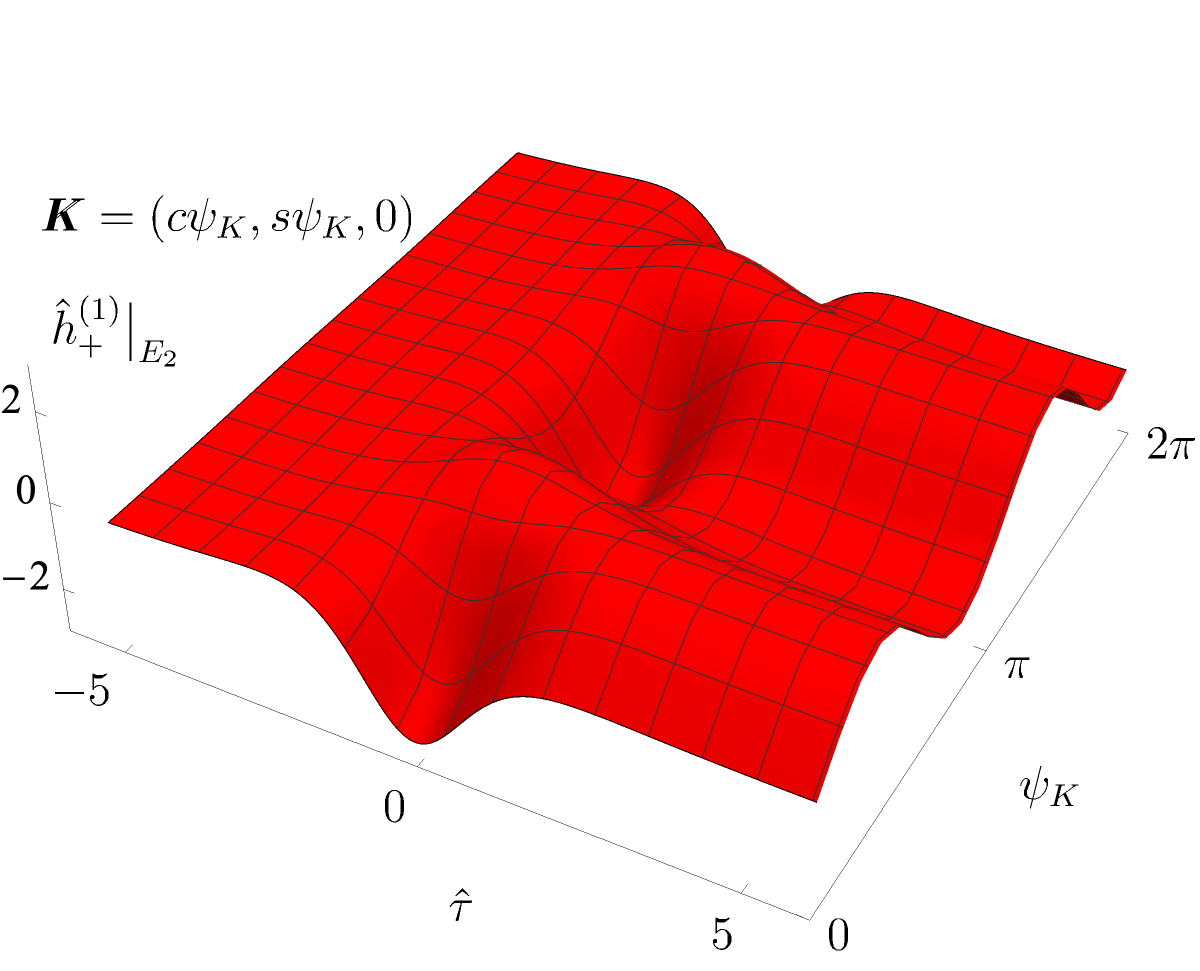}}

\includegraphics[width=2in]{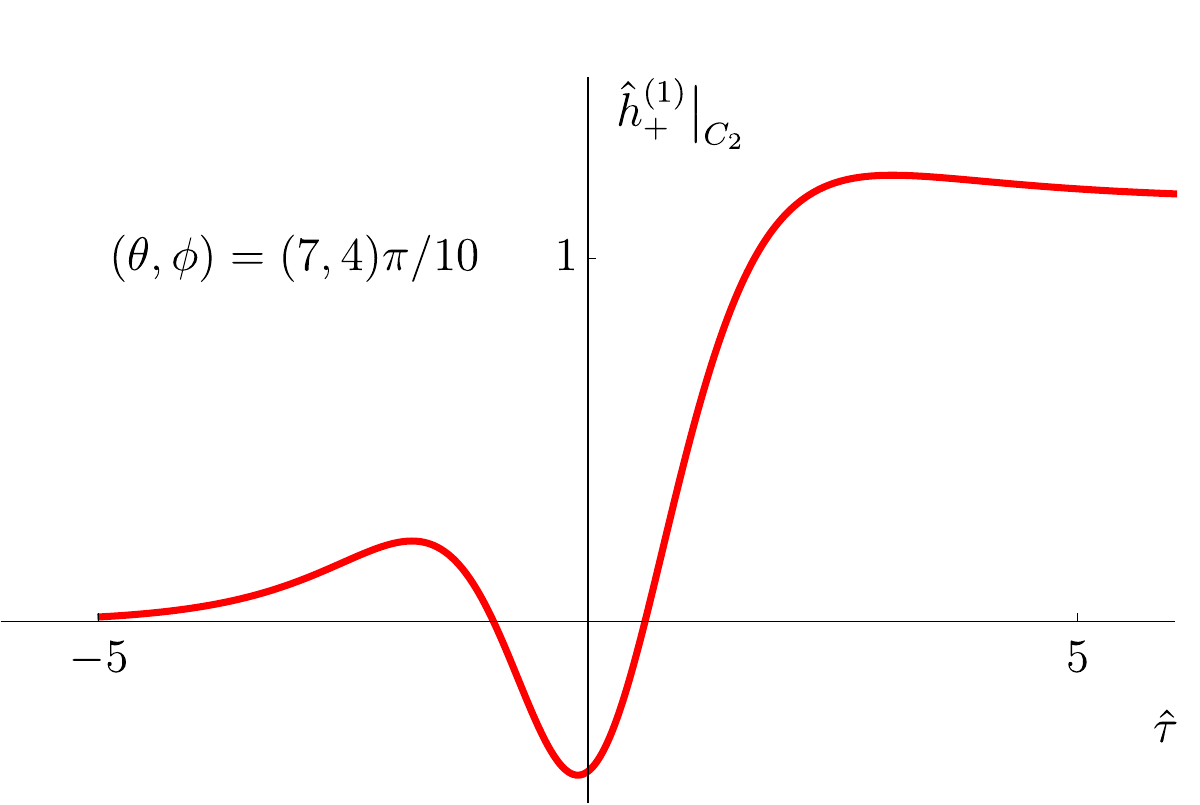}~\raisebox{-0.3cm}{\includegraphics[width=2in]{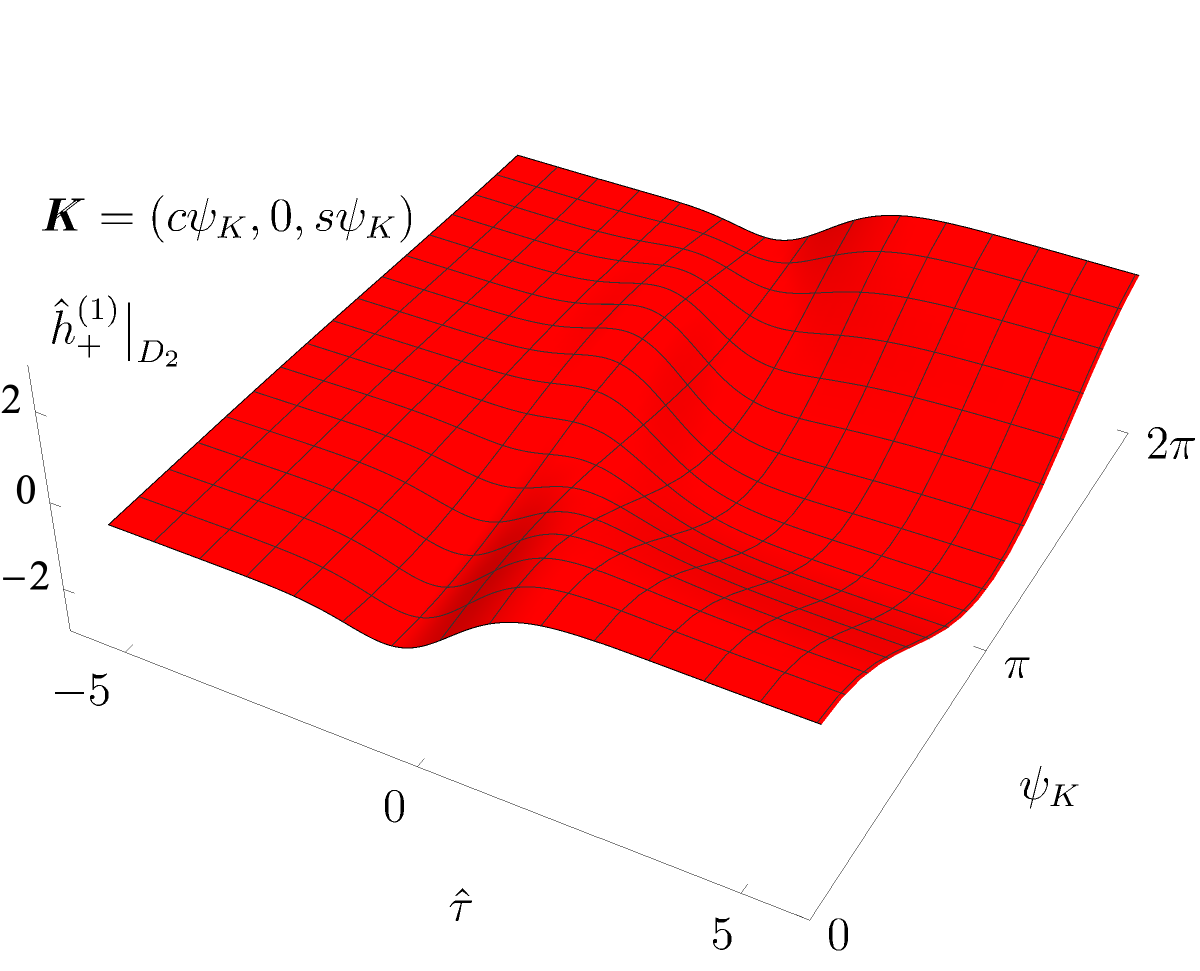}~\includegraphics[width=2in]{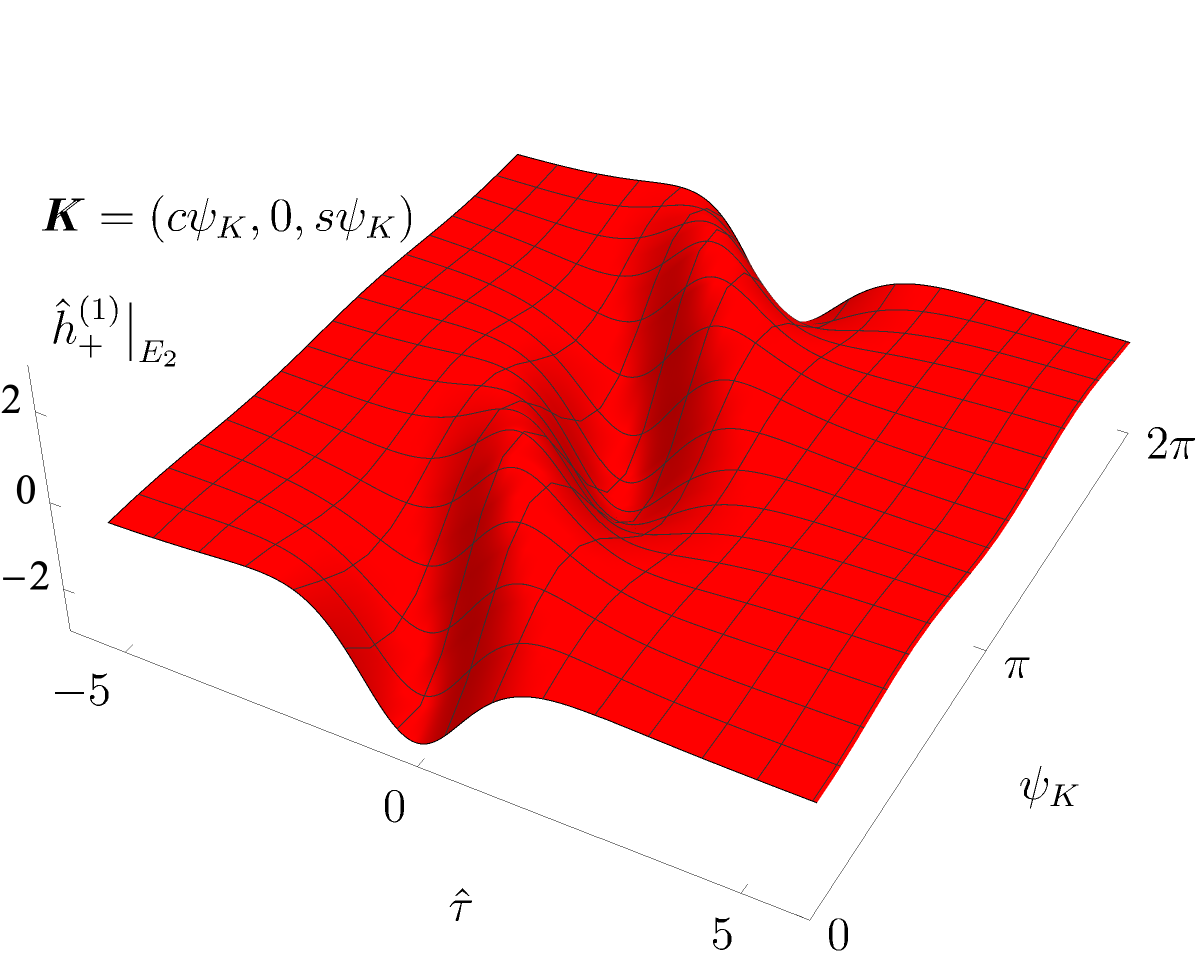}}
\caption{Plots showing various components of $32\pi \times 10^6\times {\hat h}_+$ 
      at location given by the angles $(\theta, \phi)=(7, 4)\pi/10$, for 
      $\bm S = (\cos\pi/4, \sin\pi/4, 0)$ and unit $\bm K$ vector written in terms of $c\psi_k\equiv\cos\psi_K$ and $s\psi_K\equiv\sin\psi_K$ for particles of equal masses $m_1=m_2=1$ with 
      $u_i = \frac{1}{\sqrt{24}}(5, 0, 0, \pm 1)$. We use the covariant impact parameter defined in Ref.~\cite{Bern:2020buy} and choose it to be $\bm b_{\text{cov}} = (50, 0, 0)$ in units of the inverse particles' mass. The retarded-time axis is parameterized by $\hat \tau = \tau / |\bm b_{\text{cov}}|$.
} 
\label{fig:hplus_plots}
\end{figure}

\begin{figure}[tb]
\includegraphics[width=1.9in]{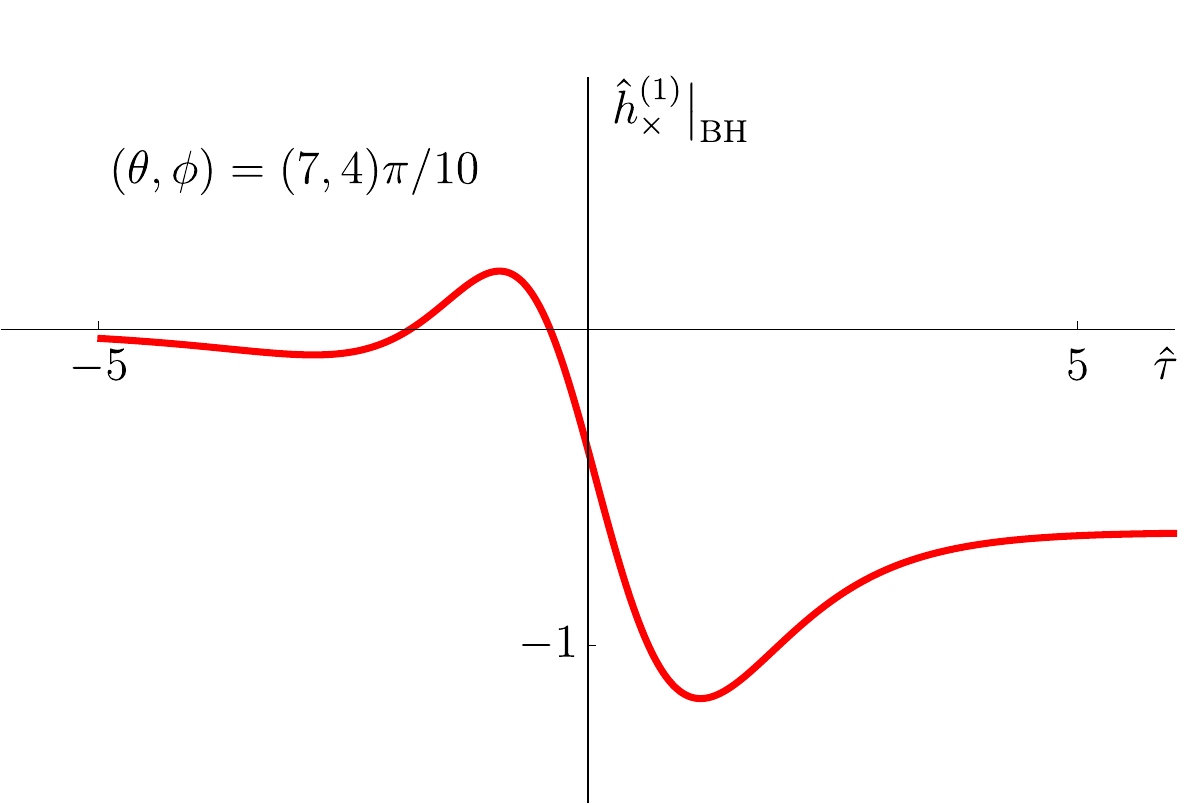}~\raisebox{-0.3cm}{\includegraphics[width=2in]{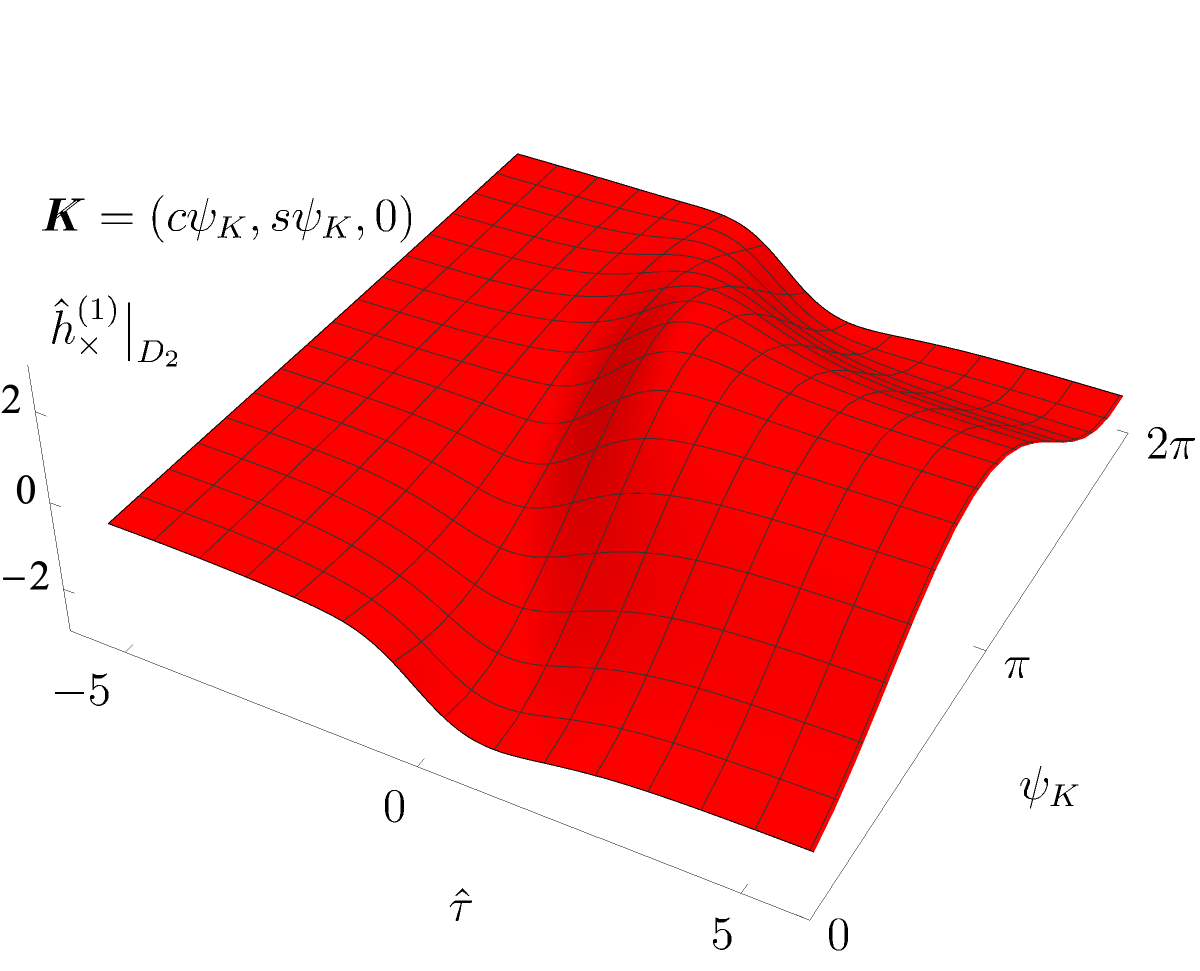}~\includegraphics[width=2in]{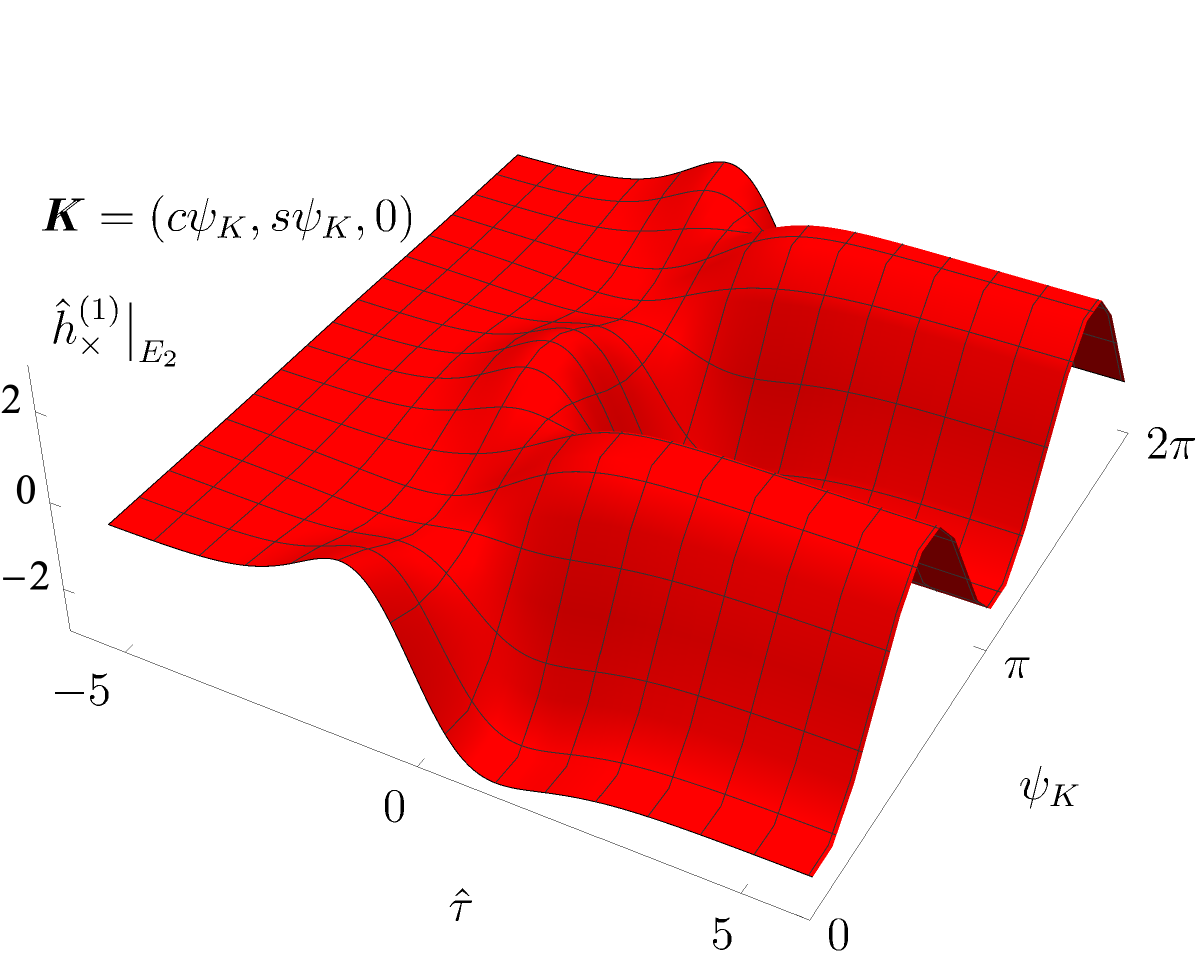}}

\includegraphics[width=1.9in]{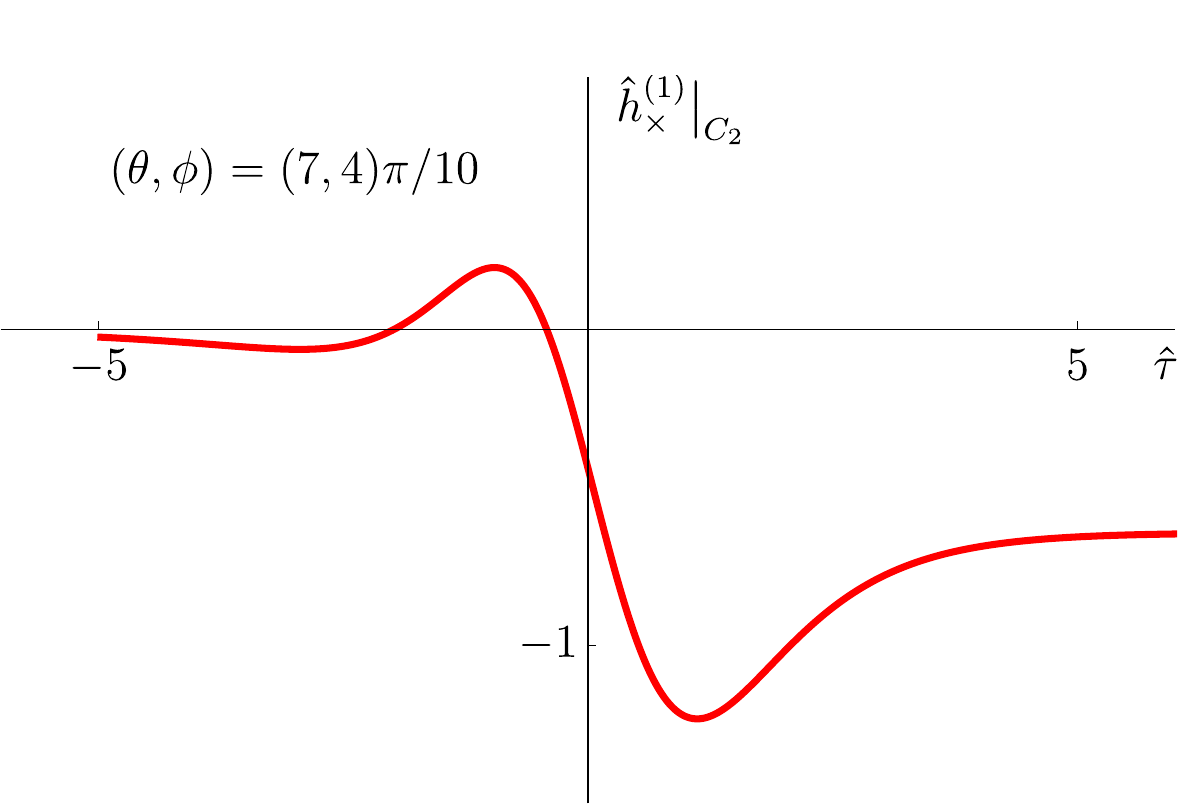}~\raisebox{-0.3cm}{\includegraphics[width=2in]{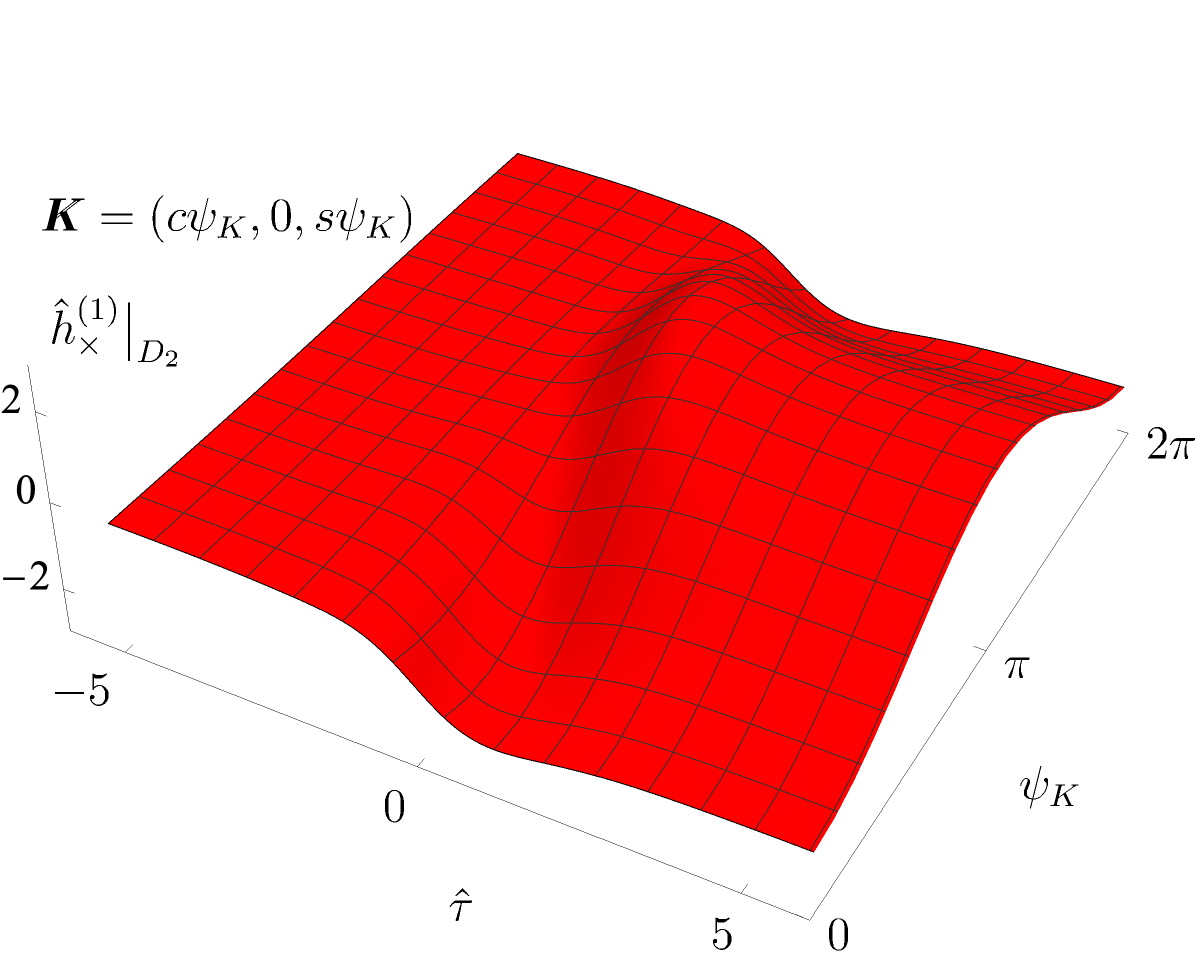}~\includegraphics[width=2in]{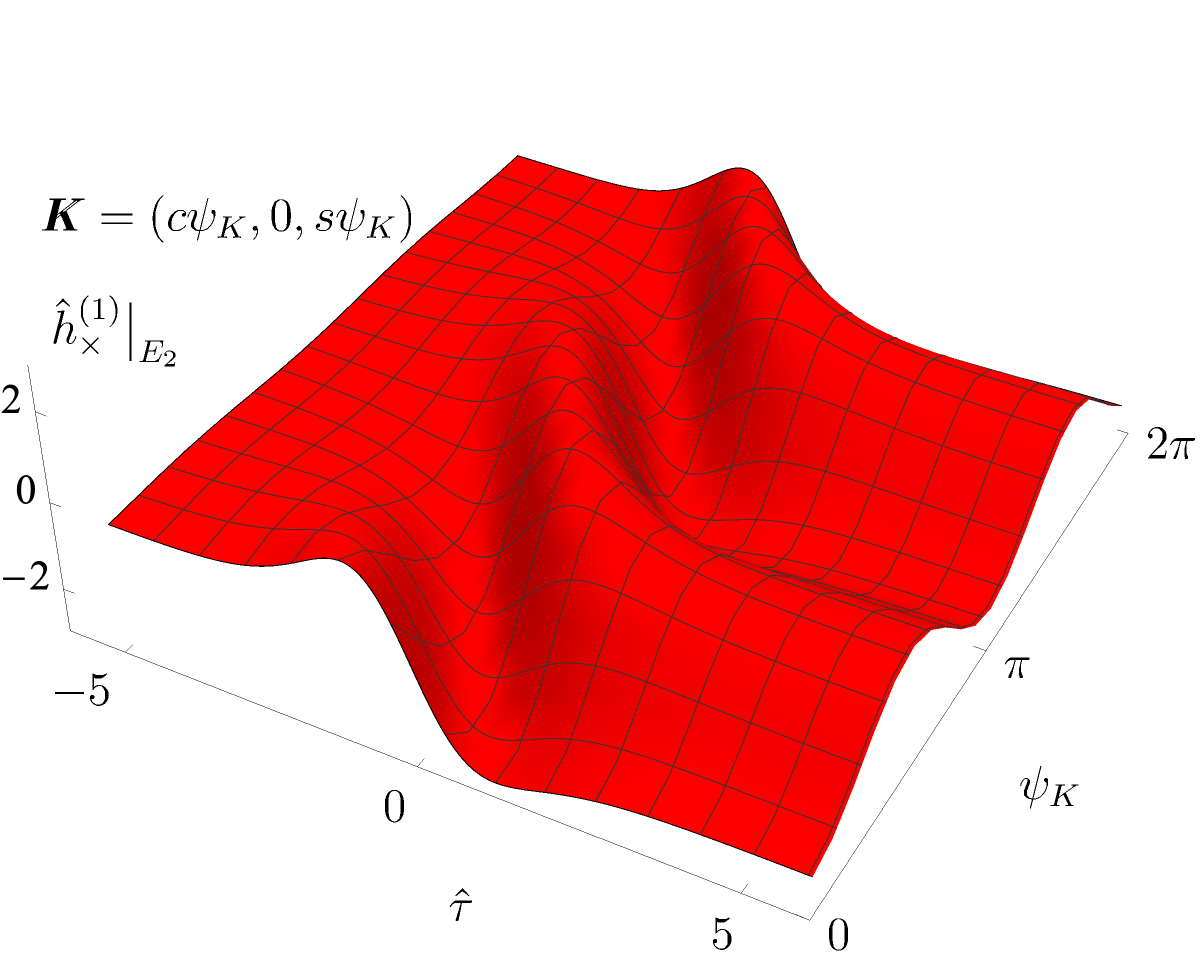}}
\caption{Plots showing various components of $32\pi \times 10^6\times {\hat h}_\times$ 
      at location given by the angles $(\theta, \phi)=(7, 4)\pi/10$, for 
      $\bm S = (\cos\pi/4, \sin\pi/4, 0)$ and unit $\bm K$ vector written in terms of $c\psi_k\equiv\cos\psi_K$ and $s\psi_K\equiv\sin\psi_K$ for particles of equal masses $m_1=m_2=1$ with $u_i = \frac{1}{\sqrt{24}}(5, 0, 0, \pm 1)$. We use the covariant impact parameter defined in Ref.~\cite{Bern:2020buy} and choose it to be $\bm b_{\text{cov}} = (50, 0, 0)$ in units of the inverse particles' mass. The retarded-time axis is parameterized by $\hat \tau = \tau / |\bm b_{\text{cov}}|$.
} 
\label{fig:hcross_plots}
\end{figure}

In \cref{fig:hplus_plots} and \cref{fig:hcross_plots} we plot the $h_{+}$ and $h_{\times}$ components\footnote{The $\hat h_+$ and $\hat h_\times$ components of the metric are defined as $\hat h_+ = \frac{1}{2}(\hat h_{11}-\hat h_{22})$ and $\hat h_\times =\hat h_{12} =\hat h_{21}$.} of the $\order(\qftS^2, \qftS\qftK, \qftK^2)$ leading-order waveform for the observation angle $(\theta, \phi) = (7, 4)\pi/10$ as given by the direction of the out-going graviton of negative helicity, $\hat k = (1, \bm{n})$ and $\bm{n} =(\sin\theta\,\cos\phi, \sin\theta\,\sin\phi,  \cos\theta)$; the background at $\tau\rightarrow-\infty$ is subtracted. 
The two-dimensional plots show the leading order ${\cal O}(S^2)$ waveform for the rest-frame spin $\qftSrf=(1, 1, 0)/\sqrt{2}$ Kerr black hole scattering off a Schwarzschild black hole and the $C_2$ contribution to this scattering waveform (i.e., the departure from Kerr towards \ccos), respectively. The three-dimensional plots show the $D_2$ and $E_2$ contributions for the same spin vector and for two one-dimensional parametrizations of the rest-frame $\qftK$ vector. 
Since the terms captured by each plot are homogeneous in $\qftS$ and $\qftK$, their relative weights can be adjusted at will, and thus, all possible choices of magnitude and orientations of $\qftKrf$ for the given $\qftSrf$ can be reconstructed from these plots.
Moreover, using \cref{eq:propC2E2}, the $\order(S^2)$ terms of a \cco for an arbitrary orientation of the rest frame spin can be obtained from the plots corresponding to the $E_2$ Wilson coefficient. 
The analytic expression for the full $\order(\gS^2)$ contribution $W^\text{LO}_{(2)}(\tau)$ to the leading-order waveform can be found in the ancillary file {\tt waveform.m}.

As highlighted in Ref.~\cite{Alaverdian:2024spu}, the amplitudes of the waveforms generated by commensurate $\qftK$ and $\qftS$ are also commensurate. This suggests that if a waveform dependent on $\qftS$ is measurable, a $\qftK$-dependent waveform can also be measured, provided $\qftK$ has a similar magnitude.
As discussed there, it is not possible to eliminate all $\qftK$ dependence at fixed order in the spin tensor by adjusting the values of the Wilson coefficients. In the next section, we extend this analysis to allow for non-linear redefinitions of $\qftK$ and $\qftS$.


\subsection{On the non-degeneracy of \texorpdfstring{$\qftK$}{K}, \texorpdfstring{$\qftS$}{S} and Wilson coefficients}

In Ref.~\cite{Alaverdian:2024spu}, we explored whether physical observables can identify the presence of a $K$ vector. To that end, we assumed that we are given a waveform signal that can be fitted by a $\qftK^\mu=0$ system and studied whether it is possible to fit the same signal by turning on $\qftK$ and then readjusting Wilson coefficients, i.e., whether under appropriate redefinitions of the coefficients the systems with and without $\qftK$ are physically equivalent.
We carried out this analysis to second order in the spin tensor and, following standard practice, we analyzed each order separately. This prevents any discrepancies from being shifted to higher orders, which were beyond our level of accuracy.
In particular, since $\qftK^\mu$ does not enter at $\order(\gS)$, the spin vector $\qftS^\mu$ remained fixed as $\qftK^\mu$ was turned on. 
Under these assumptions, in Ref.~\cite{Alaverdian:2024spu}, we concluded that $K$ cannot be absorbed by adjusting the Wilson coefficients.

With the close connection between the waveform and the three- and four-point amplitudes in \cref{eq:I1,eq:I2,eq:LOwaveformFinal}, 
here we extend the previous analysis and reach a similar conclusion while relaxing some of the previous assumptions. While we allow for nonlinear redefinitions of $\qftS$ and $\qftK$, not all redefinitions are considered legal. 
In particular, we require that possible redefinitions of $\qftS_j$ and $\qftK_j$ are independent of the process, i.e., these redefinitions can involve only variables associated with the same particle~$j$. 

Since there is no dipole contribution from $\qftK$, the redefinition of variables can only be:
\begin{align}
S_i^\mu &\rightarrow  S_i^\mu + \sum_{n\ge 2}\sum_{p+q=n} ({\widehat M}_{p, q}^\mu(p_i))_{\nu_1\dots \nu_p \rho_1\dots\rho_q} S_i^{\nu_1}\dots S_i^{\nu_p} K_i^{\rho_1} \dots K_i^{\rho_q}
\label{eq:transformationS}
\\
K_i^\mu &\rightarrow \hphantom{ S_i^\mu +} \sum_{n\ge 1}\sum_{p+q=n} ({\widehat N}_{p,q}^\mu(p_i))_{\nu_1\dots \nu_p \rho_1\dots\rho_q} S_i^{\nu_1}\dots S_i^{\nu_p} K_i^{\rho_1} \dots K_i^{\rho_q} \,,
\label{eq:transformationK}
\end{align}
with the various coefficient functions constrained so that the redefined $\qftS$ 
and $\qftK$ vectors are transverse. That is, 
\be
\label{eq:transverse_redefinition}
p_\mu {\widehat M}_{p, q}^\mu(p) = 0 \,, \qquad p_\mu {\widehat N}_{p, q}^\mu(p) = 0 \, .
\ee
With such redefinitions, together with changes in the Wilson coefficients, we attempt to remove $\qftK$ from the leading order waveform. In a worldline effective field theory approach to spinning particle dynamics, velocity-dependent redefinitions of the spin vector have been previously discussed in, e.g., Refs.~\cite{Porto:2005ac, Porto:2008tb, Porto:2008jj}.~\footnote{
Such redefinitions were used there as a means of connecting equations of motion for different SSCs.
} 

Before proceeding, let us point out that the transformations \eqref{eq:transformationS} and \eqref{eq:transformationK} do not manifestly respect the scaling properties of $\qftS$ and $\qftK$. We find that, even with such transformations, we are not able to completely remove the $\qftK$ dependence, and consequently, we do not need to restore the expected scaling of $\qftS$ and $\qftK$.

For simplicity, we consider the waveform to quadratic order in spin for the scattering of a spinning particle of momentum $p_1$ and a spinless particle of momentum $p_2$:
\begin{align}
\label{eq:splitBYspin}
W(\tau) &= W_0(\tau) + W(\tau)_{\nu_1} S^{\nu_1} 
\\
& \quad
+ W(\tau)^{(2,0)}_{\nu_1\nu_2} S^{\nu_1}S^{\nu_2} + W(\tau)^{(1,1)}_{\nu_1\rho_1} S^{\nu_1}K^{\rho_1} 
+ W(\tau)^{(0,2)}_{\rho_1\rho_2} K^{\rho_1}K^{\rho_2} 
+ \order(\gS^3) \, .
\nonumber
\end{align}
Under the transformations \eqref{eq:transformationS} and \eqref{eq:transformationK} the coefficient functions become 
\begin{align}
W(\tau)_{\nu_1}&\rightarrow W(\tau)_{\nu_1}
\\
W(\tau)^{(2,0)}_{\nu_1\nu_2}&\rightarrow W(\tau)^{(2,0)}_{\nu_1\nu_2}+W(\tau)_{\mu}({\widehat M}_{2,0}^\mu(p_1))_{\nu_1 \nu_2} + W(\tau)^{(1,1)}_{\nu_1\mu} ({\widehat N}_{1,0}^\mu(p_1))_{\nu_2}
\nonumber\\
&\qquad
+W(\tau)^{(0,2)}_{\mu_1\mu_2} ({\widehat N}_{1,0}^{\mu_1}(p_1))_{\nu_1} ({\widehat N}_{1,0}^{\mu_2}(p_1))_{\nu_2}
\label{eq:20}
\\
W(\tau)^{(1,1)}_{\nu_1\rho_1}&\rightarrow W(\tau)^{(1,1)}_{\nu_1\rho_1}+W(\tau)_{\mu}({\widehat M}_{1,1}^\mu(p_1))_{\nu_1 \rho_1} + W(\tau)^{(1,1)}_{\nu_1\mu} ({\widehat N}_{0,1}^\mu(p_1))_{\rho_1}
\nonumber\\
&\qquad
+ 2W(\tau)^{(0,2)}_{\alpha\mu} ({\widehat N}_{1,0}^\alpha(p_1))_{\nu_1} ({\widehat N}_{0,1}^\mu(p_1))_{\rho_1}
\label{eq:11}
\\
W(\tau)^{(0,2)}_{\rho_1\rho_2}&\rightarrow W(\tau)^{(0,2)}_{\rho_1\rho_2}+W(\tau)_{\mu}({\widehat M}_{0,2}^\mu(p_1))_{\rho_1 \rho_2} 
+ W(\tau)^{(0,2)}_{\sigma_1\sigma_2} ({\widehat N}_{0,1}^{\sigma_1}(p_1))_{\rho_1}({\widehat N}_{0,1}^{\sigma_2}(p_1))_{\rho_2}
\label{eq:02}
\end{align}
Since the two terms in the expression \eqref{eq:LOwaveformFinal} for the waveform have different mass dependence, these redefinitions cannot mix the two terms. Thus, to see whether $\qftK$ can be removed by a redefinition of the spin and Wilson coefficients, it suffices to inspect the transformations of the three-point and Compton amplitudes under the transformations above. 

Using a notation analogous to that in Eq.~\eqref{eq:splitBYspin}, the three-point amplitude through $\order(\gS^2)$ is given by:
\begin{align}
\mathcal{M}^{\text{3pt}} &= \mathcal{M}^{\text{3pt}}_0 + \mathcal{M}^{\text{3pt}}_{\nu_1} S^{\nu_1}+ 
\mathcal{M}^{\text{3pt},(2,0)}_{\nu_1\nu_2} S^{\nu_1}S^{\nu_2} + \mathcal{M}^{\text{3pt},(1,1)}_{\nu_1\rho_1} S^{\nu_1}K^{\rho_1} \cr
&\quad + 
\mathcal{M}^{\text{3pt},(0,2)}_{\rho_1\rho_2} K^{\rho_1}K^{\rho_2} 
+ \order(\gS^3) \,,
\\
\mathcal{M}^{\text{3pt}}_0&=-(\epsilon_1\cdot p)^2
~,\qquad
\mathcal{M}^{\text{3pt}}_\mu=\frac{(\epsilon_1\cdot p)\tilde{f}_1(p,\mu)}{m}\,,
\\
\mathcal{M}^{\text{3pt}, (2,0)}_{\nu_1\nu_2}&=-\frac{(1+C_2)(\epsilon_1\cdot p)^2k_1{}_{\nu_1}k_1{}_{\nu_2}}{2m^2}\,,
\quad
\mathcal{M}^{\text{3pt}, (1,1)}_{\nu_1\rho_1}=-\frac{D_2(\epsilon_1\cdot p)\tilde{f}_1(p,\nu_1) k_1{}_{\rho_1}}{m^2}\,,
\\
\mathcal{M}^{\text{3pt}, (0,2)}_{\rho_1\rho_2}&=-\frac{E_2(\epsilon_1\cdot p)^2k_1{}_{\rho_1}k_1{}_{\rho_2}}{2m^2}\,,
\end{align}
where $f_i^{\mu\nu} = k_i^{\mu}\epsilon_i^{\nu} -k_i^{\nu}\epsilon_i^{\mu}$ is a linearized field strength
and ${\tilde f}_i^{\mu\nu}$ is its Hodge dual.\footnote{We define Hodge dual as $\tilde{f}^{\mu\nu}=(i/2)\varepsilon^{\mu\nu\alpha\beta}f_{\alpha\beta}$.} We also use $f_i(\mathsf{v} ,\nu)\equiv f_i^{\mu\nu}\mathsf{v}_\mu$ for any vector $\mathsf{v}$, and analogously for $\tilde{f}^{\mu\nu}$.

The general structure of the coefficient functions $\widehat M$ can be inferred 
from their transversality properties and from the fact that they depend only on the momentum of the matter particle:
\begin{align}
({\widehat M}^\mu_{i,j}{})_{\nu_1 \nu_2} = A_{i,j} P^\mu_{\nu_1} p_{\nu_2} + B_{i,j} P^\mu_{\nu_2} p_{\nu_1} + C_{i,j} p_\rho \varepsilon^{\rho\mu}{}_{\nu_1 \nu_2} \,,
\end{align}
where $P^\mu_\nu$ is the projector transverse to $p$. Since the lower indices are contracted with the $\qftS$ and $\qftK$ vectors, all momentum dependence carrying these indices is projected out. Moreover, since $({\widehat M}^\mu_{2,0}){}_{\nu_1 \nu_2}$ and $({\widehat M}^\mu_{0, 2}{})_{\nu_1 \nu_2}$ are symmetric in their lower indices, it follows that ${\widehat M}_{2, 0}$ and ${\widehat M}_{0, 2}$ are effectively zero, while ${\widehat M}_{1, 1}$ is
\begin{align}
({\widehat M}^\mu_{1, 1}{})_{\nu_1 \nu_2} = C_{1, 1}\, p_\rho \varepsilon^{\rho\mu}{}_{\nu_1, \nu_2} \, .
\end{align}
Its contribution to Eq.~\eqref{eq:11}, via multiplication with $\mathcal{M}^{\text{3pt}}_\mu$, is parity-even, so it cannot contribute to the cancellation of the term $\mathcal{M}^{\text{3pt}, (1,1)}_{\nu_1\rho_1}$ which is parity-odd. Thus, for the purpose of the removal of $K$, the remaining ${\widehat M}_{1, 1}$ coefficient function is also irrelevant, so we can set $C_{1, 1}=0$.

The remaining terms on the right-hand side of relations \eqref{eq:11} and \eqref{eq:02}
are proportional to the $D_2$ and $E_2$ Wilson coefficients and the coefficient functions 
${\widehat N}_{1,0}$ and ${\widehat N}_{0, 1}$. The coefficient function $\widehat{N}$ has the same structure as $\widehat{M}$, and a similar analysis shows that no choice for ${\widehat N}_{1,0}$ and ${\widehat N}_{0, 1}$ sets these terms to zero. Thus, echoing the results of the numerical analysis in Ref.~\cite{Alaverdian:2024spu}, we conclude that $\qftK$ cannot be removed by redefining the $\qftS$ and $\qftK$ vectors as well as the Wilson coefficients governing the higher-order interactions.

\section{Two-body amplitudes and eikonal phase}
\label{sec:TwoBodyAmplitude}

\subsection{Two-body amplitudes}
The classical two-body scattering amplitude contains all information necessary for the construction of the eikonal phase in section \ref{sec:eikonal} and of the Hamiltonian in section \ref{sec:Hamiltonian}, both of which have a direct connection to scattering observables. We organize this amplitude in a perturbative expansion in Newton's constant $G$. To the first few orders, we have\footnote{Here, we only consider explicit powers of $G$. For a discussion on the so-called physical PM counting, which also tracks implicit powers of $G$ in the spin variables, see \cref{sec::constructingHam}.}
\begin{equation}
\mathcal{M}^\text{2 body} = \mathcal{M}^{G} + \mathcal{M}^{\PM{2}} + \mathcal{O}\left(G^3\right).
\label{eq:Mtwobody}
\end{equation}

\begin{figure}
\centering
\pgfmathsetmacro{\w}{1.4}
\pgfmathsetmacro{\h}{1.3}
\pgfmathsetmacro{\x}{0.3}
\pgfmathsetmacro{\r}{0.17}
\raisebox{0.65cm}{$\mathcal{M}^{G} \sim $}
\hbox{\begin{tikzpicture}[scale=.8]
    \draw [massive] (0,0) node [left]{\small $1$} -- (\w,0);
    \draw [massive] (0,\h) node [left]{\small $2$} -- (\w,\h);
    \draw [graviton] (\w/2,0) -- ++ (0,\h);
    \filldraw [fill=gray!50] (\w/2,0) circle (\r);
    \filldraw [fill=gray!50] (\w/2,\h) circle (\r);
\end{tikzpicture}}
\caption{Diagram contributing to the ${\cal O}(G)$ two-body amplitude. The exposed line is not allowed to be collapsed.}
\label{fig:LO2body}
\end{figure}

We obtain the above amplitudes by employing generalized unitarity~\cite{Bern:1994zx, Bern:1994cg, Kosmopoulos:2020pcd}. The ${\cal O}(G)$ amplitude is given by sewing together, as shown schematically in \cref{fig:LO2body}, two three-point amplitudes extracted from the Lagrangians discussed in \cref{sec:HigherOrderFT} and shown after some manipulations in \cref{eq:M3_full}. 
Similarly, we compute the ${\cal O}(G^2)$ amplitude as shown in \cref{fig:NLO2body} by adding together the two possible sewings of the Compton amplitude and two three-point amplitudes constructed from the Lagrangians discussed in \cref{sec:HigherOrderFT}, and subtracting their common terms so that they are not over-counted. We include in the ancillary file {\tt twoBody.m} the covariant form of these two-body amplitudes through cubic order in the spin. 

\begin{figure}
\centering
\pgfmathsetmacro{\w}{2.4}
\pgfmathsetmacro{\h}{1.3}
\pgfmathsetmacro{\x}{0.3}
\pgfmathsetmacro{\r}{0.17}
\raisebox{0.65cm}{$\mathcal{M}^{\text{$G^2$}} \sim $}
\hbox{\begin{tikzpicture}[scale=.8]
    \draw [massive] (0,0) node [left]{\small $1$} -- (\w,0);
    \draw [massive] (0,\h) node [left]{\small $2$} -- (\w,\h);
    \draw [graviton] (\w/2,0) -- ++ (\w/4,\h);
    \draw [graviton] (\w/2,0) -- ++ (-\w/4,\h);
    \filldraw [fill=gray!50] (\w/2,0) circle (\r);
    \filldraw [fill=gray!50] (\w/4,\h) circle (\r);
    \filldraw [fill=gray!50] (3*\w/4,\h) circle (\r);
\end{tikzpicture}}
\raisebox{0.65cm}{$+$}
\hbox{\begin{tikzpicture}[scale=.8]
    \draw [massive] (0,0) node [left]{\small $1$} -- (\w,0);
    \draw [massive] (0,\h) node [left]{\small $2$} -- (\w,\h);
    \draw [graviton] (\w/4,0) -- ++ (\w/4,\h);
    \draw [graviton] (3*\w/4,0) -- ++ (-\w/4,\h);
    \filldraw [fill=gray!50] (\w/4,0) circle (\r);
    \filldraw [fill=gray!50] (3*\w/4,0) circle (\r);
    \filldraw [fill=gray!50] (\w/2,\h) circle (\r);
\end{tikzpicture}}
\raisebox{0.65cm}{$-$}
\hbox{\begin{tikzpicture}[scale=.8]
    \draw [massive] (0,0) node [left]{\small $1$} -- (\w,0);
    \draw [massive] (0,\h) node [left]{\small $2$} -- (\w,\h);
    \draw [graviton] (\w/4,0) -- ++ (0,\h);
    \draw [graviton] (3*\w/4,0) -- ++ (0,\h);
    \filldraw [fill=gray!50] (\w/4,0) circle (\r);
    \filldraw [fill=gray!50] (3*\w/4,0) circle (\r);
    \filldraw [fill=gray!50] (\w/4,\h) circle (\r);
    \filldraw [fill=gray!50] (3*\w/4,\h) circle (\r);
\end{tikzpicture}}
\caption{The ${\cal O}(G^2)$ two-body amplitude. Each diagram represents a generalized cut, and each exposed line is cut, i.e., it cannot be collapsed. The third cut subtracts contributions with that topology present in both of the first two cuts, ensuring that no terms are double-counted.}
\label{fig:NLO2body}
\end{figure}

The resulting amplitudes are given as linear combinations of the elements of a basis of $S$- and $K$-dependent operators $\mathbb{O}^{\{i,j\}}$, 
\begin{align}
\mathcal{M}^{G} =
\frac{4\pi G}{\clq^2}
\sum_{i,j}
a^{\{i,j\}}_1 \mathbb{O}^{\{i,j\}}
\,,
\label{eq:MEFT_1PM}
\end{align}
and 
\begin{align}
\mathcal{M}^{\PM{2}} &= \mathcal{M}^{\PM{2}}_\triangle
+(4\pi G)^2\, a_{\rm iter}\int \frac{d^{D-1}\bm \ell}{(2\pi)^{D-1}}
\frac{(2\xi \en{})(4 \en{1} \en{2})}{\bm \ell^2 (\bm \ell+\clq)^2 (\bm \ell^2+2\clp\cdot \bm \ell)} \,,
\nn \\
\mathcal{M}^{\PM{2}}_\triangle &=
\frac{2\pi^2 G^2}{|\clq|}
\sum_{i,j}
a^{\{i,j\}}_2 \mathbb{O}^{\{i,j\}} \,,
\label{eq:M_2PM}
\end{align}
with $\en{} = \en{1} + \en{2}$, $\xi = \frac{\en{1} \en{2}}{\en{}^2}$ and $\en{1,2} = \sqrt{\opp^2 + m_{1,2}^2}$. The scattering process is described by the incoming center-of-mass (\com) momentum $\opp$ and the momentum transfer $\clq$. In the coefficients $a_2^{\{i, j\}}$, the first index refers to the total number of $S$ and $K$ vectors, and the second runs over the corresponding operators. The triangle subscript in $\mathcal{M}^{\PM{2}}_\triangle$ indicates that the origin of the contribution is a one-loop scalar triangle integral. We arrive at this decomposition using integration-by-parts (IBPs) identities~\cite{Chetyrkin:1981qh, Laporta:2000dsw, Smirnov:2008iw} with the software FIRE6~\cite{Smirnov:2019qkx}. We have computed one-loop amplitudes up to cubic order in combinations of the vectors $S_i$ and $K_i$ (for $i=1,2$). The precise choice of the basis operators $\mathbb{O}^{\{i,j\}}$ depends on whether or not we demand manifest covariance. The spinless part is accompanied simply by the identity\begin{equation}
\mathbb{O}^{\{0,1\}}=1.
\end{equation}
There are four linear-in-spin structures $\mathbb{O}^{\{1,j\}}$, given by
\begin{equation}
l\cdot S_i,\qquad 
q\cdot K_i,\qquad \textrm{for\ }i=1,2, 
\end{equation}
where we define the vector $l^{\mu}\equiv \varepsilon^{\mu\nu\sigma\rho}u_{1,\nu}u_{2,\sigma}q_{\rho}$. 
There are 26 quadratic-in-spin structures $\mathbb{O}^{\{2,j\}}$. They are given by the 18 structures defined by
\begin{equation}
q\cdot V_1\  q\cdot V_2,
\qquad    q^2\ V_1 \cdot V_2,
\qquad    q^2\  p\cdot V_1\  p\cdot V_2,
\end{equation}
where $(V_1,V_2)$ are the six combinations of $\{S_1,S_2,K_1,K_2\}$ which have only either $S_i$ or $K_i$, plus the 8 structures defined by
\begin{equation}
l\cdot V_1\   
q\cdot V_2\,,
\qquad l\cdot V_2\   
q\cdot V_1\,,
\end{equation}
for the four pairs of vectors containing exactly one $S_i$ and one $K_i$.  
Finally, there are 100 operators $\mathbb{O}^{\{3,j\}}$ at cubic-in-spin order. They can be generated starting from the structures
\begin{align}
\label{eq:cubicStrOdd}
q\cdot V_1\  q\cdot V_2\ 
q\cdot V_3\,,
\qquad    q^2\ 
q\cdot V_1\ 
V_2 \cdot V_3\,,
\qquad    q^2\ 
q\cdot V_1\ 
p\cdot V_2\  p\cdot V_3\,,\\
\label{eq:cubicStrEven}
l\cdot V_1\   
q\cdot V_2\ 
q\cdot V_3\,,
\qquad
q^2\, l\cdot V_1\   
V_2\cdot V_3\,,
\qquad
q^2\, l\cdot V_1\   
p\cdot V_2\ 
p\cdot V_3\,,
\end{align}
where $(V_1,V_2,V_3)$ are the permutations (with repetition) of $\{S_1,S_2,K_1,K_2\}$ which have an odd number of $S_i$ in Eq.~\eqref{eq:cubicStrOdd}, and
an even number of $S_i$ in Eq.~\eqref{eq:cubicStrEven}.
Full expressions up to 1-loop level are given in the ancillary file {\tt twoBody.m}. For illustration, we show here the tree-level amplitude through $S^3$, when only one of the particles carries spin:
\begin{align}
\nonumber
\covM^{G} =
\frac{4\pi G m_1^2m_2^2}{-q^2}
\Big[&
a_{1,\rm cov}^{\{0,1\}}+
a_{1,\rm cov}^{\{1,1\}}l\cdot \aa_1+
a_{1,\rm cov}^{\{1,3\}}q\cdot \kK_1+
a_{1,\rm cov}^{\{2,1\}}(q\cdot \aa_1)^2+a_{1,\rm cov}^{\{2,14\}}(q\cdot \kK_1)^2\\ 
\nonumber
&+a_{1,\rm cov}^{\{2,10\}}l\cdot \aa_1\ 
q\cdot \kK_1+
a_{1,\rm cov}^{\{2,11\}}l\cdot \kK_1\   
q\cdot \aa_1+
a_{1,\rm cov}^{\{3,1\}}l\cdot \aa_1  
(q\cdot \aa_1)^2
\\
 &+
a_{1,\rm cov}^{\{3,17\}}(q\cdot \aa_1)^2q\cdot \kK_1+
a_{1,\rm cov}^{\{3,34\}}l\cdot \aa_1  
(q\cdot \kK_1)^2+
a_{1,\rm cov}^{\{3,44\}}(q\cdot \kK_1)^3
\Big] \ .
\label{eq:QFTamplitude1PM}
\end{align}
We expressed this amplitude in terms of the rescaled vectors
\begin{align}
\aa_1\equiv S_1/m_1,\qquad \kK_1\equiv K_1/m_1 \,,
\label{eq:rescaledKandS}
\end{align}
which makes the mass scaling of every term in the amplitude uniform,
%
and we omit structures containing $q^2$, as their Fourier transform produces contact interactions. The coefficients are given by
\begin{align}
\nonumber 
& a_{1,\rm cov}^{\{0,1\}}=4(2\yy^2-1) \,,
& & a_{1,\rm cov}^{\{1,1\}}=-8i\yy  \,,
& & a_{1,\rm cov}^{\{1,3\}}=0\,,\\
\nonumber
& a_{1,\rm cov}^{\{2,1\}}=2(1+C_2)(2\yy^2-1) \,,
& & a_{1,\rm cov}^{\{2,10\}}=-8iD_2\yy\,,
& & \\ 
& a_{1,\rm cov}^{\{2,11\}}=0 \,,
& & a_{1,\rm cov}^{\{2,14\}}=2E_2(2\yy^2-1)\,,
& & \\ 
\nonumber
& a_{1,\rm cov}^{\{3,1\}}=-\frac{4}{3}i\yy (1+C_3) \,,
& & a_{1,\rm cov}^{\{3,17\}}=2(2\yy^2-1)(D_3-C_2)\,,
& & \\
\nonumber
& a_{1,\rm cov}^{\{3,34\}}=-4i\yy E_3 \,,
& & a_{1,\rm cov}^{\{3,44\}}=\frac{2}{3}(2\yy^2-1)F_3 \,,
& & 
\end{align}
where $\sigma$ is defined as
\begin{align}
    \sigma = \frac{p_1\cdot p_2}{m_1m_2}\,.
\end{align}
The $y$ used in \cref{sec:waveform} and the $\sigma$ here differ by $\order(q^2)$ terms, $y=\sigma+\order(q^2)$ and thus, through $\order(G^2)$, they are identical. However, for the sake of generality, we keep the notation different.
One can similarly write the terms where both particles are spinning, up to cubic order
\begin{align}
\covM^{G} =
\frac{4\pi G m_1^2 m_2^2}{-q^2}
\Big[&
a_{1,\rm cov}^{\{2,a\}}q\cdot \aa_1 q\cdot \aa_2 + a_{1,\rm cov}^{\{3,a\}}l\cdot \aa_1 (q\cdot \aa_1) q\cdot \aa_2+ a_{1,\rm cov}^{\{3,b\}}(q\cdot \aa_1)^2 l\cdot \aa_2  \nn\\
& +a_{1,\rm cov}^{\{3,c\}}q\cdot \aa_1 (q\cdot \aa_2) q\cdot \kK_1
+ a_{1,\rm cov}^{\{3,d\}}l\cdot \aa_2(q\cdot \kK_1)^2
\Big] \,,
\end{align}
and their coefficients are given by
\begin{align}
    &a_{1,\rm cov}^{\{2,a\}} = 4(2\yy^2-1) \,,
    & & a_{1,\rm cov}^{\{3,a\}} = -4i\sigma \,,
    & & a_{1,\rm cov}^{\{3,b\}}=-4i\yy C_2^{(1)}\,, \nonumber\\
    &a_{1,\rm cov}^{\{3,c\}} = 4(2\yy^2-1)D_2^{(1)} \,,
    & & a_{1,\rm cov}^{\{3,d\}} = -4i\yy E_2^{(1)}\,,
\end{align}
where the superscript $(1)$ in the Wilson coefficients denotes that they correspond to the first particle. 

For the construction of a Hamiltonian, as well as for establishing a comparison point with the toy models discussed in \sect{sec:toymodel}, we use instead the spacelike part $\bm q$ of the momentum transfer and the vectors $\bm S_i$ and $\clK_i$, which are the rest frame spatial parts of $S$ and $\covclK$  of the two particles. 
Note that the latter is the analytic continuation of the vector $K$, which was more natural to express results from field theory, as discussed in Sec. \ref{sec:basics}.
The amplitudes are expressed in the \com frame, which is defined by the kinematics, 
\begin{align}\label{eq:COMdef}
& p_1^{\mu}=-(\en{1}, \,\clp) \,,
\qquad
p_2^{\mu}=-(\en{2}, \,-\clp) \, ,
\qquad
q^{\mu}=(0, \,\clq)\,,
\qquad
\clp\cdot \clq = \clq^2/2\,, \\
& \qftS_1^{\mu} = \left(\frac{\clp\cdot\qftSrf_1}{m_1},\qftSrf_1+\frac{\clp\cdot\qftSrf_1}{m_1(m_1+\en{1})}\clp\right), \quad \qftS_2^{\mu} = \left(-\frac{\clp\cdot\qftSrf_2}{m_2},\qftSrf_2+\frac{\clp\cdot\qftSrf_2}{m_2(m_2+\en{2})}\clp\right), \nonumber\\
& \qftK_1^{\mu} = \left(\frac{\clp\cdot\qftKrf_1}{m_1},\qftKrf_1+\frac{\clp\cdot\qftKrf_1}{m_1(m_1+\en{1})}\clp\right), \quad \qftK_2^{\mu} = \left(-\frac{\clp\cdot\qftKrf_2}{m_2},\qftKrf_2+\frac{\clp\cdot\qftKrf_2}{m_2(m_2+\en{2})}\clp\right), \nonumber
\end{align}
so we can express all the Lorentz invariant products in terms of rest-frame variables. For example 
\begin{align}
    l\cdot \aa_1=\Xi_{\epsilon}\, \bs{L_q}\cdot \bs{S}_1,\qquad
    l\cdot \kK_1=\Xi_{\epsilon}\,\bs{L_q}\cdot (-i\clK_1),\qquad \Xi_{\epsilon}\equiv \frac{i\en{}}{m_1^2m_2},
    \label{eq:relationsldota}
\end{align}
where we defined $\bs{L_q}\equiv i\bs{p}\times \bs{q}$ and used the notation introduced below \cref{eq:M_2PM}.
By also taking into account the expansion of the polarization product following \cref{epdotep} to convert the covariant spin amplitude $\covM$ to the canonical spin amplitude
$\canM$, 
\begin{align}
    &\canM = \polM_1^{(s)}\cdot
\polM_4^{(s)}\polM_2^{(s)}\cdot\polM_3^{(s)} \covM\,,
\end{align}
where 
\begin{align}
&\polM_1^{(s)}\cdot\polM_4^{(s)}
	= \exp\left[-\frac{i}{m_1}\bm{q}\cdot \clK_1   \right]
	\exp\left[ \Xi_1 \bs{L_q}\cdot \bs{S}_1\right], & & \Xi_1\equiv -\frac{1}{m_1(m_1+\en{1})}\,, \nonumber\\
    &\polM_2^{(s)}\cdot\polM_3^{(s)}
	= \exp\left[\frac{i}{m_2}\bm{q}\cdot \clK_2   \right]
	\exp\left[ \Xi_2 \bs{L_q}\cdot \bs{S}_2\right], & & \Xi_2\equiv -\frac{1}{m_2(m_2+\en{2})}\,,
\end{align}
The tree-level amplitude through $\order(G\gS_1^3\gS_2^0)$ takes the form,
\begin{align}
\nonumber
\canM^{G} =
\frac{4\pi G}{\clq^2}
\Big[&
a_{1,\rm can}^{\{0,1\}}-
a_{1,\rm can}^{\{1,1\}}i\bs{L_q}\cdot \bs{S}_1+
a_{1,\rm can}^{\{1,3\}}\bs{q}\cdot \clK_1+
a_{1,\rm can}^{\{2,1\}}(\bs{q}\cdot \bs{S}_1)^2+a_{1,\rm can}^{\{2,14\}}(\bs{q}\cdot \clK_1)^2\\
\nonumber 
&-a_{1,\rm can}^{\{2,10\}}i\bs{L_q}\cdot \bs{S}_1\   
\bs{q}\cdot \clK_1-
a_{1,\rm can}^{\{2,11\}}i\bs{L_q}\cdot \clK_1\   
\bs{q}\cdot \bs{S}_1-
a_{1,\rm can}^{\{3,1\}}i\bs{L_q}\cdot \bs{S}_1  
(\bs{q}\cdot \bs{S}_1)^2\\
&
+a_{1,\rm can}^{\{3,17\}}(\bs{q}\cdot \bs{S}_1)^2\bs{q}\cdot \clK_1-
a_{1,\rm can}^{\{3,34\}}i\bs{L_q}\cdot \bs{S}_1  
(\bs{q}\cdot \clK_1)^2+
a_{1,\rm can}^{\{3,44\}}(\bs{q}\cdot \clK_1)^3
\Big],
\label{eq:QFTamplitude1PMcan}
\end{align}
where the appearing coefficients are given by
\begin{align}
\nonumber
a_{1,\rm can}^{\{0,1\}}& =a_{1,\rm cov}^{\{0,1\}}\,,\qquad 
a_{1,\rm can}^{\{1,1\}}=\Xi_{\epsilon}ia_{1,\rm cov}^{\{1,1\}}+\Xi_1ia_{1,\rm cov}^{\{0,1\}},\qquad 
a_{1,\rm can}^{\{1,3\}}=\frac{ia_{1,\rm cov}^{\{1,3\}}}{m_1}\,,\\
\nonumber
a_{1,\rm can}^{\{2,1\}}& =\frac{a_{1,\rm cov}^{\{2,1\}}}{m_1^2}+\Xi_1\Xi_{\epsilon}\bs{p}^2 a_{1,\rm cov}^{\{1,1\}}+\frac{\Xi_1^2}{2}\bs{p}^2 a_{1,\rm cov}^{\{0,1\}}\,,\\
\nonumber
a_{1,\rm can}^{\{2,10\}}& =-\frac{\Xi_{\epsilon}a_{1,\rm cov}^{\{2,10\}}}{m_1}-\frac{\Xi_1 a_{1,\rm cov}^{\{1,3\}}}{m_1}\,,\qquad
a_{1,\rm can}^{\{2,11\}}=-\frac{\Xi_{\epsilon}a_{1,\rm cov}^{\{2,11\}}}{m_1}\,,\qquad 
a_{1,\rm can}^{\{2,14\}}=-\frac{a_{1,\rm cov}^{\{2,14\}}}{m_1^2}\,,\\
\nonumber
a_{1,\rm can}^{\{3,1\}}&=\frac{\Xi_{\epsilon}ia_{1,\rm cov}^{\{3,1\}}}{m_1^2}+\frac{\Xi_1 ia_{1,\rm cov}^{\{2,1\}}}{m_1^2}+\frac{1}{2}\Xi_1^2\Xi_{\epsilon}\bs{p}^2 ia_{1,\rm cov}^{\{1,1\}}+\frac{1}{6}\Xi_1^3\bs{p}^2 ia_{1,\rm cov}^{\{0,1\}}\,,\\
\nonumber
a_{1,\rm can}^{\{3,17\}}&=\frac{ia_{1,\rm cov}^{\{3,17\}}}{m_1^3}+\frac{\Xi_1\Xi_{\epsilon}\bs{p}^2(ia_{1,\rm cov}^{\{2,10\}}+ia_{1,\rm cov}^{\{2,11\}})}{m_1}+\frac{\Xi_1^2\bs{p}^2ia_{1,\rm cov}^{\{1,3\}}}{2m_1}\,,\\ 
a_{1,\rm can}^{\{3,34\}}&=-\frac{\Xi_{\epsilon}ia_{1,\rm cov}^{\{3,34\}}}{m_1^2}-\frac{\Xi_1 ia_{1,\rm cov}^{\{2,14\}}}{m_1^2},\qquad 
a_{1,\rm can}^{\{3,44\}}=-\frac{ia_{1,\rm cov}^{\{3,44\}}}{m_1^3}\,.  
\end{align}

\subsection{Eikonal phase}
\label{sec:eikonal}
Let us now compute the eikonal phase,  which to $\order(G^2)$ is given by the two-dimensional Fourier transform (from $\bs q$ space to $\bs b$ space) of the classical part of the two-body amplitude as given in Eq.~\eqref{eq:Mtwobody}, while keeping only the triangle contribution in Eq.~\eqref{eq:M_2PM}~\cite{Bern:2020buy},
\begin{equation}
\chi=\frac{1}{4E|\clp|}\int\frac{d^2\bs q}{(2\pi)^2}e^{-i\bs q\cdot\bs b}(\mathcal M^{G}+\mathcal M^{\PM{2}}_\triangle)+\mathcal O(G^3)\, ,
\label{eq:eikonal_Fourier}
\end{equation}
while the box contribution to the amplitude is effectively included in the exponentiation of the tree-level amplitude $\mathcal M^{G}$. For concreteness, let us elaborate on the form of the eikonal phase when only one of the particles is spinning. This can be expressed as
\begin{align} 
\label{eq:LOeikonal}
\chi_1&=\frac{G}{E|\bs p|}
\left(
\alpha_1^{\{0,1\}}\log |\bs b|+
\sum_{i=1}^2\alpha_1^{\{1,i\}}\mathcal{O}^{\{1,i\}}+
\sum_{i=1}^8\alpha_1^{\{2,i\}}\mathcal{O}^{\{2,i\}}+
\sum_{i=1}^{16}\alpha_1^{\{3,i\}}\mathcal{O}^{\{3,i\}}
\right) , \\
\chi_2&=\frac{G^2\pi}{E|\bs p||\bs b|}
\left(
\alpha_2^{\{0,1\}}+
\sum_{i=1}^2\alpha_2^{\{1,i\}}\mathcal{O}^{\{1,i\}}+
\sum_{i=1}^8\alpha_2^{\{2,i\}}\mathcal{O}^{\{2,i\}}+
\sum_{i=1}^{16}\alpha_2^{\{3,i\}}\mathcal{O}^{\{3,i\}}
\right),
\end{align}
where the coefficients $\alpha_n^{\{j, i\}}$ are linear combinations of the coefficients in the amplitude (but with no dependence on the impact parameter). For illustration, we show here their explicit form at tree level 
\begin{align}
    \al{1}{0}{1}&=
    -\aC{1}{0}{1},\qquad
    \al{1}{1}{1}=
    2i\aC{1}{1}{1},\qquad
    \al{1}{1}{2}=
    -2i\aC{1}{1}{3}\,,\\
    \nonumber
    \al{1}{2}{1}&=
    \frac{2}{\bs{p}^2} \aC{1}{2}{1},
    \qquad
    \al{1}{2}{2}=
    -2\aC{1}{2}{1},
    \qquad
    \al{1}{2}{3}=0,
    \\ \nonumber
    \qquad
    \al{1}{2}{4}&=
    \frac{2}{\bs{p}^2} \aC{1}{2}{14},\qquad
    \al{1}{2}{5}=-2\aC{1}{2}{14},\qquad
    \al{1}{2}{6}=0\,,\\
    \nonumber
    \al{1}{2}{7}&=2\aC{1}{2}{10}+2\aC{1}{2}{11},\qquad
    \al{1}{2}{8}=2\aC{1}{2}{10}+2\aC{1}{2}{11}\,,\\
    \nonumber
    \al{1}{3}{1}&=\frac{4i}{\bs{p}^2} \aC{1}{3}{1},\qquad
    \al{1}{3}{2}=-12i\aC{1}{3}{1},\qquad
    \al{1}{3}{3}=0,\qquad
    \al{1}{3}{4}=\frac{4i}{\bs{p}^2} \aC{1}{3}{34}\,,\\
    \nonumber
    \al{1}{3}{5}&=-4i\aC{1}{3}{34},\qquad
    \al{1}{3}{6}=0,\qquad
    \al{1}{3}{7}=-\frac{4i}{\bs{p}^2} \aC{1}{3}{17},\qquad
    \al{1}{3}{8}=4i\aC{1}{3}{17}\,,\\
    \nonumber
    \al{1}{3}{9}&=0,\qquad
    \al{1}{3}{10}=-\frac{12i}{\bs{p}^2} \aC{1}{3}{34},\qquad
    \al{1}{3}{11}=4i\aC{1}{3}{44},\qquad
    \al{1}{3}{12}=0\,,\\
    \nonumber
    \al{1}{3}{13}&=-\frac{8i}{\bs{p}^2} \aC{1}{3}{17},\qquad
    \al{1}{3}{14}=-8i\aC{1}{3}{34},\qquad
    \al{1}{3}{15}=0,\qquad
    \al{1}{3}{16}=0 \,.
\end{align}
The following spin structures appear in the eikonal phase up to quadratic-in-spin order:
\begin{align}
\mathcal{O}^{\{0,1\}}&=1\,, \hskip 2. cm 
\mathcal{O}^{\{1,1\}}=\bsh{L}\cdot \bsh{S}\,, \hskip 1.3 cm 
\mathcal{O}^{\{1,2\}}=\bsh{b}\cdot \hat{\clK} \,,  \nn\\
\mathcal{O}^{\{2,1\}}&=(\bsh{L}\cdot \bsh{S})^2\,, \qquad\ 
\mathcal{O}^{\{2,2\}}=(\bsh{b}\cdot \bsh{S})^2 \,, \qquad\ 
\mathcal{O}^{\{2,3\}}=(\bs p\cdot \bsh{S})^2\,, \nn\\
\mathcal{O}^{\{2,4\}}&=(\bsh{L}\cdot \hat{\clK})^2\,, \qquad
\mathcal{O}^{\{2,5\}}=(\bsh{b}\cdot \hat{\clK})^2\,, \qquad
\mathcal{O}^{\{2,6\}}=(\bs p\cdot \hat{\clK})^2\,,  \nn\\
\mathcal{O}^{\{2,7\}}&= \bsh{L}\cdot \bsh{S} \,\bsh{b}\cdot \hat{\clK} \,, \hskip 2. cm 
\mathcal{O}^{\{2,8\}}= \bsh{L}\cdot \hat{\clK} \, \bsh{b}\cdot \bsh{S} \,,
\end{align}
where as usual we count orders of spin as containing either an $ \bsh{S}$ or $ \hat{\clK}$.
For cubic-in-spin, we have
\begin{align}
& \mathcal{O}^{\{3,i\}} =\mathcal{O}^{\{2,i\}} \, \bsh{L}\cdot \bsh{S} \,,\hskip 2.3 cm 
\mathcal{O}^{\{3,i+6\}}=\mathcal{O}^{\{2,i\}} \,\bsh{b}\cdot \hat{\clK} \,,
\qquad \textrm{for}\;  i=1,\ldots 6, \nn
\\
& \mathcal{O}^{\{3,13\}}=\bsh{L}\cdot \hat{\clK} \, \bsh{L}\cdot \bsh{S}\, \bsh{b}\cdot \bsh{S} \,, \qquad
\mathcal{O}^{\{3,14\}}= \bsh{L}\cdot \hat{\clK} \, \bsh{b}\cdot \hat{\clK} \, \bsh{b}\cdot \bsh{S}\,, \\ \nonumber
& \mathcal{O}^{\{3,15\}} = \bs p\cdot \hat{\clK} \, \bs p\cdot \bsh{S}\, \bsh{b}\cdot \bsh{S} \,,\qquad\ 
\mathcal{O}^{\{3,16\}}= \bs p\cdot \hat{\clK} \, \bs p\cdot \bsh{S}\, \bsh{L}\cdot \hat{\clK} \,,
\end{align} 
where the orbital angular momentum is defined as $\bs L\equiv \bs b \times \bs p$, and we normalize every vector with respect to the impact parameter, i.e.
\begin{align}
\bsh{b}=\frac{\bs b}{|\bs b|}, \hskip 1 cm 
\bsh{S}=\frac{\bs S}{|\bs b|}, \hskip 1 cm 
\hat{\clK}=\frac{\clK}{|\bs b|}, \hskip 1 cm 
\bsh{L}=\frac{\bs L}{|\bs b|}.
\end{align}

\subsection{Observables from eikonal formula}\label{sec:eikonalFormula}

The spin-dependent scattering observables are encoded in the eikonal phase~\cite{Bern:2020buy,  Gatica:2023iws, Luna:2023uwd}. Ref.~\cite{Bern:2023ity} generalized the construction in Ref.~\cite{Bern:2020buy} by including the effects of $K$ and applied it to electrodynamics.  Here, we confirm the validity of this construction through $\order(G^2 \gS^3)$ in gravity.
From Ref.~\cite{Bern:2023ity}, the transverse impulse,\footnote{Here $\clp\cdot\bs b=0$, and so all the $\bs b$-derivatives are projected orthogonal to the incoming momentum $\clp$. The contributions along $\clp$ are instead obtained using energy conservation.  See Ref.~\cite{Bern:2023ity}.
} 
spin and $\clK$-vector changes can be obtained from 
\begin{align}
    \Delta\mathbb{O}^{(1)} &= \{\chi_1,\mathbb{O}\}\,,\qquad
    \Delta\mathbb{O}^{(2)} = \{\chi_2,\mathbb{O}\}+\Delta\mathbb{O}^{(2)}_{\textrm{iter}} \,,
\label{eqDeltaO}
\end{align}
where the so-called iteration terms are given by
\begin{align}    
    \Delta\mathbb{O}^{(2)}_{\textrm{iter}} &\equiv
    \frac{1}{2}\big\{\chi_1,\{\chi_1,\mathbb{O}\}\big\}+ \mathcal{D}_L\big(\chi_1,\{\chi_1,\mathbb{O}\}\big) - \frac{1}{2}\big\{\mathcal{D}_L(\chi_1,\chi_1),\mathbb{O}\big\} \,,
\label{eqDelta1}
\end{align}
and $\mathbb{O}=(\clp_\perp,\bs S,\clK)$.
The Lorentz algebra gives the brackets of $\bs S$ and $\clK$
\begin{equation}
\{S^{i}_{1},S^{j}_{1}\}=\varepsilon^{ijk} S^{k}_{1}\,,
\quad 
\{S^{i}_{1},\mathrm{K}^{j}_{1}\}=\varepsilon^{ijk} \mathrm{K}_{1}^{k}\,,
\quad
\{\mathrm{K}^{i}_{1},\mathrm{K}^{j}_{1}\}=-\varepsilon^{ijk} S^{k}_{1}\,,
\label{eq:LorentzAlgebraPoissonBrackets}
\end{equation}
with all others vanishing. Furthermore we define $\{f,\clp_{\perp}\}\equiv\frac{\doe f}{\doe \bm b}$,
\begin{align}
\label{defineDL}
\mathcal D_L(f,g)&\equiv -\varepsilon^{ijk}\bigg(S^{i}\frac{\doe f}{\doe S^{j}}
  + \mathrm{K}^{i}\frac{\doe f}{\doe \mathrm{K}^{j}}\bigg)\frac{\doe g}{\doe L^k}\, ,
\end{align}
with an obvious generalization for two spinning particles. 
We have verified that Eqs.~\eqref{eqDeltaO}-~\eqref{eqDelta1} reproduce the impulse, spin and $\clK$-vector change from Hamilton's equations for a single spinning body.
Formulae similar to \cref{eqDeltaO,eqDelta1} have been recently obtained by direct use of the KMOC formalism in Refs.~\cite{Gatica:2023iws, Gatica:2024mur}, and in terms of Dirac brackets both from a classical-physics perspective in Ref.~\cite{Gonzo:2024zxo} and from an eikonal point of view in Ref.~\cite{Kim:2024svw,Kim:2025hpn}. Alternatively, it was shown in Ref.~\cite{Luna:2023uwd} that the so-called iteration terms
%
%
$\Delta\mathbb{O}^{(2),\mu}_{\textrm{iter}}$ can be obtained by means of
\begin{equation}
		\Delta \mathbb{O}^{(2)}_{\textrm{iter}}=
		\frac{\partial \Delta \mathbb{O}^{(1)}}{\partial p^j}
		\frac{\Delta p_{\perp}^{(1),j}}{2}+
		\frac{\partial \Delta \mathbb{O}^{(1)}}{\partial S^j}
		\frac{\Delta S^{(1),j}}{2}+
        \frac{\partial \Delta \mathbb{O}^{(1)}}{\partial \mathrm{K}^j}
		\frac{\Delta \mathrm{K}^{(1),j}}{2}+
        \frac{\partial \Delta \mathbb{O}^{(1)}}{\partial b^j}
		\frac{\Delta b^{(1),j}}{2},
		\label{2PMIter}
	\end{equation}
where the $\Delta p_{\perp}^{(1),j}$, $\Delta S^{(1),j}$, and $\Delta \mathrm{K}^{(1),j}$ are obtained from the first term in Eq.~(\ref{eqDeltaO}) while the change in impact parameter $\Delta b^{(1),j}$ is found from the requirement of conservation of total angular momentum.
This formula was obtained from an analysis of stationary phase conditions in the eikonal. We have verified that the equivalence of Eq.~(\ref{2PMIter}) with Eq.~(\ref{eqDeltaO}) holds through $\order(G^2 \gS^3)$. We can alternatively replace the impact parameter with the orbital angular momentum in all expressions, resulting in the formula:
\begin{equation}
		\Delta \mathbb{O}^{(2)}_{\textrm{iter}}=
		\frac{\partial \Delta \mathbb{O}^{(1)}}{\partial p^j}
		\frac{\Delta p_{\perp}^{(1),j}}{2}+
		\frac{\partial \Delta \mathbb{O}^{(1)}}{\partial S^j}
		\frac{\Delta S^{(1),j}}{2}+
        \frac{\partial \Delta \mathbb{O}^{(1)}}{\partial \mathrm{K}^j}
		\frac{\Delta \mathrm{K}^{(1),j}}{2}+
        \frac{\partial \Delta \mathbb{O}^{(1)}}{\partial L^j}
		\frac{\Delta L^{(1),j}}{2} \, .
		\label{2PMIterL}
	\end{equation}
Conservation of angular momentum sets 
$\Delta L^{(1),j}=-\Delta S^{(1),j}$,
bypassing the need to compute $\Delta b^{(1),j}$. 
To illustrate the results of such calculations, we list here the (transverse) impulse up to the cubic order in the spin tensor, written in a covariant form. This is a basic observable, from which the scattering angles can be immediately obtained. 
Also, similarly to the scattering of spinless bodies, we saw in \cref{sec:LOwf_and_soft_expansion} that it provides a universal characterization of the gravitational wave memory in the presence of the $S$ and $K$ vectors.
For the case of one particle spinning, we give the covariant form of the impulse up to $\order(G\gS_1^3)$,
\begin{align}\label{eq:LOimpulseToS3}
    \Delta p_1^{\mu} &= \frac{2 G m_1 m_2}{(-b^2)\sqrt{\sigma^2-1}}\bigg\{(2\sigma^2-1)b^{\mu} - \frac{2\sigma}{-b^2} \Big(b^{\mu}l_b\cdot\aa_1 + l_b^{\mu} b\cdot\aa_1 \Big) \nonumber\\
    & \quad + \frac{(2\sigma^2-1)(1+C_2)}{b^4}\left[\frac{2l_b^{\mu} (l_b\cdot\aa_1)(b\cdot\aa_1)}{\sigma^2-1}-b^{\mu}\left((b\cdot\aa_1)^2-\frac{(l_b\cdot\aa_1)^2}{\sigma^2-1}\right)\right] \nonumber\\
    & \quad +\frac{4\sigma D_2}{b^4} \left[b^{\mu}\Big(l_b\cdot \covclk_1 b\cdot\aa_1 + l_b\cdot \aa_1 b\cdot \covclk_1 \Big) + l_b^{\mu}\left(b\cdot\aa_1 b\cdot \covclk_1 - \frac{ l_b\cdot\aa_1 l_b\cdot \covclk_1}{\sigma^2-1}\right)\right] \nonumber\\
    & \quad - \frac{(2\sigma^2-1) E_2}{b^4}\left[\frac{2l_b^{\mu} (l_b\cdot \covclk_1)(b\cdot \covclk_1)}{\sigma^2-1}-b^{\mu}\left((b\cdot \covclk_1)^2-\frac{(l_b\cdot \covclk_1)^2}{\sigma^2-1}\right)\right] \nonumber\\
    & \quad + \frac{2\sigma(1+C_3)}{(-b^2)^3} \left[ b^{\mu} l_b\cdot\aa_1 \left(3(b\cdot\aa_1)^2-\frac{(l_b\cdot\aa_1)^2}{\sigma^2-1}\right) + l_b^{\mu} b\cdot\aa_1\left((b\cdot\aa_1)^2 - \frac{3(l_b\cdot\aa_1)^2}{\sigma^2-1}\right) \right] \nonumber\\
    & \quad - \frac{3(2\sigma^2-1)(C_2-D_3)}{(-b^2)^3}\left[\left(b^{\mu}b\cdot\covclk_1- \frac{l_b^{\mu}l_b\cdot\covclk_1}{\sigma^2-1}\right)\left((b\cdot\aa_1)^2 - \frac{(l_b\cdot\aa_1)^2}{\sigma^2-1}\right) \right. \nonumber\\
    & \qquad\qquad\qquad\qquad\qquad\qquad - \left. \frac{2(b^{\mu}l_b\cdot\covclk_1+l_b^{\mu}b\cdot\covclk_1)l_b\cdot\aa_1 b\cdot\aa_1}{\sigma^2-1}\right] \nonumber\\
    & \quad - \frac{6\sigma E_3}{(-b^2)^3}\left[ (b^{\mu}l_b\cdot\aa_1 + l_b^{\mu}b\cdot\aa_1)\left((b\cdot\covclk_1)^2 - \frac{(l_b\cdot\covclk_1)^2}{\sigma^2-1}\right) \right. \nonumber\\
    & \qquad\qquad\qquad + \left. 2 b\cdot\covclk_1 l_b\cdot\covclk_1\left(b^{\mu}b\cdot\aa_1 - \frac{l_b^{\mu}l_b\cdot\aa_1}{\sigma^2-1}\right)\right] \nonumber \\
    & \quad - \frac{(2\sigma^2-1)F_3}{(-b^2)^3}\left[b^{\mu} b\cdot\covclk_1\left((b\cdot\covclk_1)^2 - \frac{3(l_b\cdot\covclk_1)^2}{\sigma^2-1}\right) \right. \nonumber\\
    & \qquad\qquad\qquad\qquad - \left.\frac{l_b^{\mu}l_b\cdot\covclk_1}{\sigma^2-1}\left(3(b\cdot\covclk_1)^2 - \frac{(l_b\cdot\covclk_1)^2}{\sigma^2-1}\right)\right]\bigg\} \, . 
\end{align}
For the case of both particles spinning, the impulse up to $\order(G\gS_1^2\gS^2_2)$ is given by
\begin{align}\label{eq:LOimpulseToSsq}
\Delta p^{\mu}_{1} = \frac{2G m_1 m_2}{(-b^2)\sqrt{\sigma^2-1}} & \bigg\{\frac{2(2\sigma^2-1)}{b^4}\left[\frac{l_b^{\mu}(l_b\cdot\aa_1 b\cdot\aa_2 + b\cdot\aa_1 l_b\cdot\aa_2)}{\sigma^2-1} \right.\nonumber\\
& \qquad\qquad\qquad- \left. b^{\mu}\left(b\cdot\aa_1 b\cdot\aa_2 - \frac{l_b\cdot\aa_1 l_b\cdot\aa_2}{\sigma^2-1}\right)\right] \nonumber\\
& + \frac{6\sigma(1+C_2^{(1)})}{(-b^2)^3}\left[(b^{\mu}\aa_2\cdot l_b+l_b^{\mu}\aa_2\cdot b)\left((\aa_1\cdot b)^2-\frac{(\aa_1\cdot l_b)^2}{\sigma^2-1}\right) \right.\nonumber\\
&\qquad \qquad \qquad\qquad+ \left.2\aa_1\cdot l_b\aa_1\cdot b\left(b^{\mu}\aa_2\cdot b - \frac{l_b^{\mu}\aa_2\cdot l_b}{\sigma^2-1}\right)\right] \nonumber\\
& - \frac{6(2\sigma^2-1)D_2^{(1)}}{(-b^2)^3}\left[\frac{(b^{\mu} l_b\cdot\aa_2+ l_b^{\mu}b\cdot\aa_2)(l_b\cdot\aa_1 b\cdot\covclk_1 + b\cdot\aa_1 l_b\cdot\covclk_1)}{\sigma^2-1}\right. \nonumber\\
& \qquad\qquad\qquad - \left.\left(b^{\mu}b\cdot\aa_2 - \frac{l_b^{\mu}l_b\cdot\aa_2}{\sigma^2-1}\right)\left(b\cdot\aa_1 b\cdot\covclk_1 - \frac{l_b\cdot\aa_1 l_b\cdot\covclk_1}{\sigma^2-1}\right)\right] \nonumber\\
& - \frac{6\sigma E_2^{(1)}}{(-b^2)^3}\left[(b^{\mu}l_b\cdot\aa_2+l_b^{\mu}b\cdot\aa_2)\left((b\cdot\covclk_1)^2-\frac{(l_b\cdot\covclk_1)^2}{\sigma^2-1}\right)\right. \nonumber\\
& \qquad\qquad\qquad + \left. 2 l_b\cdot\covclk_1 b\cdot\covclk_1\left(b^{\mu}b\cdot\aa_2 - \frac{l_b^{\mu}l_b\cdot\aa_2}{\sigma^2-1}\right)\right]\bigg\}\,.
\end{align}
In both expressions, we have used the four-vector impact parameter $b^\mu=b_1^\mu-b_2^\mu$ which points from particle $2$ to $1$, along with the definition  $l_b^{\mu}\equiv \varepsilon^{\mu\nu\sigma\rho}u_{1,\nu}u_{2,\sigma}b_{\rho}$ and $\covclk^{\mu} \equiv i\kK^{\mu}$. 
With suitable relabeling, the above two expressions give the full impulse up to $\order(G\gS^3)$.
We include the one-loop two-body amplitudes up to $\order(G\gS^3)$ in the ancillary file {\tt twoBody.m}, from which one may derive the observables using \cref{eqDeltaO}.
%

\section{Effective Hamiltonian}
\label{sec:Hamiltonian}

In this section, we obtain two-body Hamiltonians whose dynamics match those of the extended field and worldline theories.  These Hamiltonians depend on dynamical degrees of freedom $\clK_i$, $i=1,2$, needed to match results from the extended theories.   We construct two Hamiltonians corresponding to two different gauge choices. The first is built to have a minimal number of terms in momentum space. While this choice is desirable, it leads to Hamiltonian coefficients that are divergent in the zero-velocity limit. This divergence, however, is an artifact of our basis choice. The second Hamiltonian removes the spurious zero-velocity singularities by including a set of operators that are redundant in that they contribute to the same amplitude terms on shell. We exploit this redundancy to remove the spurious zero-velocity singularities of the first gauge choice.
Both choices are equally valid, with one having the advantage of using a minimal set of operators and the other being free of spurious singularities.
Furthermore, the dynamics associated with $\clK_i$ can be decoupled by an appropriate choice of Wilson coefficients, preserving, respectively, the minimal number of terms in the first case and the absence of spurious singularities in the second.
Finally, we show that physical zero-velocity singularities arise in the Hamiltonian when attempting to match an extended field theory to a Hamiltonian without the $\clK$ dynamical degrees of freedom. This provides further evidence for the necessity of including $\clK$ in such cases.

\subsection{Constructing the long-range two-body Hamiltonian}
\label{sec::constructingHam}

We construct the two-body Hamiltonian that describes the long-range interactions of a binary of spinning objects in the \com frame. Our Hamiltonian contains the \com relative distance $\opr$ and relative momentum $\opp$, and the spin vectors for the two interacting bodies $\clS_i$, $i=1,2$.
We incorporate terms that allow for the spin magnitude of the objects to change, while we assume that the masses of the particles are independent from their spin magnitudes.  These terms must include tensors under intrinsic rotations, that is, quantities that do not commute with the spin vector of the object in question $\clS_i$. 
Here, we consider the simplest such case in which each object is described by an additional vector $\clK_i$, which is naturally identified with the extra degrees of freedom contained in the spin tensor, as introduced in \sect{sec:basics}.\footnote{The relation between the field-theory variables and the ones used here is $\bm S = \clS$ and $i \bm K \equiv \clK$, with the latter being a subtle identification. See the discussion below \cref{eq:SpinTensorConservation} and in Ref.~\cite{Bern:2023ity}.}
As noted in \sect{sec:MultipoleRelations}, one can go further by introducing a set of dynamical multipole-moment operators; a primary difference here is that the interactions follow from the Lorentz algebra of the underlying field or worldline theory.

To build the most general long-range two-body Hamiltonian out of these elements, we first write down all possible structures depending on the $\clS_i$ and $\clK_i$. These are built out of the above vectors with the requirement that they are invariant under the so-called classical scaling,
\begin{align}
    \opr \rightarrow \lambda^{-1}\, \opr \,, \qquad
    \opp \rightarrow \opp \,, \qquad
    \clS_i \rightarrow \lambda^{-1}\, \clS_i \,, \qquad
    \clK_i \rightarrow \lambda^{-1}\, \clK_i \,.
\end{align}
See also Eq.~(\ref{eq:soft_scale_Comp}) and Ref.~\cite{Bern:2020buy}.
We further assume that the interactions respect parity and time reversal symmetry so that the relevant spin structures are invariant under 
\begin{alignat}{5}
    &\text{Parity:} \qquad
    &&\opr \rightarrow -\opr \,, \qquad
    &&\opp \rightarrow -\opp \,, \qquad
    &&\clS_i \rightarrow \clS_i \,, \qquad
    &&\clK_i \rightarrow -\clK_i \,, \nn \\
    &\text{Time reversal:} \qquad
    &&\opr \rightarrow \opr \,, \quad
    &&\opp \rightarrow -\opp \,, \quad
    &&\clS_i \rightarrow -\clS_i \,, \quad
    &&\clK_i \rightarrow \clK_i \,, 
\end{alignat}
see, e.g., Ref.~\cite{Weinberg:1995mt}. Given these considerations, the first few spin structures are
\begin{align}
    \frac{\opL \times \clS_i}{\opr^2} \,, \qquad
    \frac{\opr \cdot \clK_i}{\opr^2} \,, \qquad
    \frac{\left(\opL \times \clS_i\right)\left(\opr \cdot \clK_j\right)}{\opr^4} \,, \qquad
    \text{etc.}
\end{align}
Once all desired spin structures $\spinStr{a}$ are listed, the Hamiltonian is given by
\begin{align}
\opH &= \sqrt{\opp ^2 + m_1^2}+\sqrt{\opp^2 + m_2^2} + \sum_{a} V^{a} \spinStr{a}
\,, \qquad
V^{a}=\sum_{n=1}^\infty 
    \left( \frac{G}{|\opr|} \right)^n c_n^{a}(\opp^2)
\,.
\label{Eq:Hamiltonian_Definition}
\end{align}

We organize our list of spin structures and corresponding Hamiltonian in increasing order in $\clS_i$ and $\clK_i$. In this way, our Hamiltonian manifests the so-called ``physical post-Minkowskian counting''~\cite{Buonanno:2024vkx},  which emphasizes the fact that $|\clS|/(G m^2) \leq 1$ for Kerr black holes. Here, we adopt an extension of this counting scheme that treats $\clS$ and $\clK$ on the same footing, as is natural from Lorentz invariance. 

Having constructed the two-body Hamiltonian, we may interpret it as a quantum mechanical operator and use it to compute the two-to-two scattering amplitude. 
We organize the EFT amplitude as a perturbative series in the gravitational coupling $G$,
\begin{equation}
\ampEFT = \ampEFT^{G} + \ampEFT^{G^2} + \ldots \,,
\qquad \text{with} \quad \ampEFT^{G^n} = \order(G^n) \,.
\label{eq:MEFT_PM}
\end{equation}
By matching this amplitude to the QFT amplitudes in \sect{sec:TwoBodyAmplitude}, we obtain the functions $c_n^{a}(\opp^2)$ that capture binary evolution in general relativity. The matching equation is 
\begin{align}
    \text{(EFT)} \qquad \ampEFT = \frac{\canM}{4 \en{1} \en{2}} \qquad \text{(QFT)}\,,
    \label{eq:matching}
\end{align}
where the factor $4 \en{1} \en{2}$ accounts for the non-relativistic normalization used in the EFT.
Matching $\ampEFT^{G}$ to the tree-level QFT amplitude yields the $\order(G)$ Hamiltonian coefficients.
For $\ampEFT^{G^n}$ with $n \ge 2$, the amplitude contains iteration pieces, which are typically infrared divergent, that cancel in the matching to the QFT amplitudes. For the case at hand, we have
\begin{align}
\ampEFT^{G^2} &= \ampEFT^{G^2}_\triangle
+(4\pi G)^2\, a_{\rm iter}\int \frac{d^{D-1}\bm \ell}{(2\pi)^{D-1}}
\frac{2\xi \en{}}{\bm \ell^2 (\bm \ell+\clq)^2 (\bm \ell^2+2\clp\cdot \bm \ell)} \,.
\label{eq:MEFT_2PM}
\end{align}
The notation follows that of \cref{eq:M_2PM} and we have manifested the fact that the iteration pieces are identical to classical order between the QFT and EFT amplitudes.
From the first term in Eq.~(\ref{eq:MEFT_2PM}), we extract the $\mathcal{O}\left(G^2\right)$ Hamiltonian coefficients through \cref{eq:matching}.
For a detailed discussion of the matching process see Ref.~\cite{Bern:2023ity}.

\subsection{A minimal basis for the Hamiltonian}

The Hamiltonian constructed above contains redundancies, in that different choices of the functions $c_n^{a}(\opp^2)$ lead to identical dynamics.
We employ standard methods from effective field theory and scattering amplitudes to arrive at a Hamiltonian with a minimal number of spin structures.
While having a minimal basis is not necessary, it is useful since one needs to consider fewer terms when carrying out calculations.
In this subsection, we provide three examples of increasing complexity that demonstrate this procedure.

In the usual Hamiltonian description of a system, one may perform canonical transformations that alter the Hamiltonian without changing the dynamics. 
Analogously, effective field theories contain redundancies due to the ability to perform field redefinitions without changing physical observables. 
We may obtain a minimal basis for our effective Hamiltonian by first computing the scattering amplitude. 
The scattering amplitude, being an on-shell quantity, is free of these redundancies. 
Indeed, the fact that scattering amplitudes obtained from the effective Hamiltonian should be invariant under canonical transformation was already exploited in Refs.~\cite{Cheung:2018wkq, Bern:2019nnu, Bern:2019crd, Bern:2020buy, Kosmopoulos:2021zoq}. There, Hamiltonians in different coordinate choices were shown to be equivalent by verifying that the scattering amplitudes they produce are identical.

We proceed to demonstrate our process in a series of examples. For the present discussion, it is sufficient to study the tree-level contribution of the Hamiltonian terms to the amplitude.
They are related to the Hamiltonian via a Fourier transform up to a sign (see \cref{eq:FTHamiltonian}). For example, 
\begin{align}
    \opH = \ldots &+ G \, c^{(1,1)}_{1}\left(\opp^2\right)  \,\frac{\opL \cdot \clS_1}{|\opr|^3} + \ldots \Rightarrow \nn \\
    \ampEFT &= \ldots + 4 \pi G \, c^{(1,1)}_1\left(\opp^2\right) \,  \frac{\Lq \cdot \clS_1}{\clq^2} + \ldots \,,
\end{align}
where in the second line $\clq$ is the momentum transfer in the scattering process, while $\opp$ and $\clS_1$ correspond to the incoming \com momentum and spin of particle 1 respectively. These are to be contrasted with the corresponding off-shell quantities in the first line. For more details, see Refs.~\cite{Cheung:2018wkq,Bern:2020buy}.

The on-shell paradigm for finding a minimal basis for the Hamiltonian is to compute the amplitude and keep only non-redundant terms that contribute. 
For instance, the combination $\opr \cdot \opp$ could a priori appear both in the spin structures and in the potential functions $V^a$. However, after the Fourier transform this combination leads to $\opp \cdot \clq = \clq^2/2$, where the equality holds only on shell. Since this term is subleading in the classical limit, terms containing $\opr \cdot \opp$ may be safely neglected from our construction; this choice is referred to as isotropic gauge.\footnote{We emphasize that we are not claiming that in explicit Hamiltonians, one may neglect terms containing $\opr \cdot \opp$. Instead, one may build a different Hamiltonian without these terms that captures the same dynamics.}

As a second example, consider the following two spin structures,
\begin{align}
    \spinStr{(2,2)} = \frac{\clS_1^2}{\opr^2}\,, \quad 
    \spinStr{(2,3)} = \frac{(\opr \cdot \clS_1)^2}{\opr^4}\,.
\end{align}
At $\mathcal{O}\left(G\right)$ their Fourier transform contains terms of the form $\ln(\clq^2/\mu^2)$ for some energy scale $\mu$. Employing the fact that the interactions are mediated by massless particles at tree level in the underlying field theory, we may demand that no such terms appear. Only the following combination leads to the expected tree-level analytic behavior:
\begin{align}
    \frac{G}{|\opr|}\left(3\spinStr{(2,3)} - \spinStr{(2,2)}\right) \quad \Rightarrow \quad
    4\pi G \, \frac{(\clq \cdot \clS_1)^2}{\clq^2}\,.
\end{align}
This demonstrates that there are fewer independent Hamiltonian functions than spin structures (one instead of two, in this case), as shown by the computation of the scattering amplitude.

As a final example, consider the spin structures,
\begin{alignat}{2}
    &\spinStr{(3,4)} =  \frac{(\opr\cdot\clS_1)^2 (\opL\cdot\clS_2)}{\opr^6} \,, \quad 
    &&\spinStr{(3,5)} = \frac{\clS_1^2 (\opL\cdot\clS_2)}{\opr^4} \,, \nn \\
    &\spinStr{(3,6)} = \frac{(\opr\cdot\clS_1) ((\opp \times \clS_1) \cdot \clS_2)}{\opr^4} \,, \quad 
    &&\spinStr{(3,7)} = \frac{(\opp \cdot \clS_1)(\opp \cdot \clS_2)(\opL \cdot \clS_1)}{\opr^4}\,, \nn \\
    &\spinStr{(3,8)} = \frac{(\opp \cdot \clS_1)^2(\opL \cdot \clS_2)}{\opr^4}\,, \quad  
    &&\spinStr{(3,9)} = \frac{(\clS_1\cdot\clS_2) (\opL\cdot\clS_1)}{\opr^4} \,.
\end{alignat}
By demanding that the Fourier transform has the expected analytic behavior, we find
\begin{align}
    \frac{G}{|\opr|} \sum_{\alpha=4}^9 c^{(3,\alpha)}_1 \spinStr{(3,\alpha)} \quad \Rightarrow \quad
    -4\pi G \, c^{(3,4)}_1 \, \frac{(\clq \cdot \clS_1)^2 (\Lq \cdot \clS_2)}{15\clq^2} \,,
    \label{eq:exSpurious1}
\end{align}
together with the constraints,
\begin{align}
    &c_1^{(3, 5)} = -\frac{3}{5} c_1^{(3, 4)} + c_1^{(3, 6)}\,, \quad
    c_1^{(3, 7)} = -c_1^{(3, 8)} =\frac{c_1^{(3, 9)}}{\clp^2}\,, \quad
    c_1^{(3, 9)} = \frac{2}{5} c_1^{(3, 4)} - c_1^{(3, 6)} \,.
    \label{eq:exSpurious2}
\end{align}
The $1/\clp^2$ appears after the application of the Schouten identity,
\begin{align}
    (\clq \cdot \clS_1) ((\opp \times \clS_1) \cdot \clS_2) &= - (\clS_1 \cdot \clS_2) ((\opp \times \clq) \cdot \clS_1) + \clS_1^2 ((\opp \times \clq) \cdot \clS_2) \\
    &+ \frac{(\opp \cdot \clS_1) (\opp \cdot \clS_2) ((\opp \times \clq) \cdot \clS_1)}{\opp^2} - \frac{(\opp \cdot \clS_1)^2 ((\opp \times \clq) \cdot \clS_2)}{\opp^2} + \mathcal{O}\left( \opp \cdot \clq \right), \nn
\end{align}
which arises from the fact that any four vectors, in the present case $\clq$, $\opp$, $\clS_1$, and $\clS_2$, are linearly dependent in 3 dimensions. Note, also, that we made use of the on-shell condition $\opp \cdot \clq = \clq^2/2$.
We observe that for these six spin structures, we only need one function, and here we chose $c_1^{(3, 4)}$. Furthermore, we observe that $c_1^{(3, 6)}$ is a gauge redundancy, i.e., we are free to choose any value for this function without changing the amplitude. A natural choice is to simply set $c_1^{(3, 6)} = 0$; indeed this was the choice made in Ref.~\cite{Bern:2022kto}. This choice implies that the coefficients $c_1^{(3, 7)}$ and $c_1^{(3, 8)}$ have a spurious zero-velocity singularity. We refer to this singularity as spurious since it is an artifact of the gauge choice made, but it does not affect physical observables.  In the present paper we instead choose $c_1^{(3, 6)} = (2/5) c_1^{(3, 4)}$, which leads to $c_1^{(3, 7)} = c_1^{(3, 8)} = c_1^{(3, 9)} = 0$, since this removes the spurious singularity.

We observe that the interplay of Schouten identities and the on-shell conditions may lead to spurious zero-velocity singularities in the effective Hamiltonian. 
The above example demonstrates that some care is needed before concluding that the appearance of a zero-velocity singularity indicates a problem.
This subtlety led to the claim~\cite{Levi:2022rrq} that  even the tree-level Hamiltonian of Ref.~\cite{Bern:2022kto} is problematic and does not reproduce the earlier results of Ref.~\cite{Levi:2014gsa}. However, the tree-level Hamiltonian of Ref.~\cite{Bern:2022kto} and the tree-level Hamiltonians presented here give amplitudes that match those computed from the Hamiltonian of Ref.~\cite{Levi:2014gsa} where overlapping up to terms that do not contain the graviton pole. This suggests that these Hamiltonians define equivalent classical dynamics.
In contrast, as we discuss in \sect{sec:PhysicsalSingulartiesSubsection}, similar singularities can also arise if one attempts to match the dynamics from a QFT with an unconstrained spin tensor to a Hamiltonian that only contains the spin-vector degrees of freedom  (assuming the Wilson coefficients are not specially chosen to decouple the extra degrees of freedom).

\subsection{Spurious zero-velocity singularities and different bases for the Hamiltonian}

Although it amounts to a gauge redundancy, removing the spurious singularities from the Hamiltonian is often beneficial. This may be achieved by employing non-minimal bases for the Hamiltonian. In the present work, we obtain the effective Hamiltonian in two distinct gauges that yield equivalent dynamics. The first has a small number of spurious singularities and a minimal number of spin structures in momentum space. The second has a non-minimal number of spin structures and no spurious singularities.
We emphasize that the Hamiltonians obtained may be restricted, so that the $\clK$ dynamics decouple while preserving the minimal number of terms and the absence of spurious singularities.

\paragraph{Gauge 1 for the effective Hamiltonian.}
Many of the spurious singularities encountered during the prescription described above exist in the position-space Hamiltonian coefficients and not the amplitude. This can be seen in the example of Eqs.~(\ref{eq:exSpurious1}) and (\ref{eq:exSpurious2}), and it is the case for all $\mathcal{O}(G)$ terms we consider here. We may refer to these as spurious singularities introduced by the Fourier transform.
A simple way to remove them is to build a minimal Hamiltonian basis in terms of on-shell momentum-space spin structures instead of position-space spin structures, i.e., build directly the ones that appear in the amplitude. This is natural in the on-shell paradigm since this is the space where all on-shell constraints are manifest. The position-space spin structures are then given by the complete list of spin structures that appear in the inverse Fourier transform. 
The momentum-space on-shell spin structures are minimal; hence, we can obtain a unique Hamiltonian by matching to the QFT scattering amplitudes. At the same time, by relaxing the requirement of minimality on the position-space off-shell spin structures, we do not introduce spurious singularities during the Fourier transform.

\paragraph{Gauge 2 for the effective Hamiltonian.}
The gauge-1 Hamiltonian has spurious singularities only at $\mathcal{O}\left(G^2\right)$ and third order in $\clS$ and $\clK$. We define the gauge-2 Hamiltonian by including the following set of additional spin structures,
\begin{alignat}{2}
    &\spinStr{(3,111)} = \frac{(\opp \cdot \clS_1) \left( (\opr \times \clS_1) \cdot \clS_2 \right)}{\opr^4} \,, \quad 
    &&\spinStr{(3,112)} = \frac{(\opp \cdot \clS_2) \left( (\opr \times \clS_1) \cdot \clS_2 \right)}{\opr^4} \,, \nn \\
    &\spinStr{(3,113)} = \frac{(\opp \cdot \clK_1) \left( (\opr \times \clK_1) \cdot \clS_1 \right)}{\opr^4} \,, \quad 
    &&\spinStr{(3,114)} = \frac{(\opp \cdot \clK_1) \left( (\opr \times \clK_1) \cdot \clS_2 \right)}{\opr^4} \,, \nn \\
    &\spinStr{(3,115)} = \frac{(\opp \cdot \clK_2) \left( (\opr \times \clK_2) \cdot \clS_1 \right)}{\opr^4} \,, \quad 
    &&\spinStr{(3,116)} = \frac{(\opp \cdot \clK_2) \left( (\opr \times \clK_2) \cdot \clS_2 \right)}{\opr^4} \,,
\end{alignat}
with
\begin{align}
    c^{(3,\alpha)}_1\left(\opp^2\right) = 0\,, \quad 
    c^{(3,\alpha)}_2\left(\opp^2\right) = \Big[\text{constant independent of } \opp^2\Big] \,, \quad
    \alpha = 111, \ldots, 116 \,,
\end{align}
and demanding the absence of spurious singularities. In the notation of the previous subsection, we have,
\begin{align}
    \frac{G^2}{\opr^2} \, \frac{(\opp \cdot \clS_1) \left( (\opr \times \clS_1)\cdot \clS_2\right)}{\opr^4} \quad \Rightarrow \quad
    \frac{i \pi^2 G^2 |\clq|}{4} \, (\opp \cdot \clS_1) ((\clq \times \clS_1) \cdot \clS_2) \,,
\end{align}
and similarly for the other spin structures listed. Using 
\begin{align}
    ((\clq \times \clS_1) \cdot \clS_2) = 
    \frac{(\opp \cdot \clS_2)((\opp \times \clq) \cdot \clS_1)}{\opp^2} - 
    \frac{(\opp \cdot \clS_1)((\opp \times \clq) \cdot \clS_2)}{\opp^2} + 
    \mathcal{O}\left( \opp \cdot \clq \right),
\end{align}
and similar identities, we may remove the spurious singularities present in the gauge-1 Hamiltonian.

\subsection{Physical zero-velocity singularities}
\label{sec:PhysicsalSingulartiesSubsection}

In this subsection, we comment on the consequences of using a two-body Hamiltonian containing only $\clS$ to match the extended systems whose classical description should include both $\clS$ and $\clK$, again leading to a conclusion that the systems with and without $\clK$ are physically distinct.   Prior to the understanding developed in Refs.~\cite{Bern:2023ity, Alaverdian:2024spu}, demonstrating that there are extra degrees of freedom, such a construction was carried out~\cite{Bern:2022kto} by artificially setting $\clK$ to vanish in the initial and final states without taking into account that the extra degrees of freedom can propagate in intermediate stages.  One might wonder whether such a procedure leads to inconsistencies.
As discussed in detail in Ref.~\cite{Bern:2023ity}, the QFT framework for compact spinning objects, whose classical description includes both $\clS$ and $\clK$, encompasses states with different spin magnitudes and allows for transitions between them. 
A question that arises is whether it is possible to integrate out the states of different spin magnitudes to arrive at a description in terms of states of a single spin magnitude, which corresponds to a classical description in terms of $\clS$ alone. In our model, all states have the same mass irrespective of their spin magnitude. When there is no energy separation between states that are integrated out and states that are maintained in the effective description, this is often imprinted in the EFT by \textit{non-localities}.
It is straightforward to see that an EFT in terms of $\clS$ alone develops zero-velocity singularities, which we interpret as a manifestation of this effect.

We present an explicit example of this phenomenon in the case where only one particle is spinning. Since Hamiltonian coefficients that correspond to terms linear in $\clK$ vanish in general relativity, the first instance of this phenomenon is at 3\textsuperscript{rd} order in $\clS$. Given the complexity of terms at this order, we choose to instead discuss an example from electrodynamics, where we can demonstrate this phenomenon already at linear order in $\clS$ and $\clK$. 
We have,
\begin{align}
    \opH &= \ldots + V^{(0,1)}\left(\opp^2,\opr^2\right) + V^{(1,1)}\left(\opp^2,\opr^2\right) \, \frac{\opL \cdot \clS_1}{|\opr|^2} + V^{(1,3)}\left(\opp^2,\opr^2\right) \, \frac{\opr \cdot \clK_1}{|\opr|^2} + \ldots \Rightarrow \nn \\[5pt]
    &\ampEFT = \ldots +a^{(0,1)} \left(\opp^2,\clq^2\right) +a^{(1,1)} \left(\opp^2,\clq^2\right) \,  (\Lq \cdot \clS_1) + 
    a^{(1,3)}\left(\opp^2,\clq^2\right) (\clq \cdot \clK_1) +  \ldots \,,
\end{align}
with
\begin{align}
    a^{(n,\alpha)}\left(\opp^2,\clq^2\right) =  & 
    \frac{4\pi G}{\clq^2} a^{(n,\alpha)}_1\left(\opp^2\right) + 
    \frac{2\pi^2 G^2}{|\clq|} a^{(n,\alpha)}_2\left(\opp^2\right)  \\
    &+(4\pi G)^2\, a^{(n,\alpha)}_{\rm iter}\left(\opp^2\right)\int \frac{d^{D-1}\bm \ell}{(2\pi)^{D-1}}
\frac{2\xi \en{}}{\bm \ell^2 (\bm \ell+\clq)^2 (\bm \ell^2+2\clp\cdot \bm \ell)} + \mathcal{O}\left(G^3\right),\nn
\end{align}
mirroring Eqs.~(\ref{eq:MEFT_PM}) and~(\ref{eq:MEFT_2PM}).
As already discussed in Ref.~\cite{Bern:2022kto}, 
we have for the tree-level coefficient functions
\begin{align}
 a^{(0,1)}_1 = -c^{(0,1)}_1 \,, \quad a^{(1,1)}_1 = c^{(1,1)}_1 \,, \quad a^{(1,3)}_1 = -c^{(1,3)}_1 \,,
\label{eq:as_1PM}
\end{align}
and for the one-loop coefficient functions
\begin{align}
\label{eq:as_2PM}
a^{(0,1)}_2 &= -c^{(0,1)}_2 + 2 \en{} \xi  c^{(0,1)}_1 \pder c^{(0,1)}_1+\frac{(1-3 \xi ) \left(c^{(0,1)}_1\right)^2}{2 \en{} \xi }\,, \\
a^{(1,1)}_2 &= \frac{c^{(1,1)}_2}{2} - \en{} \xi  c^{(1,1)}_1 \pder c^{(0,1)}_1 - \en{} \xi  c^{(0,1)}_1 \pder c^{(1,1)}_1 + \frac{(3 \xi -1) c^{(0,1)}_1 c^{(1,1)}_1}{2 \en{} \xi } \nn \\ & + \frac{\en{} \xi \left(\left(c^{(1,3)}_1\right)^2-2 c^{(0,1)}_1 c^{(1,1)}_1\right)}{2 \clp^2} \,, \nn \\
a^{(1,3)}_2 &= -\frac{c^{(1,3)}_2}{2} -\frac{1}{2} \en{} \xi  c^{(1,3)}_1 \left(c^{(1,1)}_1-2 \pder c^{(0,1)}_1 \right) + \en{} \xi c^{(0,1)}_1 \pder c^{(1,3)}_1 + \frac{(1-3 \xi ) c^{(0,1)}_1 c^{(1,3)}_1}{2 \en{} \xi } \, , \nn
\end{align}
where we used the shorthands $c^{(n,\alpha)}_i \equiv c^{(n,\alpha)}_i\left( \clp^2 \right)$ and $\pder \equiv \frac{d}{d \clp^2}$. 
We observe that $a^{(1,1)}_2$ develops zero-velocity singularities. These cancel in the matching against identical singularities in the QFT amplitude, leading to singularity-free Hamiltonian coefficients $c^{(n,\alpha)}_i$.

Attempting to match, instead, to an effective Hamiltonian without $\clK$ would lead to Hamiltonian coefficients with zero-velocity singularities. Indeed, considering external states of fixed spin magnitude but allowing different-spin-magnitude states to propagate only in the intermediate stages would lead to $a^{(1,3)} = 0$,
while $a^{(0,1)}$ and $a^{(1,1)}$ would remain unchanged. Since these conditions would give $c^{(1,3)}_1 = 0$, the amplitude coefficient $a_2^{(1,1)}$ would not develop the appropriate zero-velocity singularity to cancel the corresponding one from the QFT amplitude. This would result in a zero-velocity singularity for the Hamiltonian coefficient $c_2^{(1,1)}$. We interpret this zero-velocity singularity as physical and signaling the off-shell propagation of a mode that should also be included as an on-shell state in the effective description. 
The fact that there exist no Schouten identities in this case, since there are only 3 available vectors $\clp$, $\clq$ and $\clS$, indicates that this zero-velocity singularity may not be removed.

If the zero-velocity singularity in the Hamiltonian points to a pathology, it should manifest in the observables. Indeed, already at linear order in $\clS$ and $\clK$, we find that it is impossible to match the impulse from a Hamiltonian with $c^{(1,3)}_1 \neq 0$ to that from a Hamiltonian with $c^{(1,3)}_1 = 0$. This mirrors the discussion of \cref{sec:waveform}, reinforcing the conclusion that the dynamics of a system described by $\clS$ and $\clK$ are distinct from those of a system with $\clS$ alone.


\section{Conclusions and Outlook}
\label{sec:conclusion}


In this paper we provided further details showing that extended systems where no SSC is imposed (\gco) are physically distinct from those where one is imposed (\cco). 
The body-fixed tetrad has six degrees of freedom. 
In the minimal effective-field-theory description of the body, three of them are removed by imposing the SSC. An extended effective field theory, in which no SSC is imposed, promotes them to the dynamical variables described by the $K$ vector. 
The couplings of these new dynamical variables are governed by Wilson coefficients {\em beyond} those that describe conventional systems.
In Ref.~\cite{Alaverdian:2024spu} we showed that, through 2\textsuperscript{nd} order in $S$ and $K$, these degrees of freedom affect both inclusive and point-like gravitational observables and, under certain assumptions, they cannot be compensated by adjusting the spin vector and the Wilson coefficients describing conventional compact objects. 
Here we relaxed these assumptions, allowing for nonlinear redefinitions of the $S$ and $K$ vectors, and demonstrated that the same conclusion continues to hold through 3\textsuperscript{rd} order in $S$ and $K$.

We continued the exploration of the physics of these degrees of freedom by analyzing in detail the Compton amplitudes from both the QFT and worldline perspective, as well as the QFT two-body amplitudes. We showed that the features observed through ${\cal O}(\gS^2)$ persist through 4\textsuperscript{th} order in the spin tensor.
In particular, we showed that for special values of the additional Wilson coefficients, that is of the Wilson coefficients beyond those describing a \cco, the $K$ degrees of freedom decouple even in the absence of an SSC.
Thus, even though an SSC is not imposed, our formalism allows us to recover it {\em a posteriori}. It therefore provides a path to carrying out higher-order spin-dependent calculations in a clean, unconstrained setting, relating the spin tensor and the expectation value of Lorentz generators in spin coherent states. It will be interesting to apply it to spin-dependent calculations at $\order(G^3)$ and above.

In Ref.~\cite{Alaverdian:2024spu} and in this paper we treat the interactions of spinning particles in a formal expansion in the $S$ and $K$ vectors. It was however pointed out in Ref.~\cite{Brandhuber:2023hhl} that the leading-order $K=0$ waveform with exact spin dependence exhibits new features. It would be interesting to carry out a similar analysis for $K\ne 0$. The dependence on $K$ originating from the polarization tensors in \cref{epdotep}
can naturally be absorbed into a coordinate transformation.
An analysis involving the contribution of the Lagrangian interaction vertices requires understanding the choices of Wilson coefficients 
for which an all-order pattern involving the $K$ vector can be identified.

We constructed a worldline theory describing the same degrees of freedom as the four-dimensional field theory. As in the latter, the $K$ vector is introduced by relaxing the SSC; there is a one-to-one correspondence between the interaction terms and Wilson coefficients of this worldline theory and those of the field theory. For the same special values of additional Wilson coefficients the time evolution of the $K$ vector decouples from that of the spin and coordinates, leading effectively to the same observables as in the worldline theory with SSC imposed. As in the field-theory approach, this offers a path to carrying out calculations in an unconstrained framework while obtaining results for spins obeying an SSC. 
We also discussed the parallels between our worldline theory and worldline theories with dynamical multipole moments proposed previously~\cite{Porto:2007qi, Goldberger:2020fot, Steinhoff:2021dsn}, the contribution of $K$ to these multipoles, and its interpretation in terms of additional massless degrees of freedom.

While the $K$ vector was initially identified~\cite{Bern:2023ity} in the intricacies of the representations and the coherent states of the Lorentz group, we demonstrated here that some physical systems, one rather common and one exotic, can exhibit it. For the case of a Newtonian bound state, these degrees of freedom are described by the Laplace-Runge-Lenz vector, which describes the shape and orientation of the orbit. 
This example, which we discussed in some detail and matched it with our field theory, captures the general property of the $K$ vector to allow for a conservative change in the magnitude of the spin and thus, from a field-theory perspective, to describe massless degrees of freedom. 
In \cref{eq:AvsK}, {\em mapping} the field-theory $K$ to the Laplace-Runge-Lenz vector $\bm A$ of the Newtonian bound state as $\bm K \mapsto (\text{constant})\times i \bm A$ is 
effectively Wick rotating the boost operator (whose expectation value is $i\bm K$~\cite{Bern:2023ity}). This mapping, hence, transforms the field-theory $\text{SO}(1,3)$ symmetry to the well-known $\text{SO}(4)$ symmetry of the Newtonian bound state. 

We also studied the multipoles of the Rasheed-Larsen black hole~\cite{Rasheed:1995zv,Larsen:1999pp} by probing it with a scalar probe particle. Ref.~\cite{Bena:2020uup} obtained the multipole moments of the Rasheed-Larsen black hole and showed that they are qualitatively different from those of the Kerr black hole. Employing the formalism of Ref.~\cite{Kosmopoulos:2023bwc}, we computed the amplitude for the scattering of an uncharged probe particle off the Rasheed-Larsen black hole to leading order in $G$. 
We matched this amplitude to that of our field theory, establishing that, for certain values of the Wilson coefficients, our field theory for \gcos describes the Rasheed-Larsen black hole to the order considered. We concluded that, through that order, the limit to the Kerr black hole is controlled by the Wilson coefficients alone. This echoes our general discussion that the dynamics of a \cco can always be recovered from that of \gcos by appropriate choices of Wilson coefficients. 
The example of the Rasheed-Larsen black hole suggests that, even if the dynamics of the $K$ vector do not impact Kerr black hole mergers, they are potentially crucial for accurately interpreting signals from mergers of exotic black holes and other astrophysical objects.

While the above examples shed light on the physical meaning of the $K$ vector, it would be interesting to also develop a more fundamental understanding.
A standard approach to massless degrees of freedom, 
at least from a worldline perspective, is in terms of Goldstone particles associated to some broken symmetry. 
The fundamental theory describing a relativistic compact body should respect the space-time Poincar\'e symmetry $\text{ISO}(1,3)$ and a certain internal symmetry $\mathscr{S}$, so that the total symmetry group is $\mathscr{G}=\text{ISO}(1,3)\times \mathscr{S}$; for rigid bodies, $\mathscr{S}$ is a subgroup of $\text{SO}(3)$.
The relevant degrees of freedom in a conventional worldline description can be identified as the Goldstone bosons resulting from the spontaneous breaking of $\mathscr{G}$ to time translation $P^0$ and a single copy of the rotation group or a subgroup thereof.
The worldline and angular degrees of freedom can be realized as Goldstone bosons of the broken translation and rotation generators, respectively~\cite{Delacretaz:2014oxa}. 

To fit the $K$ vector into this picture it is necessary to have a larger internal symmetry. Algebraically, $K$ behaves like a boost generator, such that in principle there are two ways to embed $K$ into $\mathscr{G}$:
\begin{enumerate*}[label=(\arabic*)]
    \item $K^{i}\sim L^{0i}$ is the space-time boost generator in the Poincar\'e group;
    \item $K^{i}$ is part of the internal symmetry group $\mathscr{S}$, for example, $\mathscr{S}=\text{SO}(4)$ while $\frac{1}{\sqrt{2}}(S^{i}\pm K^{i})$ generate the two $\text{SU}(2)$ subgroups, respectively.
\end{enumerate*}
For the first case, the Goldstone bosons associated with the broken space-time boost generators are removed by the inverse Higgs mechanism~\cite{Ivanov:1975zq}. This is because the commutator between $L^{0i}$ and the unbroken time translation leads to broken space translations, $[L^{0i},P^0]\sim P^{i}$, and consequently a Goldstone boson of $L^{0i}$ does not generate independent perturbations to the vacuum~\cite{Low:2001bw}. 
Therefore, the Goldstone bosons of $L^{0i}$ represent a redundancy of the system, and the inverse Higgs constraint works as a gauge fixing. 
One may still solve the system of equations of motion without completely eliminating these Goldstone bosons by explicitly solving the inverse Higgs constraint. 
This is common in gauge theories, where commonly used gauge fixings, like the Feynman gauge, do not remove all the redundancies.
In such cases, one may expect that the remaining Goldstone degrees of freedom automatically decouple from physical observables.
In contrast, if $K^{i}$ belongs to the internal group $\mathscr{S}$, then $[K^{i},P^{0}]=0$ and the corresponding Goldstone bosons are independent.
The Newtonian bound state discussed in \sect{sec:toymodel} exhibits (a slightly constrained version of) this structure, with the role of the $K$ vector being played by the Laplace-Runge-Lenz vector. We leave for the future the interesting task of understanding if more involved physical systems, perhaps described by continuous mass distributions,  exhibit a similar extended internal symmetry group.

Another possible fundamental interpretation of the $K$ vector builds on its close relation to the boost generator while avoiding the need for an extended internal symmetry group. Indeed, it was shown in Ref.~\cite{Alberte:2020eil} and reemphasized in Ref.~\cite{Komargodski:2021zzy} that,\footnote{We thank Ian Low for pointing out these references to us.} while spontaneous breaking of boost symmetry does not lead to a fundamental Goldstone field, it leads however to massless (multi-particle) states. These states saturate a certain sum rule stemming from the relation 
\begin{align}
\langle\Omega |\delta_{K^i} T^{0 j}|\Omega\rangle = (e+P)\delta^{ij} \, ,
\end{align}
where $\delta_{K^i}$ is a boost generated by $K^i$, $T^{\mu\nu}$ is the stress tensor, $e$ and $P$ are the energy density and the pressure, respectively, and $\Omega$ is a state that breaks boost invariance. See  Ref.~\cite{Komargodski:2021zzy} for details. It was further pointed out there that the sound mode in hydrodynamics, which is a collective excitation, yields a spectral density~\cite{Komargodski:2021zzy}
\begin{align}
\rho_{T^{00}, T^{0j}}(\omega, k) = (e+P) k^j \delta(\omega) \,
\end{align}
for the time-ordered correlator $\langle {\rm T}_t\, T^{00}(t, x) T^{0j}(0)\rangle$ which has exactly this property. 
The detailed analysis of Ref.~\cite{Delacretaz:2020nit} shows that, although there is no unique set of states that directly generates this spectral density, all states close in energy to $|\Omega\rangle$ contribute to it. 
It would be interesting to understand if the dynamics of our $K$ vector may be interpreted as an effective description of the dynamics of such gapless states, especially in the context of a hydrodynamic approach to the mass distribution of compact objects.

Starting from the two-body amplitudes, we constructed a two-body Hamiltonian describing the dynamics of the extended field and worldline theories. 
The expectation values $\clS$ and $\clK$ of the Hamiltonian operators obeying a Lorentz algebra are {\em identified} with the field-theory quantities $\bm S$ and $i\bm K$, which are themselves expectation values of the generators of the field-theory Lorentz algebra. As we discussed, the imaginary unit combined with the reality of $\clK$ and ${\bm K}$ reflect the nonunitarity of the finite-dimensional representations of the Lorentz group, which are used in the field-theory construction.

In building the Hamiltonian, we observed that writing a minimal ansatz may lead to an answer that involves spurious zero-velocity singularities, which may be readily removed by including certain redundant terms. Given this observation, we obtained the Hamiltonian in two different gauge choices, which yield identical dynamics. In the first, the number of momentum-space terms is minimal but the Hamiltonian coefficients exhibits spurious zero-velocity singularities. In the second, we include a set of redundant terms and fix them to remove all spurious zero-velocity singularities.

For the case of electrodynamics, where $K$ appears at a lower order, we explored the possibility of matching an extended system to a Hamiltonian without the dynamical $\clK$ vector. We found that in this case the Hamiltonian develops physical zero-velocity singularities. These singularities suggest possible discrepancies in observables that may not be removed in the above fashion. This serves as further evidence of the necessity to include a dynamical $\clK$ vector to capture the evolution of extended systems.
With the understanding developed in this paper and Refs.~\cite{Bern:2023ity,Alaverdian:2024spu}, we conclude that the Hamiltonian of Ref.~\cite{Bern:2022kto}, while valid for the special values of Wilson coefficients for which $\clK$ decouples and the system is described by $\clS$ alone, cannot capture the rich dynamics of a \gco. Since the future gravitational-wave experiments are expected to reach accuracy of the 7\textsuperscript{th} order in the physical post-Minkowskian counting scheme (see, e.g., Ref.~\cite{Buonanno:2022pgc}), it is very interesting to obtain the complete Hamiltonian valid at $\mathcal{O}\left(G^2 \, \clS^a \clK^b\right)$, with $a+b=5$.

The two-body Hamiltonian allowed us to verify that the $K$-dependent spinning eikonal formula, connecting the eikonal phase of the two-body amplitude and conservative observables, yields the correct impulse, $S$ and $K$ change through $\order(G^2\, \gS^3)$.
It will be interesting to explore the covariant organization of this formula along the lines of  \cite{Luna:2023uwd}, see also \cite{Gatica:2024mur} for recent results in this direction.
It will also be interesting to carry out tests of the direct relation between amplitudes and observables at higher orders in Newton's constant. 

The approach to gravitational interactions of compact spinning bodies described here, which follows from our previous work, identified possible new light degrees of freedom and their effective dynamics --- and hints towards a richer phenomenology for these bodies than previously thought.
Regarding conventional compact objects that satisfy an SSC, our approach allows us to extract their dynamics through unconstrained calculations.
Our formalism should help produce observables useful for the analysis of the dynamics of conventional astrophysical spinning bodies, meeting the needs of increasingly more accurate gravitational-wave experiments. 
We believe that it may also assist in the identification of exotic astrophysical bodies, characterized by the additional degrees of freedom we analyzed in this paper.

\begin{acknowledgments}
We thank Juan Pablo Gatica, Donal O'Connell, Rafael Porto, and Mikhail Solon for useful discussions.  
We especially thank Justin Vines for many important and valuable discussions and for collaborating on earlier work~\cite{Bern:2023ity}, which is the basis for much of this paper.
Z.~B. and T.~S.  are supported by the U.S. Department of Energy
(DOE) under award number DE-SC0009937. 
D.~K. is supported by the Swiss National Science Foundation under grant no. 200021-205016.
A.~L. is supported by funds from the European Union’s Horizon 2020 research and innovation program under the Marie Sklodowska-Curie grant agreement No.~847523 ‘INTERACTIONS’.  
R.~R. and F.~T.~are supported by
the U.S.  Department of Energy (DOE) under award number~DE-SC00019066.
F.~T.~ is also supported by the startup grant from the department of physics at Fudan University.
We also thank the Mani L. Bhaumik Institute for Theoretical Physics for support.  
This paper was completed while Z.~B., R.~R. and F.~T. participated in the ``What is Particle Theory?'' program at the Kavli Institute for Theoretical Physics (KITP) and supported in part by grant NSF PHY-2309135.
\end{acknowledgments}

\newpage

\appendix

\section{Conventions}
\label{sec:conventions_appendix}

In this appendix, we present our conventions, correcting an inconsistency in the conventions used in Refs.~\cite{Bern:2023ity, Alaverdian:2024spu}, amounting to a sign flip in the definition of $K$ for the covariant amplitudes that are expressed in terms of the vectorial variables $S$ and $K$: $\mathcal{M}(S,K) \rightarrow \mathcal{M}(S,-K)$. 
In our field-theory construction, the massive higher-spin particles are described by symmetric traceless tensor fields,
\begin{align}
    \phis \equiv \phi_{a_1a_2\ldots a_s} = \phi_{(a_1a_2\ldots a_s)}\,,\qquad \eta^{a_1a_2} \phi_{a_1a_2\ldots a_s} = 0\,,
\end{align}
where $s$ is a large integer, and the field is written in the local Lorentz frame spanned by the tetrad $e^a_{\mu}$,
\begin{align}
    g_{\mu\nu} = e^{a}_{\mu} e^{b}_{\nu} \eta_{ab}\,,\quad \eta_{ab} = \text{diag}(1,-1,-1,-1)\,.
\end{align}
We adopt the standard definition for the local Lorentz covariant derivative
\begin{align}
    & \nabla_{\mu}\phi^{a_1a_2\ldots a_s} = \partial_{\mu} \phi^{a_1a_2\ldots a_s} + \sum_{i=1}^{s} \omega_{\mu}{}^{a_i}{}_{b} \phi^{a_1 \ldots a_i \ldots a_s} \,,\nonumber\\
    & \nabla_{\mu}\phi_{a_1a_2\ldots a_s} = \partial_{\mu} \phi_{a_1a_2\ldots a_s} + \sum_{i=1}^{s} \omega_{\mu}{}^{b}{}_{a_i} \phi_{a_1 \ldots a_i \ldots a_s} \,.
\end{align}
It is straightforward to show that the above equations can be rewritten as
\begin{align}
    & \nabla_{\mu}\phi^{a_1a_2\ldots a_s} = \partial_{\mu} \phi^{a_1a_2\ldots a_s} + \frac{i}{2} \omega_{\mu a b} (M^{ab})^{a_1a_2\ldots a_s}{}_{b_1b_2\ldots b_s} \phi^{b_1b_2\ldots b_s} \,,\nonumber\\
    & \nabla_{\mu}\phi_{a_1a_2\ldots a_s} = \partial_{\mu} \phi_{a_1a_2\ldots a_s} + \frac{i}{2} \omega_{\mu a b} (M^{ab})^{b_1b_2\ldots b_s}{}_{a_1a_2\ldots a_s} \phi^{b_1b_2\ldots b_s} \,,
\end{align}
where the Ricci rotation coefficient $\omega_{\mu a b}$ is defined as 
\begin{align}
    \omega_{\mu a b} = e^{\nu}_{b} D_{\mu} e_{\nu a} \,,
\end{align}
and $M^{ab}$ is given by
\begin{align}
    (M^{ab})_{c_1c_2\ldots c_s}{}^{d_1d_2\ldots d_s} = 2 i s \delta_{(c_1}^{[a}\eta^{b](d_1}_{\vphantom{(c_1}}\delta^{d_2\vphantom{]}}_{c_2\vphantom{(}} \ldots \delta^{d_s)}_{c_s)} \,.
\end{align}
They satisfy the usual Lorentz algebra with the mostly minus signature,
\begin{align}\label{eq:Lorentz_Alg_M}
    [M^{a_1a_2}, M^{a_3a_4}] = i (\eta^{a_3a_1}M^{a_4a_2} +\eta^{a_2 a_3} M^{a_1 a_4} - \eta^{a_4 a_1} M^{a_3 a_2} - \eta^{a_2 a_4} M^{a_1 a_3}) \,.
\end{align}
Our formalism will not explicit depend on the integer $s$, and this rewriting allows us to suppress the Lorentz indices and simply use $\phis$ throughout this work. Note that the traceless condition can be trivialized by using the spinor representation.

The Poisson brackets of the classical spin tensor $\gS^{ab}$ are also given by the Lorentz algebra
\begin{align}\label{eq:Lorentz_Alg_S}
    \{\gS^{a_1a_2}, \gS^{a_3a_4}\} = \eta^{a_3a_1}\gS^{a_4a_2} +\eta^{a_2 a_3} \gS^{a_1 a_4} - \eta^{a_4 a_1} \gS^{a_3 a_2} - \eta^{a_2 a_4} \gS^{a_1 a_3} \,,
\end{align}
in parallel with \cref{eq:Lorentz_Alg_M}. One can show that the spin vector and boost vector defined through the decomposition~\eqref{eq:gStensor}, 
\begin{align}
\gS(p)^{\mu\nu} &= - \frac{1}{m}\varepsilon^{\mu\nu\rho\sigma}p_\rho \qftS_\sigma + \frac{i}{m}(p^{\mu} \qftK^{\nu}-p^{\nu} \qftK^{\mu}) \nonumber\\
&= - \frac{1}{m}\varepsilon^{\mu\nu\rho\sigma}p_\rho \qftS_\sigma + \frac{1}{m}(p^{\mu} \covclK^{\nu}-p^{\nu} \covclK^{\mu}) \,,
\label{eq:SpinDecAppendix}
\end{align}
with $\varepsilon^{0123} = 1$ satisfy the usual Lorentz algebra
\begin{align}
    & \{\qftS^{i},\qftS^{j}\} = \varepsilon^{ijk} \qftS^{k}\,,\qquad \{\qftS^{i},\covclK^{j}\} = \varepsilon^{ijk} \covclK^{k} \,,\qquad \{\covclK^{i},\covclK^{j}\} = - \epsilon^{ijk} \qftS^{k} \,.
\end{align}
In the above, we assumed that the energy component of $p$ is positive. When computing amplitudes in the all-outgoing convention, we take  \cref{eq:SpinDecAppendix}, the mass $m$ and the vectors $S^\mu$ and $K^\mu$ as unaffected by the crossing of the momentum to its physical, positive-energy value.

When specifying kinematics in the center-of-mass frame, we use
\begin{align}
& p_1^{\mu}=-(\en{1}, \,\clp_1) \,,
\hspace{5pt}
p_2^{\mu}=-(\en{2}, \,\clp_2) \,,
\hspace{5pt}
p_3^{\mu}=(\en{3}, \,\clp_3) \,,
\hspace{5pt}
p_4^{\mu}=(\en{4}, \,\clp_4) \,,
\hspace{5pt}
q = (0, \clq)\,,
\end{align}
with $q = p_2+p_3$. Hence, we have 
\begin{align}
\clp_1 = -\clp_2 = \clp \,,
\quad
\clp_4 = -\clp_3 = \clp - \clq \,, 
\quad
\en{1} = \en{4}\,,
\quad
\en{2} = \en{3}\,.
\label{eq:qConvention}
\end{align}
This choice for the vector $\clq$ is reflected in our conventions for the Fourier transform. In particular, in \cref{sec: metricWK}, $\delta \tilde{g}^{\mu\nu}(\bm q)$ is defined as 
\begin{align}
    \delta \tilde{g}^{\mu\nu}(\bm q) = \int d^{d-1} \bm r \, e^{-i \bm q \cdot \bm r} \delta g^{\mu\nu}(\bm r) \,,
\label{eq:FTMetric}
\end{align}
since the metric is evaluated at $\bm x_2$ in the corresponding interaction term in the Lagrangian and $\bm p_3 = \bm p_2 + \bm q$ in our amplitudes. Similarly, in \cref{sec:Hamiltonian} we use
\begin{align}
    \tilde{f}(\bm q) = \int d^{d-1} \bm r \, e^{i \bm q \cdot \bm r}  f(\bm r) \,.
\label{eq:FTHamiltonian}
\end{align}
The difference in the sign in the exponent follows from the fact that, here, $\bm r = \bm x_1-\bm x_2$. 
To perform the integrals we employ dimensional regularization with $d=4-2\epsilon$.

\section{\texorpdfstring{$R^2$}{R2} interactions involving \texorpdfstring{$K$}{K} vectors}\label{sec:LcontactK}

In this Appendix, we collect the contact terms needed so that, for the values of the Wilson coefficients given in Eq.~\eqref{eq:NS_cont_cond}, the $K$-dependent Compton amplitude reduces to the Compton amplitude for a \cco. 
For these values, the SSC is effectively imposed, even though no constraints
are imposed at the Lagrangian level.

Contact interactions for the $\order(S^3 K)$ sector:
\begin{align}\label{eq:LS3K_cont}
\mathcal{L}^{(4)}_{S^3K} &= \frac{2iD_{4a}}{3m^7}R_{af_1bf_2}\widetilde{R}_{cf_3df_4}\nabla^a\nabla^c\phis \PLS^{f_3}\PLS^{f_4}\PLS^{f_1}M^{f_2e}\nabla_e\nabla^{b}\nabla^{d}\phis \\
&\quad - \frac{5 i D_{4b}}{3m^7}R_{af_1bf_2}\widetilde{R}_{cf_3df_4}\nabla^a\nabla^c\phis \PLS^{f_1}\PLS^{f_2}\PLS^{f_3}M^{f_4e}\nabla_e\nabla^{b}\nabla^{d}\phis \nonumber\\
&\quad - \frac{i D_{4c}}{3m^7}R_{af_1bf_2}\widetilde{R}^{f_2}{}_{cdf_4}\nabla^a\nabla^c\phis \PLS_{e}\PLS^{e}\PLS^{f_1}M^{f_4g}\nabla_g\nabla^b\nabla^d\phis \nonumber\\
&\quad + \frac{4iD_{4d}}{3m^7}R_{af_1bf_2}\widetilde{R}^{f_2}{}_{cdf_4}\nabla^a\nabla^c\phis\PLS^{f_1}\PLS^{f_4}\PLS_{e}M^{eg}\nabla_g\nabla^b\nabla^d\phis \nonumber\\
&\quad - \frac{iD_{4e}}{12m^7}R_{af_1bf_2}\widetilde{R}^{f_2}{}_{cdf_4}\nabla^a\nabla^c\phis \PLS_{e}\PLS^{e}\PLS^{f_4}M^{f_1g}\nabla_g\nabla^b\nabla^d\phis \nonumber\\
&\quad - \frac{iD_{4f}}{6m^7}R_{af_1bf_2}\widetilde{R}^{f_2}{}_{c}{}^{f_1}{}_{d}\nabla^a\nabla^c\phis\PLS_{e}\PLS^{e}\PLS_{g}M^{gh}\nabla_h\nabla^b\nabla^d\phis \, .
\nonumber
\end{align}

Contact interactions for the $\order(S^2 K^2)$ sector:
\begin{align}\label{eq:LS2K2_cont}
\mathcal{L}^{(4)}_{S^2K^2} &= \left[\frac{2E_{4a}+3E_{4b}}{4m^8}R_{af_1bf_2}R_{cf_3df_4}+\frac{2E_{4a}-3E_{4b}}{4m^8}\widetilde{R}_{af_1bf_2}\widetilde{R}_{cf_3df_4}\right] \\
&\hspace{6cm} \times\nabla^{a}\nabla^{c}\phis\PLS^{f_1}\PLS^{f_2}M^{f_3g}M^{f_4h}\nabla_{g}\nabla_{h}\nabla^{b}\nabla^{d}\phis \nonumber\\
&\quad + \left[\frac{2E_{4c}-E_{4d}}{4m^8}R_{af_1bf_2}R_{cf_3df_4}+\frac{2E_{4c}+E_{4d}}{4m^8}\widetilde{R}_{af_1bf_2}\widetilde{R}_{cf_3df_4}\right] \nonumber\\
&\hspace{6cm} \times\nabla^{a}\nabla^{c}\phis\PLS^{f_1}\PLS^{f_3}M^{f_2g}M^{f_4h}\nabla_{g}\nabla_{h}\nabla^{b}\nabla^{d}\phis \nonumber\\
&\quad + \left[\frac{2E_{4e}-E_{4f}}{12m^8}R_{af_1bf_2}R^{f_2}{}_{cdf_4}+\frac{2E_{4e}+E_{4f}}{12m^8}\widetilde{R}_{af_1bf_2}\widetilde{R}^{f_2}{}_{cdf_4}\right] \nonumber\\
&\hspace{6cm} \times\nabla^{a}\nabla^{c}\phis\PLS^{f_5}\PLS_{f_5}M^{f_1g}M^{f_4h}\nabla_{g}\nabla_{h}\nabla^{b}\nabla^{d}\phis \nonumber\\
&\quad + \left[\frac{4E_{4g}+E_{4h}}{12m^8}R_{af_1bf_2}R^{f_2}{}_{cdf_4}+\frac{4E_{4g}-E_{4h}}{12m^8}\widetilde{R}_{af_1bf_2}\widetilde{R}^{f_2}{}_{cdf_4}\right] \nonumber\\
&\hspace{6cm} \times\nabla^{a}\nabla^{c}\PLS^{f_1}\PLS^{f_4}M^{f_5g}M_{f_5h}\nabla_g\nabla^{h}\nabla^{b}\nabla^{d}\phis \nonumber\\
&\quad + \left[\frac{2E_{4i}-E_{4j}}{24m^8}R_{af_1bf_2}R^{f_2}{}_{c}{}^{f_1}{}_{d}+\frac{2E_{4i}+E_{4j}}{24m^8}\widetilde{R}_{af_1bf_2}\widetilde{R}^{f_2}{}_{c}{}^{f_1}{}_{d}\right] \nonumber\\
&\hspace{6cm} \times\nabla^{a}\nabla^{c}\phis\PLS^{f_3}\PLS_{f_3}M^{f_4g}M_{f_4h}\nabla_{g}\nabla^{h}\nabla^{b}\nabla^{d}\phis\nonumber\\
&\quad + \left[\frac{2E_{4k}-E_{4l}}{24m^8}R_{af_1bf_2}R^{f_2}{}_{c}{}^{f_1}{}_{d}+\frac{2E_{4k}+E_{4l}}{24m^8}\widetilde{R}_{af_1bf_2}\widetilde{R}^{f_2}{}_{c}{}^{f_1}{}_{d}\right] \nonumber\\
&\hspace{6cm} \times\nabla^{a}\nabla^{c}\phis\PLS^{f_3}\PLS^{f_4}M_{f_3g}M_{hf_4}\nabla^{g}\nabla^{h}\nabla^{b}\nabla^{d}\phis \, .
\nonumber
\end{align}

Contact interactions for the $\order(S K^3)$ sector:
\begin{align}\label{eq:LSK3_cont}
\mathcal{L}^{(4)}_{SK^3} &= -\frac{iF_{4a}}{3m^9}R_{af_1bf_2}\widetilde{R}_{cf_3df_4}\nabla^{a}\nabla^{c}\phis\PLS^{f_1}M^{f_2e_1}M^{f_3e_2}M^{f_4e_3}\nabla_{e_1}\nabla_{e_2}\nabla_{e_3}\nabla^{b}\nabla^{d}\phis \\
&\quad -\frac{5iF_{4b}}{3m^9}R_{af_1bf_2}\widetilde{R}_{cf_3df_4}\nabla^{a}\nabla^{c}\phis\PLS^{f_4}M^{f_2e_1}M^{f_3e_2}M^{f_1e_3}\nabla_{e_1}\nabla_{e_2}\nabla_{e_3}\nabla^{b}\nabla^{d}\phis \nonumber\\
&\quad + \frac{iF_{4c}}{3m^9}R_{af_1bf_2}\widetilde{R}^{f_2}{}_{cdf_4} \nabla^{a}\nabla^{c}\phis\PLS^{f_1}M^{e_1g}M_{g}{}^{e_2}M^{f_4e_3}\nabla_{e_1}\nabla_{e_2}\nabla_{e_3}\nabla^{b}\nabla^{d}\phis \nonumber\\
&\quad + \frac{iF_{4d}}{12m^9}R_{af_1bf_2}\widetilde{R}^{f_2}{}_{cdf_4} \nabla^{a}\nabla^{c}\phis \PLS^{f_4}M^{e_1g}M_{g}{}^{e_2}M^{f_1e_3}\nabla_{e_1}\nabla_{e_2}\nabla_{e_3}\nabla^{b}\nabla^{d}\phis \nonumber\\
&\quad + \frac{iF_{4e}}{6m^9}R_{af_1bf_2}\widetilde{R}^{f_2}{}_{cdf_4} \nabla^{a}\nabla^{c}\phis\PLS^{g}M^{f_1e_1}M^{f_4e_2}M_{g}{}^{e_3}\nabla_{e_1}\nabla_{e_2}\nabla_{e_3}\nabla^{b}\nabla^{d}\phis \nonumber\\
&\quad + \frac{5iF_{4f}}{3m^9}R_{af_1bf_2}\widetilde{R}^{f_2}{}_{c}{}^{f_1}{}_{d}\nabla^{a}\nabla^{c}\phis\PLS^{h}M^{e_1g}M_{ge_2}M_{he_3}\nabla_{e_1}\nabla^{e_2}\nabla^{e_3}\nabla^{b}\nabla^{d}\phis \, .
\nonumber
\end{align}

Contact interaction for the $\order(K^4)$ sector:
\begin{align}\label{eq:LK4_cont}
\mathcal{L}^{(4)}_{K^4} &= \frac{G_{4a}}{48m^{10}}R_{af_1bf_2}R_{cf_3df_4}\nabla^a\nabla^c\phis M^{f_1i}M^{f_2j}M^{f_3k}M^{f_4l} \nabla_i\nabla_j\nabla_k\nabla_l\nabla^b\nabla^d\phis \\
&\quad + \frac{G_{4b}}{48m^{10}}\widetilde{R}_{af_1bf_2}\widetilde{R}_{cf_3df_4}\nabla^a\nabla^c\phis M^{f_1i}M^{f_2j}M^{f_3k}M^{f_4l} \nabla_i\nabla_j\nabla_k\nabla_l\nabla^b\nabla^d\phis \nonumber\\
&\quad + \frac{G_{4c}}{48m^{10}}R_{af_1bf_2}R^{f_2}{}_{cdf_4}\nabla^a\nabla^c\phis M^{f_1i}M^{jk}M_{kl}M^{f_4n} \nabla_i\nabla_j\nabla^{l}\nabla_n\nabla^b\nabla^d\phis \nonumber\\
&\quad + \frac{G_{4d}}{48m^{10}}\widetilde{R}_{af_1bf_2}\widetilde{R}^{f_2}{}_{cdf_4}\nabla^a\nabla^c\phis M^{f_1i}M^{jk}M_{kl}M^{f_4n} \nabla_i\nabla_j\nabla^{l}\nabla_n\nabla^b\nabla^d\phis \nonumber\\
&\quad + \frac{G_{4e}}{48m^{10}}R_{af_1bf_2}R^{f_2}{}_{c}{}^{f_1}{}_{d}\nabla^a\nabla^c\phis M^{pi}M^{kj}M_{kl}M_{pn} \nabla_i\nabla_j\nabla^l\nabla^n\nabla^b\nabla^d\phis \nonumber\\
&\quad + \frac{G_{4f}}{48m^{10}}\widetilde{R}_{af_1bf_2}\widetilde{R}^{f_2}{}_{c}{}^{f_1}{}_{d}\nabla^a\nabla^c\phis M^{pi}M^{kj}M_{kl}M_{pn} \nabla_i\nabla_j\nabla^l\nabla^n\nabla^b\nabla^d\phis \, .
\nonumber
\end{align}



\bibliographystyle{JHEP}
\bibliography{ref}{}

\end{document}